\documentclass[11pt]{article}

\usepackage{subcaption}
\usepackage{caption}
\usepackage{adjustbox}

\usepackage[nohyperref]{jmlr2e}

\usepackage{breqn} 
\usepackage{mathtools}
\usepackage{mathrsfs}
\usepackage{bm}
\usepackage{centernot}
\usepackage{dsfont}

\newtheorem{defn}[theorem]{Definition} 

\usepackage{enumerate}

\usepackage{accents}

\usepackage[ruled,lined,nofillcomment]{algorithm2e}
\usepackage{setspace}
\usepackage{xcolor}

\SetCommentSty{mycommfont} 

\newcommand{\G}{\mathcal{G}}
\newcommand{\E}{\mathcal{E}}
\newcommand{\Eb}{\bm{\mathcal{E}^*}}
\newcommand{\I}{\mathcal{I}}
\newcommand{\V}{\mathcal{V}}
\newcommand{\iS}{\mathcal{S}}

\newcommand{\iSb}{\bm{\mathcal{S}}^* }

\newcommand{\X}{\mathcal{X}}
\newcommand{\U}{\mathcal{U}}
\newcommand{\Ib}{\bm{\mathcal{I}}^*}
\newcommand{\Pb}{\mathbf{P}}
\newcommand{\M}{\mathcal{M}}
\newcommand{\Mb}{\bm{\mathcal{M}^*}}

\newcommand{\data}{\{\iS^{(i)}\}_{i=1}^n}
\newcommand{\edata}{\{\E^{(i)}\}_{i=1}^n}
\newcommand{\adata}{\{\iS^*_i\}_{i=1}^n}
\newcommand{\eadata}{\{\E^*_i\}_{i=1}^n}
\newcommand{\gdata}{\{\G^{(i)}\}_{i=1}^n}
\newcommand{\gadata}{\{\G^*_i\}_{i=1}^n}

\newcommand{\modeprop}{[\iS^m]^\prime}
\newcommand{\modecurr}{\iS^m}
\newcommand{\emodeprop}{[\E^m]^\prime}
\newcommand{\emodecurr}{\E^m}

\DeclareMathOperator*{\argmax}{argmax}
\DeclareMathOperator*{\argmin}{argmin}

\usepackage{hyperref}
\usepackage{cleveref}

\usepackage[title, titletoc]{appendix}
\crefname{supp}{Supplement}{Supplements} 
\Crefname{supp}{Supplement}{Supplements} 

\Crefname{defn}{definition}{definition}
\Crefname{defn}{Definition}{Definition}

\begin{document}

\title{Modelling Populations of Interaction Networks via Distance Metrics}

\author{\name George Bolt \email g.bolt@lancaster.ac.uk\\
	\addr STOR-i Centre for Doctoral Training\\
	Lancaster University, Lancaster, UK, LA1 4YF
	\AND
	\name Sim\'{o}n Lunag\'{o}mez \email simon.lunagomez@itam.mx\\
	\addr Department of Statistics\\
	Instituto Tecnol\'{o}gico Aut\'{o}nomo de México (ITAM)\\
	R\'{i}o Hondo 1, Altavista, Álvaro Obreg\'{o}n, 01080 Ciudad de M\'{e}xico, CDMX, Mexico
	\AND
	\name Christopher Nemeth \email c.nemeth@lancaster.ac.uk\\
	\addr Department of Mathematics and Statistics\\
	Lancaster University, Lancaster, UK, LA1 4YF
}

\ShortHeadings{Modelling Populations of Interaction Networks}{Bolt, Lunag\'{o}mez and Nemeth}

\editor{}

\maketitle

\begin{abstract}
	Network data arises through observation of relational information between a collection of entities. Recent work in the literature has independently considered when (i) one observes a sample of networks, connectome data in neuroscience being a ubiquitous example, and (ii) the units of observation within a network are edges or paths, such as emails between people or a series of page visits to a website by a user, often referred to as interaction network data. The intersection of these two cases, however, is yet to be considered. In this paper, we propose a new Bayesian modelling framework to analyse such data. Given a practitioner-specified distance metric between observations, we define families of models through location and scale parameters, akin to a Gaussian distribution, with subsequent inference of model parameters providing reasoned statistical summaries for this non-standard data structure. To facilitate inference, we propose specialised Markov chain Monte Carlo (MCMC) schemes capable of sampling from doubly-intractable posterior distributions over discrete and multi-dimensional parameter spaces. Through simulation studies we confirm the efficacy of our methodology and inference scheme, whilst its application we illustrate via an example analysis of a location-based social network (LSBN) data set. 
\end{abstract}


\section{Introduction}
\label{sec:intro}

\begin{figure}
    \centering
	\begin{subfigure}{0.32\linewidth}
	    \centering
	    \includegraphics[page=1,width=0.8\textwidth]{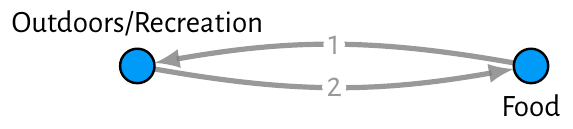}
	\end{subfigure}
	\begin{subfigure}{0.32\linewidth}
	    \centering
	    \includegraphics[page=2,width=0.8
	    \textwidth]{img/intro_plots.pdf}
	\end{subfigure}
	\begin{subfigure}{0.32\linewidth}
	    \centering
	    \includegraphics[page=3,width=0.65\textwidth]{img/intro_plots.pdf}
	\end{subfigure}
	\begin{subfigure}{0.32\linewidth}
	    \centering
	    \includegraphics[page=4,width=0.8\textwidth]{img/intro_plots.pdf}
	\end{subfigure}
	\begin{subfigure}{0.32\linewidth}
	    \centering
	    \includegraphics[page=5,width=0.8\textwidth]{img/intro_plots.pdf}
	\end{subfigure}
	\begin{subfigure}{0.32\linewidth}
	    \centering
	    \includegraphics[page=6,width=0.8\textwidth]{img/intro_plots.pdf}
	\end{subfigure}
	\caption{Data of a single user from the Foursquare data set \citep{Yang2015}. Going from left to right, the first five subplots show the observed interactions sequence $\iS$, with each path representing a single day of check-ins, whilst the final plot visualises the aggregate multigraph $\G_{\iS}$, where the thickness of an edge is proportional to the number of times  it appears in $\G_\iS$.}
	\label{fig:ex_interaction_seq}
\end{figure}

Network data, defined to be information regarding relations amongst some collection of entities, can appear in various guises, each with subtle idiosyncrasies warranting particular methodological considerations. This work considers the intersection of two such sub-types of network data. On the one hand, we consider when a sample of independent networks is observed. Such data, often assumed to be drawn from some population of networks, is becoming increasingly prevalent in the neuroscience literature \citep{Behrens2012,chung2021}, and has motivated recent methodological developments on multiple-network models \citep{Lunagomez2021,Le2018,Durante}. On the other hand, we assume that paths or edges represent the units of observation within each network. Typically referred to as interaction networks,
work has similarly been done towards defining methodologies suitable to their analysis, most notably with the proposal of so-called edge-exchangeable models \citep{Crane2018,Cai2016,caron2017}. As far as we are aware, the intersection of these two cases, that is, where one observes multiple independent interaction networks, is yet to be considered in the literature. 
As a motivating example, consider the Foursquare check-in data set of \cite{Yang2015}. Foursquare is a location-based social network (LSBN) where users share places they have visited with their friends by `checking-in' to locations they visit, such as restaurants or music venues. The data set of \cite{Yang2015} contains historical check-ins of users in New York and Tokyo. Considering a single user, we can see a day of check-ins as a path through the set of venue categories, as illustrated in \Cref{fig:ex_interaction_seq}. Over an extended time period, we expect to observe check-ins on multiple days, leading to a series of paths being observed. In this way, the data of a single user can be seen as an interaction network. Furthermore, there are data for not one but possibly hundreds of users, that is, we have multiple interaction networks.

With a single observation, we are typically interested in analysing its structure. When faced with an independent sample of networks the following more familiar statistical questions are raised

\begin{enumerate}
	\item[(a)] What is an average network?
	\item[(b)] How variable are observations about this average?
	\item[(c)] Is there heterogeneity in the observations?
\end{enumerate}

These questions are consistent with those arising more generally in object orientated data analysis \citep{marron2021}, which considers the statistical analysis of populations of complex objects. In this work, we focus on (a) and (b) in particular.


\subsection{Related Work}

We now review closely related work appearing in the literature. Firstly, there has been work on models suitable for multiple-network data, where observations are typically represented via graphs. Here, \cite{Lunagomez2021} construct models through graph distances, using the Fr\'{e}chet mean and entropy as notions of the mean and variance respectively. \cite{Le2018}, \cite{peixoto2018me} and \cite{Newman2018} propose measurement error models which view observed networks as noisy realisations of an unknown ground truth. Along similar lines, \cite{Mantziou2021} and \cite{young2022} have recently extended the measurement error models to capture heterogeneity, providing model-based approaches to clustering networks. 

Others have considered adapting models originally proposed to analyse a single network. This includes the latent space model (LSM) \citep{Hoff2002}, which has been extended by \cite{sweet2013}, who assume a hierarchical model in which each observation is drawn from an LSM with its own parameter, with these parameters being linked via a prior, \cite{Gollini2016}, who assume observations share the same latent coordinates, and \cite{Durante}, who take a non-parametric approach, using a mixture of LSMs combined with shrinkage priors which induce removal of redundant components and unnecessary dimensions in latent coordinates. The random dot product graph (RDPG) model \citep{young2007} has also been extended. Here \cite{levin2017} assume observations are drawn i.i.d. from the same RDPG model, whilst \cite{Nielsen2018}, \cite{wang2019} and \cite{arroyo2021} consider relaxing this i.i.d. assumption, constructing their models to permit variation in the RDPG parameters across observations, better capturing heterogeneity. The exponential random graph model (ERGM) \citep{holland1981,frank1986} has similarly been extended, where \cite{Lehmann2021} consider a hierarchical model \cite[in similar spirit to][]{sweet2013}, whilst \cite{Yin2022} consider a finite mixture of ERGMs. Finally, others have adapted the stochastic blockmodel (SBM) \citep{Nowicki2001}, with \cite{sweet2014} building upon their earlier work \citep{sweet2013}, assuming a hierarchical model where each observation is drawn from an SBM with its own parameterisation, whilst \cite{stanley2016} and \cite{reyes2016} consider mixtures of SBMs. 

There has also been work on hypothesis testing for network-valued data \citep{Ginestet2014,durante2018testing,ghoshdastidar2020,chen2021hypothesis}, where we note in particular \cite{Ginestet2014} make use of a distance between graphs, in similar spirit to \cite{Lunagomez2021} and this present work. 

All of these works are connected by a desire to answer standard statistical questions in the context of network-valued data. However, none consider paths being the units of observation within each network. Instead, all are designed to analyse network data represented via graphs. As such, to analyse data which is truly path-observed, their use would require first aggregating observations to graphs in a pre-processing step; an operation which will often not be injective and hence lead to a potential loss of information. 


In another direction, there is related work which has considered edge or path-observed network data. In particular, there has been development of so-called edge-exchangeable network models \citep{Cai2016,Crane2018,Williamson2016,ghalebi2019dynamic,ghalebi2019dnnd}. Of these we note that only \cite{Crane2018} allow paths as the observational units. Closely related to the edge-exchangeable models are those based on exchangeable random measures \citep{caron2017,veitch2015}. These two streams of work are connected in so far as they deviate from more traditional models which are based upon assumptions of vertex exchangeability, and in doing so produce graphs which exhibit sparsity and heavy-tailed degree distributions; features often observed empirically. In another direction, others have considered models based upon higher-order Markov chains \citep{Scholtes2017,peixoto2017sequences}, where in line with this present work \cite{Scholtes2017} considered paths as the units of observation. 

The common theme in all these works is a focus on models which can capture a particular structure within a single network, such as sparsity, heavy-tailed degree distributions or high-order dependence in paths or edges. They are not, however, designed to analyse multiple observations. As such, they can only provide answers to (a)-(c) via post-hoc analysis of parameters inferred for each observation.

\subsection{Summary of Contributions}

In the literature, it appears there is a present lack of methods to analyse samples of interaction networks which fully respect the structure of the data; either one converts observations to graphs, possibly disregarding information, or one models each observation individually. We look to address this gap. To this end, we propose a new Bayesian modelling framework. Namely, through use of distance metrics between observations, we construct families of models via location and scale parameters, akin to a Gaussian distribution. The location parameter, itself an interaction network, admits an interpretation analogous to the mean, whilst the scale parameter can be seen as a notion of variance or precision. Conducting inference of these parameters thus provides a reasoned approach to answering questions (a) and (b). Our methodology is intended to work with any distance metric of the practitioners choosing, leading to a flexible framework which can be tailored to suit different questions of interest. We supplement our methodology with specialised Markov chain Monte Carlo (MCMC) algorithms which facilitate parameter inference. In particular, by merging the recently proposed involutive MCMC (iMCMC) framework of \cite{Neklyudov2020} with the exchange algorithm of \cite{Murray2006}, we are able to not only circumvent issues pertaining to doubly-intractable posteriors, but also navigate a multi-dimensional
discrete parameter space.

The remainder of this paper will be structured as follows. In \Cref{sec:background}, we provide background details regarding the data structure and notation. Then, in \Cref{sec:modelling} we formally introduce our methodology, stating proposed models and providing examples of distance metrics. In \Cref{sec:inference}, we outline our Bayesian approach to inference, discussing prior specification, stating our assumed hierarchical model and detailing our proposed MCMC scheme. In \Cref{sec:simulations}, we detail simulation studies undertaken to confirm the efficacy of our methodology and posterior inference scheme, whilst in \Cref{sec:data_analysis} we illustrate its applicability via an analysis of the Foursquare check-in data. We finalise with conclusions and discussion in \Cref{sec:discussion}.

\section{Data Representation}

\label{sec:background}

Due to the nature of the data, observed paths often arrive in a known order and, depending on the questions one would like to ask, it may or may not be desirable to encode this in our representation. For example, consider question (a) of \Cref{sec:intro}. When looking to find an average, do we want to take the observed order into account, finding an average \textit{sequence} of paths? Or do we want to disregard the order information, and instead find an average \textit{set} of paths? To cover both situations, we propose two data representations which will be used within our framework. Namely, what we call \textit{interaction sequences} and \textit{interaction multisets}, covering the ordered and un-ordered cases respectively. 

Adopting the vernacular of network data analysis, we formally denote the collection of entities under consideration via a \textit{vertex set} $\V$, assumed to be some discrete set (typically the set of integers $\V = \{1,\dots,V\}$) where we let $V= |\V|$ denote the number of vertices. For example, in the Foursquare check-in data (\Cref{fig:ex_interaction_seq}), $\V$ would represent the venue categories. An interaction sequence will be denoted as follows
\begin{equation*}
    \begin{aligned}
        \iS = (\I_1,\dots,\I_N)
    \end{aligned}
\end{equation*}
where the $\I_i$ will be referred to as \textit{interactions}. We consider the case where interactions are represented via \textit{paths}, that is
\begin{equation*}
    \begin{aligned}
        \I_i = (x_{i1},\dots,x_{in_i}) 
    \end{aligned}
\end{equation*}
where $x_{ij}\in\V$, and thus $\iS$ is a sequence of paths. Returning to the Foursquare example, $\I_i$ would denote a single day of check-ins, where this user started at a venue of category $x_{i1}$ then moved to category $x_{i2}$ and so on, with $\iS$ denoting all the observed check-ins of a single user over some fixed time period. Furthermore, the ordering of interactions reflects the order in which they were observed, for example, $\I_1$ appears before $\I_2$ and so on.

When the order of interaction arrival is not of interest, we instead represent the data via an interaction \textit{multiset}. A multiset is a set with multiplicities, that is, a set where elements can occur more than once, and is the natural order-invariant generalisation of a sequence. An interaction multiset will be denoted as follows
\begin{equation*}
    \begin{aligned}
        \E = \{\I_1,\dots,\I_N\},
    \end{aligned}
\end{equation*} 
where the $\{ \}$ parenthesis signify this is a multiset, where $\I_i$ similarly denote paths. For example, regarding the Foursquare data, this would simply represent the collection of observed check-ins for a single user, with the order of interactions as written above implying nothing with regards to the order of interacton arival, for example, $\I_1$ was not necessarily observed before $\I_2$.
We note this representation has strong similarities with that used in \cite{Crane2018}. In particular, what is defined therein as an interaction network can be seen as a countably infinite interaction multiset. 


In this work, we consider the case where \textit{multiple} interaction sequences or multisets are observed. For example, in the Foursquare dataset we have check-in information on not one but many users, and thus observe a sample of interaction sequences or multisets, one for each user. Representing the $i$th observation via a interaction sequence $\iS^{(i)}$ or multiset $\E^{(i)}$, and letting $n$ denote the sample size, we therefore observe 
\begin{equation*}
    \begin{aligned}
        \iS^{(1)},\dots,\iS^{(n)} && \text{or} && \E^{(1)},\dots,\E^{(n)},
    \end{aligned}
\end{equation*}
where the choice of representation depends on ones interest in order. Our methodology will provide a means to analyse such samples of data. 

We finish this subsection by discussing aggregation. Both multisets and sequences of paths can be aggregated to form graphs, and use of any currently proposed multiple-network methodology would actually necessitate this as a pre-processing step. A \textit{graph} $\G = (\E_\G, \V_\G)$ consists of a set $\V_\G$ of \textit{vertices} and a set $\E_\G$ of \textit{edges} where $e = (i,j) \in \E_\G$ if there is an edge from vertex $i\in \V_\G$ to $j \in \V_\G$. Graphs can be un-directed, where $(i,j) \in \E_\G \iff (j,i) \in \E_\G$, or they can be directed, where $(i,j) \in \E_\G$ need not imply $(j,i) \in \E_\G$ (and \textit{vice versa}). Any interaction sequence $\iS$ over vertices $\V$ can be aggregated to form a directed graph $\G_\iS=(\E_\iS, \V)$ as follows: let $(i,j) \in \E_\iS$ if the traversal from vertex $i$ to vertex $j$ was observed at least once in $\iS$, that is, if $x_{kl}=i$ and $x_{k(l+1)}$ for some $1 \leq k \leq N$ and $1 \leq l \leq n_k$. 

Within an interaction sequence, a traversal between two vertices may occur more than once. Thus we can also aggregate to form a so-called \textit{multigraph}, where edges can appear any number of times. We denote a multigraph similarly via $\G = (\E_\G, \V_\G)$, where in this case $\E_\G$ is a \textit{multiset} of edges, that is, any given edge $(i,j)\in \V\times \V$ can appear in $\E_\G$ multiple times. Any interaction sequence $\iS$ over vertices $\V$ can be aggregated to form a directed multigraph $\G_\iS=(\E_\iS, \V)$ by including the edge $(i,j)$ each time the traversal from vertex $i$ to vertex $j$ is observed, or more formally one can let
$$\E_\iS = \left\{ \left(x_{kl},x_{k(l+1)}\right) \, : \, 1 \leq k \leq N, \, 1 \leq l \leq (n_k-1)\right\},$$
for example, an aggregate multigraph for the Foursquare check-in data of a single user can be seen in \Cref{fig:ex_interaction_seq}.


One can also aggregate an interaction mutliset $\E$ over vertex set $\V$ to form a graph or multigraph in the same manner. Suppose $\tilde{\iS}$ is an interaction sequence obtained by placing the paths of $\E$ in arbitrary order, then we let $$\G_\E = \G_{\tilde{\iS}} = (\E_{\tilde{\iS}}, \V),$$
which applies equally to the definition of the graph or multigraph, where in the former case $\E_{\tilde{\iS}}$ will be a set of edges, whilst in the latter it will be a multiset of edges.

Observe that aggregation is not injective, that is, one may have $\iS \not = \iS^\prime$ (respectively $\E \not = \E^\prime)$ whilst $\G_\iS = \G_{\iS^\prime}$ (respectively $\G_\E = \G_{\E^\prime}$). For interaction sequences, a trivial case would be a re-ordering of paths. This, however, is not the only example, and there can instead be interaction sequences or multisets which are structurally dissimilar but nonetheless have equivalent aggregate graphs. In this way, aggregation incurs a loss of information that will be avoided with our methodology.

\section{Metric-Based Interaction-Network Models}
\label{sec:modelling}

We now introduce our proposed models for interaction sequences and multisets respectively. The core idea behind both is an assumption that observed data points are `noisy' realisations of some (unknown) ground truth, where quantification of this noise is facilitated via a pre-specified distance metric. Equivalently, they can be seen as Gaussian-like distributions over their respective discrete metric spaces, controlled by a location parameter, itself an interaction sequence or multiset, and a real-valued scale parameter. 

\subsection{Interaction-Sequence Models}
\label{sec:sequences}

We let $\iSb$ denote the infinite discrete space consisting of all interaction sequences over the fixed vertex set $\V$, further details of which can be found in \Cref{sec_sup:sample_spaces}. Towards eliciting a probability distribution over $\iSb$, we first endow it with a distance metric $d_S : \iSb \times \iSb \to \mathbb{R}_+$, defining a metric space $(\iSb, d_S)$. We then select an element of the space $\iS^m \in \iSb$, referred to as the \textit{mode}, upon which to center the distribution, before choosing $\gamma >0$, referred to as the \textit{diffusion}, controlling the concentration of probability mass in $\iSb$ about the mode $\iS^m$. In this way, $\iS^m$ and $\gamma$ can be seen as location and scale parameters, respectively, analogous to the mean and variance of a Gaussian distribution. A family of probability distributions can now be defined as follows.

\begin{defn}[Spherical Interaction Sequence Family]
    \label[defn]{def:SIS}
	For a given metric $d_S(\cdot, \cdot)$ on $\iSb$, a mode $\iS^m \in \iSb$ and a diffusion parameter $\gamma > 0$, the probability of observing $\iS$ is given by
	\begin{equation}
	\label{eq:SIS}
		p(\iS \, | \, \iS^m, \gamma) \propto \exp\{ -\gamma d_S(\iS, \iS^m)\},
	\end{equation}
	and we write 
	\begin{equation*}
	    \label{eq:SIS_sim}
	    \iS \sim \text{\normalfont SIS}(\iS^m, \gamma)
	\end{equation*}
	if we assume $\iS$ was sampled via (\ref{eq:SIS}). This we refer to as the Spherical Interaction Sequence (SIS) family of probability distributions over $\iSb$ with parameters $\iS^m$ and $\gamma$.
\end{defn}

\subsection{Interaction-Multiset Models}
\label{sec:multisets}

For interaction multisets, the family of models we propose are more-or-less a carbon copy of the SIS models, albeit with slightly different notation. In this case, $\Eb$ will denote the sample space; here consisting of all interaction multisets over the fixed vertex set $\V$ (further details provided in \Cref{sec_sup:sample_spaces}). Again, we endow $\Eb$ with a distance metric $d_E \, : \Eb \times \Eb \to \mathbb{R}_+$, similarly defining a metric space $(\Eb, d_E)$. We then construct a distribution over $\Eb$ via location and scale in exactly the same way, where in this case our location parameter will be an interaction \textit{multiset} $\E^m \in \Eb$. The resultant family of probability distributions is defined as follows.

\begin{defn}[Spherical Interaction Multiset Family]
    \label[defn]{def:SIM}
	Given a metric $d_E(\cdot, \cdot)$ on $\Eb$, a mode $\E^m \in \Eb$,  and a diffusion parameter $\gamma > 0$, the probability of observing $\E$ is given by
	\begin{equation}
	\label{eq:SIM}
		p(\E \, | \, \E^m, \gamma) \propto \exp\{ -\gamma d_E(\E, \E^m)\},
	\end{equation}
	and we write 
	\begin{equation*}
	    \label{eq:SIM_sim}
	    \E \sim \text{\normalfont SIM}(\E^m, \gamma)
	\end{equation*}
	if we assume $\E$ was sampled via (\ref{eq:SIM}). This we refer to as the Spherical Interaction Multiset (SIM) family of probability distributions over $\Eb$ with parameters $\E^m$ and $\gamma$.
\end{defn}

Note that in (\ref{eq:SIS_sim}) and (\ref{eq:SIM_sim}) no reference is made to $d_S(\cdot, \cdot)$ or $d_E(\cdot, \cdot)$, even though the respective distributions clearly depend on them. The reasoning here is these values are not intended to be model parameters but instead subjective choices made by the practitioner prior to any analysis.

\subsection{Properties of Interaction-Network Models}
\label{sec:models_properties}

We now highlight some key aspects of both model families. Being almost identical, these will be shared between the two, and thus for brevity we here provide details regarding the interaction-sequence models only, noting that the same properties will hold for the multiset models, up to a change of notation. 

Firstly, distributions of the form (\ref{eq:SIS}) agree with the intuition of location and scale. In particular, for fixed $d_S(\cdot,\cdot)$, an observation $\iS$ has the highest probability when $\iS = \iS^m$, and as $\iS$ moves further away from $\iS^m$, that is, as $d_S(\iS, \iS^m)$ increases, we see a decay in probability. Thus, $\iS^m$ indeed represents the mode of the distribution and can be seen as controlling the location of probability within $\iSb$. Furthermore, the rate at which the probability decreases with respect to distance is controlled by $\gamma$, with larger values leading to a faster decrease, implying a greater concentration of probability mass about the mode. In this way, $\gamma$ is analogous to the inverse of the variance, often referred to as the \textit{precision}. 

This latter aspect, the control of variance by $\gamma$, can be formalised as in \cite{Lunagomez2021} via the \textit{entropy}, defined to be 
\begin{equation}
    \begin{aligned}
        H(\iS^m ,\gamma) &:= - \mathbb{E}\left[\log p(\iS \, | \, \iS^m, \gamma)\right],
    \end{aligned}
\end{equation} 
which can be seen to quantify the uniformity of $p(\iS|\iS^m, \gamma)$, whereby larger values of $H(\iS^m, \gamma)$ imply this distribution is `more uniform' over $\iSb$, with a minimum value of $H(\iS^m, \gamma)=0$ attained by a pointmass. The entropy also has an interpretation with regards to randomness or variance, whereby distributions with a higher entropy are more random, that is more variable. It can be shown that with any $d_S(\cdot, \cdot)$ and $\iS^m$, the entropy $H(\iS^m,\gamma)$ is a monotonic function of $\gamma$ (\Cref{sec_sup:monotonicity}). Consequently, one can say $\gamma$ controls the variance of the distribution, as formalised via the entropy.

The distribution as stated in \Cref{def:SIS} is normalised as follows
\begin{equation}
    p(\iS \, | \, \iS^m, \gamma) = Z(\iS^m, \gamma)^{-1} \exp\{-\gamma d_S(\iS, \iS^m)\}
    \label{eq:SIS_normalised}
\end{equation}
where
\begin{equation*}
    \label{eq:sis_norm_const}
    Z(\iS^m, \gamma) = \sum_{\iS \in \iSb} \exp\{-\gamma d_S(\iS, \iS^m)\}
\end{equation*}
is the normalising constant, often referred to as the \textit{partition function}. In general, this summation is intractable; an aspect which will come into play significantly when we consider the computational aspects of the methodology in later sections. In fact, with $\iSb$ being infinite, there is no guarantee that (\ref{eq:sis_norm_const}) will even exist for a given $\gamma$. Consequently, for some parameterisations \eqref{eq:SIS_normalised} may be an improper distribution. Pragmatically, this can cause issues when sampling from these models, whereby one can observe a divergence in dimension for low values of $\gamma$. However, this will generally not be an issue provided one chooses $\gamma$ sufficiently large. In any case, it can be helpful to constrain the sample space to be finite to ensure existence of normalising constants, details of which can be found in \Cref{sec_sup:sample_spaces_bounded}.

We conclude this section by highlighting similarities of our models with those seen in other applications. Firstly, as already touched upon, both families are a direct extension of the spherical network family (SNF) proposed by \cite{Lunagomez2021} for graphs. Also, our assumption that observed data points are `noisy' realisations of an unknown ground truth is very much in keeping with the measurement error models for graphs \citep{Le2018,peixoto2018me,Newman2018}. Finally, there are connections with the complex Watson distribution \citep{Mardia1999}, used in the context of statistical shape analysis, and the Mallows model \citep{Vitelli2018a} used for preference learning.

\subsection{Distance Metrics}

\label{sec:metrics}

A fundamental aspect of our modelling framework is the use of distance metrics. Not only are these key in facilitating the construction of both model families but they also represent an important modelling choice. In this section, examples of two distances measures usable within our framework will be given for interaction sequences and multisets, respectively.

\begin{figure}
    \centering
    \begin{subfigure}{0.4\linewidth}
        \centering
        \includegraphics[page=3]{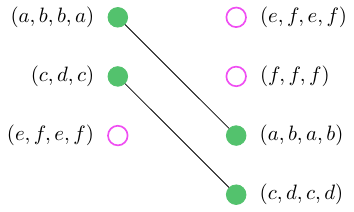}
        \caption{monotone matching}
    \end{subfigure}
    \begin{subfigure}{0.45\linewidth}
        \centering
        \includegraphics[page=2]{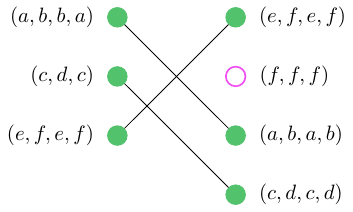}
        \caption{general matching}
    \end{subfigure}
    \caption{A comparison of matchings. In both (a) and (b) we have two interaction sequences displayed displayed top-down. Here (a) shows a \textit{monotone} matching of interactions, where the order is preserved, whilst (b) shows a general matching, where any two interactions can be paired together.}
    \label{fig:distance_examples}
\end{figure}

For interaction sequences, one can consider the so-called \textit{edit distance} \cite[Definition 4.1]{Bolt2022}. This admits two interpretations: the minimum cost of transforming one sequence into the other, or the cost of an optimal pairing of paths across the two sequences.
It is the latter interpretation that we formally define. Suppose $\iS$ and $\iS^\prime$ denote two interaction sequences, then a pairing of their paths can be encoded via a \textit{matching}, defined to be a multiset of pairs
$$\M = \{(\I, \I^\prime) \, : \, \I\in \iS, \I^\prime \in \iS^\prime\},$$
satisfying certain properties. For example, each interaction of one observation may be paired with at most one of the other and \textit{vice versa}. For comparison of sequences, it makes to sense to further constrain the structure of matchings so as to ensure the preservation of order. This is achieved by allowing only so-called \textit{monotone} matchings (see \Cref{fig:distance_examples}). 

The edit distance between $\iS$ and $\iS^\prime$ is now defined as follows
\begin{equation}
    \begin{aligned}
        d_{\text{Edit}}(\mathcal{S},\mathcal{S}^\prime ) := \min_{\mathcal{M} \in \Mb_m }\left\{ \sum_{(\I, \I^\prime) \in \mathcal{M}} d_I(\mathcal{I}, \mathcal{I}^\prime) + \sum_{\I \in \mathcal{M}^c_{\mathcal{S}}  }d_I(\mathcal{I}, \Lambda) +  \sum_{\I^\prime \in \mathcal{M}^c_{\mathcal{S}^\prime}  }d_I(\Lambda, \mathcal{I}^\prime)\right\}
    \end{aligned}
    \label{eq:edit_distance}
\end{equation}
where $\Mb_m$ is the set of monotone matchings between the two sequences; $\M_{\iS}$ is the restriction of $\M$ to $\iS$, that is, the set of interactions in $\iS$ included in the matching, with $\M^c_\iS:=\iS \setminus \M_\iS$ the interactions of $\iS$ \textit{not} included in $\M$; $d_I$ is some distance metric between interactions, that is, paths; and $\Lambda$ denotes the empty interaction. That is, we sum the pairwise distances of matched interactions (equivalent to the cost of transforming interactions to one another), and then sum penalisations for unmatched interactions, where an unmatched interaction $\I$ incurs penalty $d_I(\I,\Lambda)$ (equivalent to the cost of inserting or deleting $\I$). Computation of $d_{\text{Edit}}$ can be achieved via dynamic programming \citep{Bolt2022} at a complexity of $\mathcal{O}(N M)$, where $N$ and $M$ are the lengths of $\iS$ and $\iS^\prime$ respectively. 

For interaction multisets, we consider the \textit{matching distance} \cite[Definition 3.1]{Bolt2022}. This has close connections with the edit distance, with the interpretation being more-or-less identical: on the one hand, the minimum cost of editing one multiset to get the other, whilst on the other an optimal matching of interactions across the two observations. The differential factor is an absence of order information, making consideration of monotonicity in the matching unnecessary. Given two interaction multisets $\E$ and $\E^\prime$, the matching distance is defined as follows
\begin{equation}
    \begin{aligned}
        d_{\text{M}}(\mathcal{E},\mathcal{E}^\prime) := \min_{\mathcal{M}\in \Mb}\left\{ \sum_{(\I, \I^\prime) \in \M} d_I(\mathcal{I}, \mathcal{I}^\prime) + + \sum_{\I \in \mathcal{M}^c_{\E}  }d_I(\mathcal{I}, \Lambda) +  \sum_{\I^\prime \in \mathcal{M}^c_{\E^\prime}  }d_I(\Lambda, \mathcal{I}^\prime) \right\}
    \end{aligned}
\end{equation}
where $\Mb$ is the set of matchings between the two multisets (with no monotonicity constraints), whilst $\M_{\E}$ is the restriction of $\M$ to $\E$, that is, the subset of interactions from $\E$ included in $\M$, with $\M^c_\E := \E \setminus \M_\E$ again denoting the interactions not included in the matching $\M$. Computation of $d_{\text{M}}$ can be achieved via the Hungarian algorithm \citep{Kuhn1955}, with a complexity of $\mathcal{O}(\max(N,M)^4 + N M)$, where $N$ and $M$ denote the cardinalities of $\E$ and $\E^\prime$ respectively \citep{Bolt2022}.

Use of both $d_{\text{Edit}}$ and $d_{\text{M}}$ requires specification of a distance $d_I$ between paths. Here we provide examples of two such distances, namely the longest common subpath (LSP) and longest common subsequence (LCS) distances. The core idea of both is to quantify the dissimilarity of two paths by considering how much they have (or do not have) in common.

\begin{figure}
    \centering
    \begin{subfigure}{0.48\linewidth}
        \centering
        \includegraphics[width=0.6\textwidth,page=2]{img/distances_paths.pdf}
        \caption{Common subpath}
        \label{fig:path_distances_subpath}
    \end{subfigure}
    \begin{subfigure}{0.48\linewidth}
        \centering
        \includegraphics[width=0.6\textwidth,page=1]{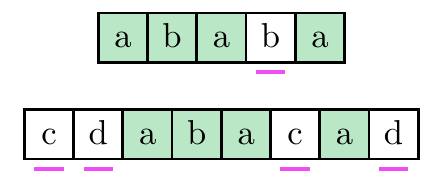}
        \caption{Common subsequence}
        \label{fig:path_distances_subseq}
    \end{subfigure}
    \caption{A comparison of common subsequences and subpaths. In (a) and (b) we see the same pair of paths, with (a) highlighting a common subpath, as indicated by shaded (green) entries, whilst (b) shows a common subsequence. In both cases, these are in fact maximal.}
    \label{fig:path_distances}
\end{figure}

For $\I = (x_1,\dots, x_n)$ a \textit{subpath} of $\I$ from index $i$ to $j$ is denoted as follows
$$\I_{i:j} = (x_i, \dots, x_j) $$
where $1 \leq i \leq j \leq n$. Given another path $\I^\prime=(y_1,\dots,y_m)$ one can consider finding subpaths that both $\I$ and $\I^\prime$ have in common, that is
$$\I_{i : j} = \I^\prime_{k:l}$$
where $1 \leq i \leq j \leq n$ and $1 \leq k \leq l \leq m$ (\Cref{fig:path_distances_subpath}). Alternatively, one can consider common subsequences. Assuming that $\bm{v} = (v_1, \dots, v_s)$ with $1 \leq v_1 < v_2 < \dots < v_s \leq n$, then the \textit{subsequence} of $\I$ indexed by $\bm{v}$ is given by
$$\I_{\bm{v}} = (x_{v_1}, \dots, x_{v_s}),$$
and one can similarly consider finding subsequences that both $\I$ and $\I^\prime$ have in common, that is 
$$\I_{\bm{v}} = \I^\prime_{\bm{u}}$$
where $\bm{u}=(u_1,\dots,u_r)$ is such that $1 \leq u_1 < u_2 < \dots < u_r \leq m$ and $\bm{v}$ is as above (\Cref{fig:path_distances_subseq}). 


Towards defining a distance, one can consider finding a maximal subpath or subsequence, that is, one such that there exists no other subpath or subsequence of greater length. For example, the common subpaths and subsequences of \Cref{fig:path_distances} are maximal. Intuitively, a distance is now defined by counting the number of entries not included in the common subpath (or subsequence) from either $\I$ or $\I^\prime$. 

More formally, the two distances can be defined as follows
\begin{equation*}
    \begin{aligned}
        d_{\text{LSP}}(\I, \I^\prime) := n + m - 2\delta_{\text{LSP}} && && &&  d_{\text{LCS}}(\I, \I^\prime) := n + m - 2\delta_{\text{LCS}}
    \end{aligned}
\end{equation*}
where $\delta_{\text{LSP}}$ and $\delta_{\text{LCS}}$ denote the maximum common subpath and subsequence lengths respectively. This equates to counting the underlined entries of \Cref{fig:path_distances}, whereby we consider the $n+m$ total entries (of both paths), and then remove the double counting made for the entries of the common subpath or subsequence. As an example, in \Cref{fig:path_distances} we have $\delta_{\text{LSP}}=3$ and $\delta_{\text{LCS}}=4$, and thus $d_{\text{LSP}}(\I,\I^\prime)=7$ whilst $d_{\text{LCS}}(\I,\I^\prime)=5$.

Computation of these distances reduces to finding the maximum length common subpaths and subsequences, both of which can be achieved via dynamic programming with a complexity $\mathcal{O}(n m)$. Further details on the implementation can be found in \cite{Bolt2022}, Appendix C.

\subsection{Modelling Choices}

\label{sec:modelling_choices}

A key modelling choice is whether to opt for an SIS or SIM model. This should be framed around ones interest in order information, whereby if one would like to take order into account then an SIS model should be assumed, choosing an SIM model in the converse case. This can be equated with making a form of exchangeability assumption reminiscent of edge-exchangeable network models \citep{Cai2016,Crane2018}, where in assuming an SIM model one is considering observations equal up to a permutation of path order as having the same probability of being sampled. 

Another important modelling decision is the choice of distance metric. This will imply certain assumptions regarding the structure of noise in the model, in turn altering the interpretations of inference. For example, taking $d_I = d_{\text{LSP}}$ within either the edit or matching distances of \Cref{sec:metrics} will imply samples drawn from the SIS or SIM models will contain paths which share common subpaths with those of the mode, whilst if one took $d_I=d_{\text{LCS}}$ then samples would instead share common subsequences with the mode. As a consequence, when inferring model parameters, in particular the mode, the interpretation thereof will differ. In particular, in assuming an LSP distance between paths the inferred mode will likely consist of a multiset or sequence of subpaths often seen together in the observed data, whilst if we assume an LCS distance we will instead uncover a multiset or sequence of often-observed subsequences.


\section{Bayesian Inference}

\label{sec:inference}

Given an assumed model, the goal of inference is to discern which parameters are likely to have generated the observed data. In our case, this amounts to inferring the mode and dispersion parameters. We approach this task via a Bayesian perspective, first assuming prior distributions for model parameters before incorporating observed data to form the posterior. Through a specialised MCMC algorithm, we subsequently obtain samples from the posterior upon which to base our inference. In this section, we provide details regarding each of these aspects.

For brevity, we give details regarding the interaction-sequence models (\Cref{sec:sequences}) only, noting the approach taken for the multiset models is almost identical, albeit with a change of notation and some minor alterations to the MCMC algorithms. Full details of inference for the interaction-multiset models analogous to those given below can be found in \Cref{sec_sup:inference_sim}.

\subsection{Priors, Hierarchical Model and Posterior}

\label{sec:inference_priors}
In specifying a prior for the mode we follow \cite{Lunagomez2021} and assume it was itself sampled from an SIS model, that is
\begin{equation}
    \label{eq:SIS_mode_prior}
    \begin{aligned}
        \iS^m \sim \text{SIS}(\iS_0, \gamma_0)
    \end{aligned}
\end{equation}
where $(\iS_0, \gamma_0)$ are specified hyperparameters. For the dispersion $\gamma$ we simply require a distribution $p(\gamma)$ whose support is a subset of the non-negative reals. For example, we typically take
$\gamma \sim \text{Gamma}(\alpha_0, \beta_0)$ with $(\alpha_0,\beta_0)$ being hyperparameters. Given these specifications, an observed sample $\data$ is thus assumed to be drawn via
\begin{equation}
    \begin{aligned}
        \iS^{(i)} \, | \, \iS^m , \gamma &\sim \text{SIS}(\iS^m, \gamma) & (\text{for }i=1,\dots,n)\\
        \iS^m &\sim \text{SIS}(\iS_0, \gamma_0) & \\
        \gamma &\sim p(\gamma). &
    \end{aligned}
\end{equation}
The likelihood of the sample $\data$ is given by
\begin{equation*}
    \begin{aligned}
        p(\{\iS^{(i)}\}_{i=1}^n \, | \, \iS^m, \gamma) &= \prod_{i=1}^n p(\iS^{(i)} \, | \, \iS^m, \gamma) \\ 
        &= Z(\iS^m, \gamma)^{-n}\exp\left\{ - \gamma \sum_{i=1}^n d_S(\iS^{(i)}, \iS^m)\right\},
    \end{aligned}
\end{equation*}
and we have the following posterior, up to a constant of proportionality
\begin{equation}
    \label{eq:SIS_posterior}
    \begin{aligned}
        p(\iS^m, \gamma \, | \, \{\iS^{(i)}\}_{i=1}^n ) &\propto p(\{\iS^{(i)}\}_{i=1}^n \, | \, \iS^m, \gamma) p(\iS^m) p(\gamma) \\
        &\propto Z(\iS^m, \gamma)^{-n}\exp\left\{ - \gamma \sum_{i=1}^n d_S(\iS^{(i)}, \iS^m)\right\} \\
        &\qquad\times \exp\{-\gamma_0 d_S(\iS^m, \iS_0)\} p(\gamma).
    \end{aligned}
\end{equation}


\subsection{Sampling from the Posterior}
\label{sec:inference_joint}
\begin{figure}
    \centering
    \includegraphics[width=0.9\linewidth, page=2]{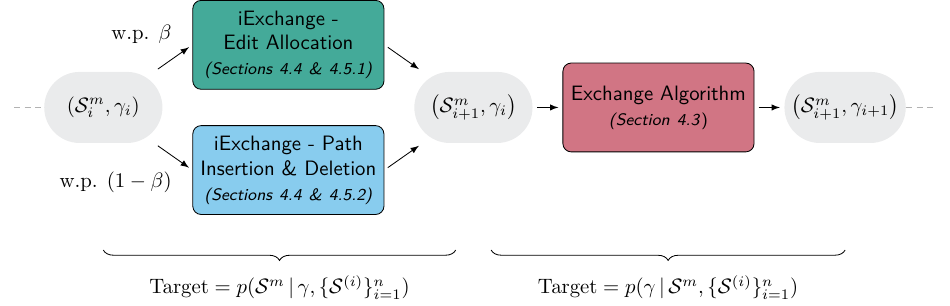}
    \caption{Summary of our MCMC scheme to sample from the SIS posterior. We first update the mode via the iExchange algorithm, doing an edit allocation move with probability $\beta$, or a path insertion and deletion move otherwise. We then update the dispersion via the exchange algorithm.}
    \label{fig:posterior_sampler_summary}
\end{figure}

To sample from the posterior \eqref{eq:SIS_posterior}, we use a component-wise MCMC algorithm which alternates between sampling from the two conditional distributions
\begin{equation}
    \label{eq:SIS_posterior_cond}
    \begin{aligned}
        p(\iS^m \, | \, \gamma, \data) && \text{and} && p(\gamma \, | \, \iS^m, \data).
    \end{aligned}
\end{equation}
Since the normalising constant $Z(\iS^m, \gamma)$ depends on the parameters of interest this implies \eqref{eq:SIS_posterior} is doubly intractable \citep{Murray2006}. Such terms will also persist in both conditionals above, making them also doubly intractable. This precludes the use of standard MCMC algorithms such as Metropolis-Hastings and necessitates use of the exchange algorithm proposed by \cite{Murray2006}. 

A high-level summary of our scheme is visualised in \Cref{fig:posterior_sampler_summary}. For the dispersion conditional, being a distribution over the real line, we can apply the exchange algorithm directly. In contrast, the mode $\iS^m$ is a discrete object, the dimensions of which can vary both in terms of the number of paths and their lengths. This makes the sample space for the mode conditional far less trivial, and so we consider merging the exchange algorithm with the involutive MCMC (iMCMC) framework of \cite{Neklyudov2020}; defining what we call the iExchange algorithm. To fully explore the sample space, we mix together two iExchange moves. In particular, with probability $\beta$, we enact a move perturbing the paths currently present, whilst with probability $(1-\beta)$ we attempt a move which varies the number of paths.



\subsection{Updating the Dispersion}

\label{sec:inference_dispersion_cond}

Here we outline our MCMC scheme to sample from the dispersion conditional. In this instance, we suppose $\iS^m$ is fixed and $q(\gamma^\prime|\gamma)$ is some proposal density. In a single iteration, given current state $\gamma$ we do the following
\begin{enumerate}
    \item Sample a proposal $\gamma^\prime$ from $q(\gamma^\prime|\gamma)$
    \item Sample an auxiliary dataset $\adata$ of size $n$ (same as observed data) where $$\iS_i^* \overset{\text{i.i.d.}}{\sim} \text{SIS}(\iS^m, \gamma^\prime),$$
    \item Evaluate the following probability 
    \begin{equation}
        \begin{aligned}
            \alpha(\gamma, \gamma^\prime) &= \min\left\{1,\frac{p(\gamma^\prime|\iS^m, \data)p(\adata|\iS^m, \gamma)q(\gamma|\gamma^\prime)}{p(\gamma|\iS^m, \data)p(\adata|\iS^m, \gamma^\prime) q(\gamma^\prime| \gamma)}\right\}
        \end{aligned}
        \label{eq:post_gamma_acc_prob}
    \end{equation}
    \item Move to state $\gamma^\prime$ with probability $\alpha(\gamma, \gamma^\prime)$, staying at $\gamma$ otherwise.
\end{enumerate}
For the proposal $q(\gamma^\prime|\gamma)$ we consider sampling $\gamma^\prime$ uniformly over a $\varepsilon$-neighbourhood of $\gamma$ with reflection at zero. More specifically, we first sample $\gamma^* \sim \text{Uniform}(\gamma-\varepsilon, \gamma+\varepsilon)$ and then let $\gamma^\prime = \gamma^*$ if $\gamma^* > 0$ and let $\gamma^\prime = -\gamma^*$ otherwise. 

This is a direct application of the exchange algorithm \citep{Murray2006} and as such the resultant Markov chain admits $p(\gamma \, | \, \iS^m , \data)$ as its stationary distribution. Moreover, this is what one might call an ``exact-approximate'' MCMC algorithm, in the sense that (asymptotically) samples drawn thereof will be distributed according to the desired target, meaning that one could in theory obtain exact samples given infinite resource. A closed form of \eqref{eq:post_gamma_acc_prob} and derivation thereof can be found in \Cref{sec_sup:inference_dispersion_cond}.

\subsection{Updating the Mode}

\label{sec:inference_mode_cond}

We now outline our MCMC scheme to sample from the mode conditional. The key difference here is in the proposal generation mechanism, which follows the iMCMC algorithm \citep{Neklyudov2020} in using a combination of random sampling and deterministic maps. Here we assume the dispersion $\gamma$ is fixed and $\iS^m$ denotes our current state. Instead of specifying a proposal density, one defines auxiliary variables $u\in \U$, a deterministic function $f \, : \, \iSb \times \U \to \iSb \times \U$ and a conditional distribution $q(u \, | \, \iS^m)$ over auxiliary variables. The function $f$ must also be an \textit{involution}, meaning that is acts as its own inverse, that is, $f^{-1}=f$. A single iteration now consists of the following
\begin{enumerate}
    \item Sample $u \sim q(u|\iS^m)$
    \item Invoke involution $f(\iS^m, u) = ([\iS^m]^\prime, u^\prime)$, obtaining proposal $\modeprop$
    \item Sample auxiliary dataset $\adata$ of size $n$ where 
    $$\iS^*_i \overset{\text{i.i.d.}}{\sim} \text{SIS}([\iS^m]^\prime, \gamma)$$
    \item Evaluate the following probability 
    \begin{equation}
    \label{eq:acc_prob_post_mode}
        \begin{aligned}
            \alpha(\iS^m,[\iS^m]^\prime) &= \min\left\{1,\frac{p([\iS^m]^\prime \, | \,\gamma, \data)p(\adata \, | \,\iS^m, \gamma)q(u^\prime \, | \, [\iS^m]^\prime)}{p(\iS^m \, | \,\gamma, \data)p(\adata \, | \,[\iS^m]^\prime, \gamma) q(u \, | \, \iS^m)} \right\}
        \end{aligned}
    \end{equation}
    \item  Move to state $[\iS^m]^\prime$ with probability $\alpha(\iS^m, [\iS^m]^\prime)$, staying at $\iS^m$ otherwise.
\end{enumerate}
Much like the proposal density of a Metropolis-Hasting or exchange algorithm, the $u$, $f(\iS^m,u)$ and $q(u \, | \, \iS^m)$ represent free choices. We consider mixing together two such specifications, details of which we provide in the next section. 

This scheme represents an instance of what we call the iExchange algorithm (\Cref{alg_sup:iexchange}, \Cref{sec_sup:iexchange_alogrithm}). As shown in \Cref{sec_sup:iexchange_alogrithm}, this can be seen as a special case of the iMCMC algorithm. As such, this represents an exact-approximate MCMC algorithm with the resultant Markov chain admitting $p(\iS^m|\gamma, \data)$ as its stationary distribution. Note the iExchange algorithm as defined in \Cref{sec_sup:iexchange_alogrithm} includes a Jacobian term in the acceptance probability which we do not include above. The reasoning being that since both $\iSb$ and $\U$ are discrete spaces and $f(\iS,u)$ is a one-to-one function (since it is invertible) such terms are not required. 

\subsection{Mode Update Moves}

\label{sec:inference_mode_moves}

We now give details regarding two iExchange specifications for the mode conditional updates. In the first, we keep the number of paths fixed, varying only the path lengths or what we call the \textit{inner dimension}. For example, in the context of the Foursquare data, this would amount to altering a particular sequence of check-ins. In the second, we look to vary the number of paths or what we call the \textit{outer dimension}. For example, in the Foursquare data this would equate to introducing or removing a whole day of check-ins. 

\subsubsection{Edit Allocation}

\label{sec:inference_edit_alloc}

\begin{figure}
    \centering
    \includegraphics[width=0.85\linewidth]{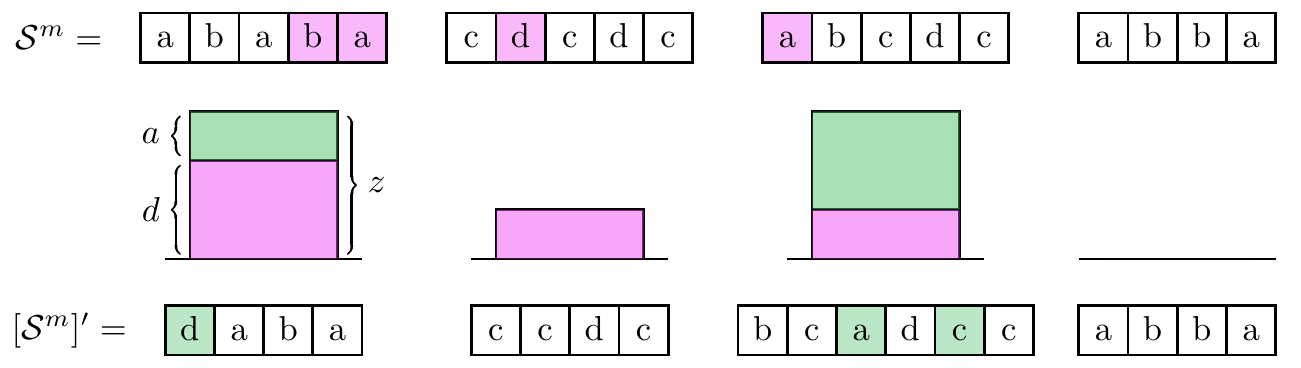}
    \caption{Illustrating the edit allocation move. Shaded entries indicate deletions and insertions, whilst bars visualise allocation of edits to paths. Bar height is proportional to the number of edits allocated to a path $z$, whilst the green (top) portion of the bar denotes the number of insertions $a$ and the pink (bottom) portion represents the number of deletions $d$. }
    \label{fig:edit_alloc}
\end{figure}

Supposing $\iS^m = (\I_1,\dots,\I_N)$ is our current state, the main idea of this move is to allocate a number of ``edits'' to each path in $\iS^m$. These edits consist of inserting and deleting entries, where if the number of insertions and deletions are unbalanced, paths of smaller or larger sizes relative to the current state will be proposed, thus varying the inner dimension. For an illustration, see \Cref{fig:edit_alloc}. 

We now give descriptive details of this proposal generation mechanism and show how it can be cast in the light of iMCMC. First, we specify the total number of edits to be made, denoting this $\delta \in \mathbb{Z}_{\geq1}$. Next, we specify an allocation of these edits to the paths of $\iS^m$, denoting this $\bm{z} = (z_1,\dots,z_N)$, where $z_i \in \mathbb{Z}_{\geq0}$ denotes the number of edits allocated to the $i$th path such that $\sum_{i=1}^N z_i = \delta$. For example, in \Cref{fig:edit_alloc} we have $\delta=7$ and $\bm{z} = (3,1,3,0)$. 

Given $z_i$ we edit the $i$th path $\I_i$ to obtain a corresponding proposal $\I_i^\prime$ in the following manner. First, we partition the $z_i$ edits between deletions and insertions, letting $d_i \in \{0,\dots,\min(n_i, z_i)\}$ denote the number of deletions,  where $n_i$ denotes the length of the $i$th path, with $a_i = z_i - d_i$ then denoting the number of insertions. Note, we cannot delete more entries than are present, hence the restriction $d_i \leq \min(n_i,z_i)$. 

The penultimate step is to specify which entries to delete and where to insert new entries, which we denote via \textit{subsequences}. Introducing the notation $[n] = (1,\dots, n)$, we define subsequence of $[n]$ of size $m$ to be a vector $\bm{v} = (v_1, \dots, v_m)$ such that $1 \leq v_1 < v_2 < \dots < v_m \leq n$. Now, we let $\bm{v}_i$ be a subsequence of $[n_i]$ of size $d_i$ denoting the entries of $\I_i$ to be deleted, whilst $\bm{v}^\prime_i$ is subsequence of $[m_i]$ of size $a_i$, denoting the location of entries to be inserted in $\I^\prime_i$, where $m_i = n_i-d_i+a_i$ denotes the length of $\I^\prime_i$. For example, considering the first path in \Cref{fig:edit_alloc} we have $\I_1=(a,b,a,b,a)$ and $\I^\prime_1=(d,a,b,a)$ with $\bm{v}_1=(4,5)$ and $\bm{v}^\prime_1=(1)$ indexing the deletions and insertions respectively. The final step is to specify entries to insert, which we denote $\bm{y}_i=(y_{i1},\dots,y_{ia_i})$ where $y_{ij} \in \V$. For example, in \Cref{fig:edit_alloc} we have $\bm{y}_1=(d)$.

Given the information above, one can enact the specified deletions and insertions, mapping to a proposal $[\iS^m]^\prime = (\I^\prime_1,\dots,\I^\prime_N)$. This can be viewed in the iMCMC framework as follows. First, collate all this information into the auxiliary variable $u = (\delta, \bm{z},u_1,\dots,u_N)$ where $u_i = (d_i, \bm{v}_i, \bm{v}^\prime_i, \bm{y}_i)$. Now, if we write the required involution as follows
$$f(\iS^m, u) = (f_1(\iS^m, u), f_2(\iS^m, u)) = ([\iS^m]^\prime, u^\prime),$$
then in enacting the specified edit operations we have effectively defined the first component $f_1(\iS^m, u) = [\iS^m]^\prime$. Specification of the second component is more involved, and so we delegate these details to \Cref{sec_sup:inference_edit_alloc}. Regarding the auxiliary distribution $q(u|\iS^m)$, we consider the following
\begin{equation*}
    \begin{aligned}
        \delta &\sim \text{Uniform}\{1,\dots, \nu_{\text{ed}}\} \\
        \bm{z} \, | \, \delta &\sim \text{Multinomial}(\delta \, ; \, 1/N, \dots, 1/N)\\
        d_i \, | \,  z_i &\sim \text{Uniform}\{0,\dots,\min(z_i, n_i)\} & (\text{for } i=1,\dots,N)\\
    \end{aligned}
\end{equation*}
whilst $\bm{v}_i$ and $\bm{v}^\prime_i$ are drawn uniformly and the entry insertions $\bm{y}_i$ are assumed to be sampled from some general distribution $q(\bm{y}_i |\I_i)$, which we typically take to be the uniform distribution over $\V$. The only tuning parameter here is $\nu_{\text{ed}}$, which controls the aggressiveness of proposals, with larger values leading to more edits being attempted on average. 

Further details, including full definition of the involution $f$, examples of possible insertion distributions $q(\bm{y}_i|\I_i)$ and derivations of key terms appearing the acceptance probability \eqref{eq:acc_prob_post_mode}, can be found in \Cref{sec_sup:inference_edit_alloc}.

\subsubsection{Path Insertion and Deletion}

\label{sec:inference_trans_dim}

\begin{figure}[b]
    \centering
    \includegraphics[width=0.9\linewidth]{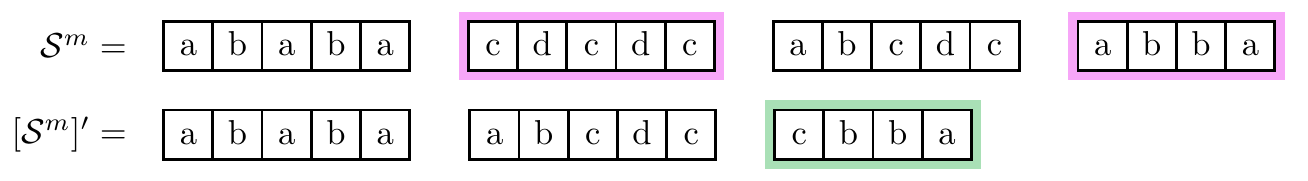}
    \caption{Illustrating path insertion and deletion move, where given current state $\iS^m$ the proposal $[\iS^m]^\prime$ is obtained by deleting and inserting the highlighted paths.}
    \label{fig:trans_dim}
\end{figure}

With this move we look to vary the outer dimension, that is, the number of paths. Similar to \Cref{sec:inference_edit_alloc}, we consider doing so by random deletion and insertion. The difference in this case is that we delete and insert whole paths (see \Cref{fig:trans_dim}). 

In particular, with $\iS^m = (\I_1,\dots,\I_N)$ denoting our current state, we first choose a total number of insertions and deletions $\varepsilon \in \mathbb{Z}_{\geq1}$. Next, we partition these, letting $d \in \{0,\dots,\min(N, \varepsilon)\}$ denote the number of deletions, leaving $a = \varepsilon - d$ insertions. For example, in \Cref{fig:trans_dim} we have $\varepsilon=3$, $d=2$ and $a=1$. Next, we choose locations of deletions and insertions. In particular, we let $\bm{v}$ be a length $d$ subsequence of $[N]$ denoting which paths of $\iS^m$ are to be deleted, whilst $\bm{v}^\prime$ is a length $a$ subsequence of $[M]$, where $M=N-d+a$, denoting where inserted paths will be located in our proposal $[\iS^m]^\prime$. For example, in \Cref{fig:trans_dim} we have $\bm{v} = (2,4)$ and $\bm{v}^\prime = (3)$. Finally, for each $i=1,\dots,a$ we choose some path $\I^*_i$ to insert into entry $v^\prime_i$ of $[\iS^m]^\prime$. For example, in \Cref{fig:trans_dim} we have a single path $\I^*_1 = (c,b,b,a)$ which we insert to the third entry.

As in \Cref{sec:inference_edit_alloc}, given the information above we can insert and delete the corresponding paths to obtain a proposal $[\iS^m]^\prime$. Collating this into the auxiliary variable $u=(\varepsilon, d, \bm{v}, \bm{v}^\prime, \I^*_1,\dots,\I^*_a)$ this can similarly be seen as defining the first component of the required involution, with details of the second component found in \Cref{sec_sup:inference_trans_dim}. Regarding sampling of auxiliary variables we consider the following
\begin{equation*}
    \begin{aligned}
        \varepsilon & \sim \text{Uniform}\{1,\dots,\nu_{\text{td}}\} \\
        d \, | \, \varepsilon &\sim \text{Uniform}\{0,\dots,\min(N, \varepsilon)\} \\
    \end{aligned}
\end{equation*}
whilst we sample $\bm{v}$ and $\bm{v}^\prime$ uniformly and assume path insertions $\I^*_i$ are drawn from some general distribution over paths $q(\I|\iS^m)$. This leaves two tuning parameters, $\nu_{\text{td}}$ and $q(\I|\iS^m)$, which in combination facilitate control over the aggressiveness of proposals. In particular, $\nu_{\text{td}}$ controls the number of deletions and insertions attempted, whilst $q(\I|\iS^m)$ affects how impactful each of these insertions and deletions are. Again, further details can be found in \Cref{sec_sup:inference_trans_dim}.

\subsection{Sampling Auxiliary Data} 

\label{sec:inference_model_sampling}

Both algorithms to target the conditionals outlined in \Cref{sec:inference_dispersion_cond,sec:inference_mode_cond} require exact sampling of auxiliary data from appropriate interaction-sequence models. Unfortunately, we cannot do this in general. Instead, we consider replacing this with approximate samples obtained via an iMCMC algorithm. 

In particular, suppose we would like to obtain samples from an $\text{SIS}(\iS^m, \gamma)$ model. Assuming that $\iS$ denotes the current state, and auxiliary variables $u$, involution $f(\iS,u)$ and auxiliary distribution $q(u|\iS)$ have be defined, in a single iteration we do the following
\begin{enumerate}
    \item Sample $u \sim q(u|\iS)$
    \item Invoke involution $f(\iS, u) = (\iS^\prime, u^\prime)$ 
    \item Evaluate the following probability 
    \begin{equation}
        \begin{aligned}
            \alpha(\iS,\iS^\prime) &= \min\left\{1,\frac{p(\iS^\prime \, | \,\iS^m, \gamma)q(u^\prime \, | \, \iS^\prime)}{p(\iS^\, | \, \iS^m, \gamma) q(u \, | \, \iS)} \right\}
        \end{aligned}
        \label{eq:sis_model_sampling_imcmc_acc_prob}
    \end{equation}
    \item  Move to state $\iS^\prime$ with probability $\alpha(\iS, \iS^\prime)$, staying at $\iS$ otherwise.
\end{enumerate}
where $p(\iS|\iS^m, \gamma)$ denotes the likelihood as given in \eqref{eq:SIS}. Towards specifying $u$, $f(u,\iS)$ and $q(u|\iS)$, we now recycle the moves of \Cref{sec:inference_mode_cond}, again mixing these together with some proportion $\beta \in (0,1)$. Note, as in \Cref{sec:inference_mode_cond}, we omit the Jacobian term in the acceptance probability above since we are working with discrete spaces.

In sampling auxiliary data in this manner, we now have two MCMC-based elements: what one might call the \textit{outer} MCMC algorithm, navigating the parameter space, and the \textit{inner} MCMC algorithm, sampling auxiliary data. We note this approach has been considered by others. In particular, \cite{Liang2010} proposed the so-called double Metropolis-Hastings algorithm which replaces the exact samples of the exchange algorithm with those obtained via a Metropolis-Hasting scheme. The difference in our case is use of the more general iMCMC framework, be that in the outer MCMC
scheme (as in the iExchange algorithm), or the inner MCMC scheme (as outlined above).

A consequence of using approximate auxiliary samples within the algorithms of \Cref{sec:inference_dispersion_cond,sec:inference_mode_cond} is the resulting schemes will become approximate, as opposed to exact-approximate.
That is to say, even in the theoretical limit, samples will not necessarily be distributed according to the desired target but instead an approximation thereof. However, as the auxiliary samples look more like an i.i.d. sample one will get closer to the respective exact-approximate algorithm. Thus, one can in theory get arbitrarily close to an exact-approximate scheme by taking steps to reduce the bias of the MCMC-based auxiliary samples, such as introducing a burn-in period or taking a lag between samples.


\section{Simulation Studies}

\label{sec:simulations}

In this section we outline simulation studies undertaken to confirm the efficacy of our methodology. In the first two, we examine the posterior concentration, exploring how this is affected by variability of observed data and structural features of the mode. In the third, we assess convergence of the posterior predictive via a missing data problem. In each, we will be working with the interaction-sequence models.

\subsection{Posterior Concentration}


 \begin{figure}
    \centering
    \includegraphics[width=0.95\linewidth]{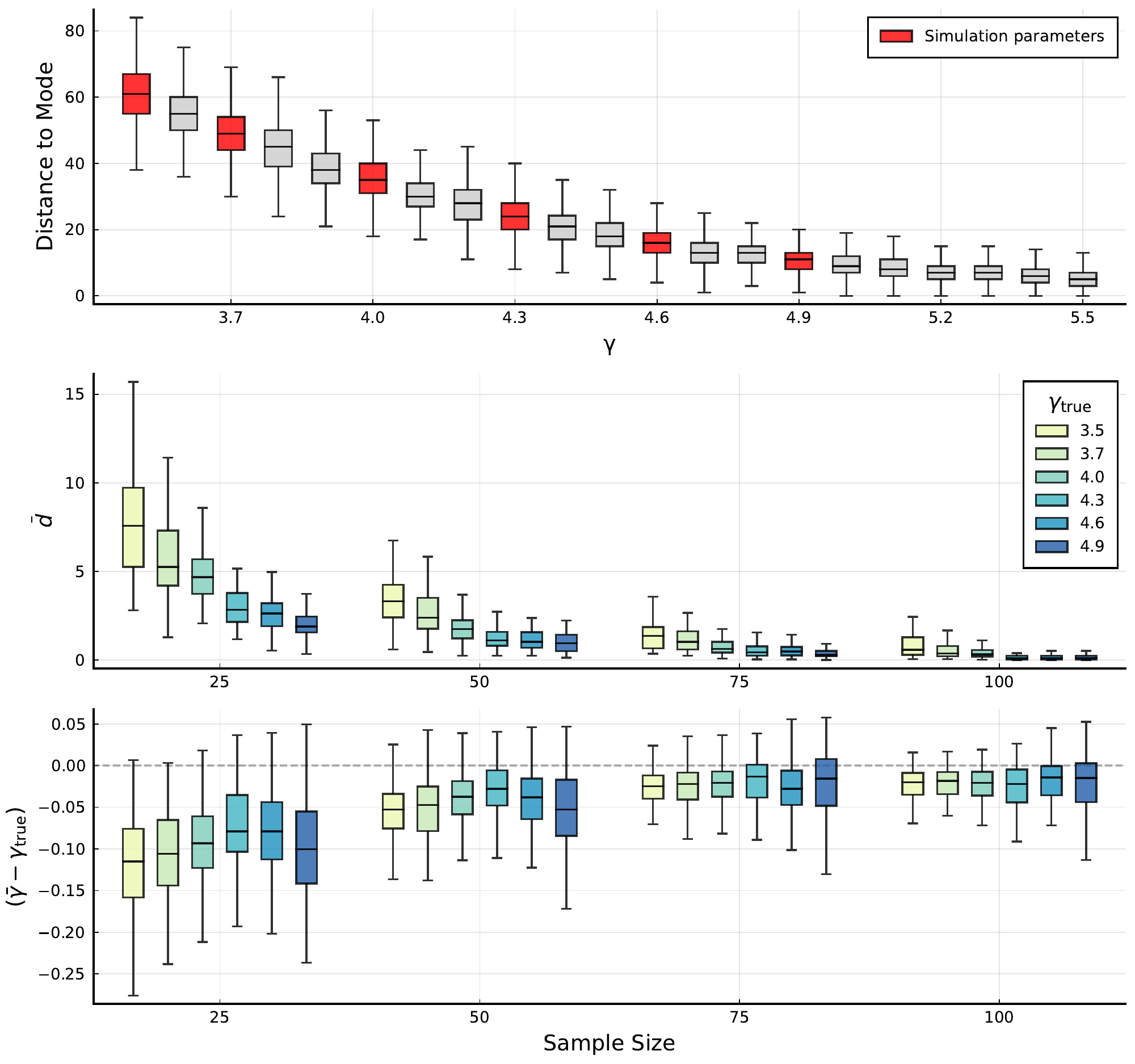}
    \caption{A summary of our first simulation (\Cref{sec:sim_post_conc_gamma}), where the top plot visualises the scale of the SIS model used therein, in particular, for different values of $\gamma$ it shows $\{d_S(\iS^{(i)},\iS^m_{\text{true}})\}_{i=1}^{1000}$ where $\iS^{(i)} \sim \text{SIS}(\iS^m_{\text{true}},\gamma)$, sampled via the iMCMC scheme of \Cref{sec:inference_model_sampling}. The remaining two plots summarise simulation outputs for each pair $(\gamma_{\text{true}},n)$, where the middle shows distributions of $\bar{d}$, the average distance to the true mode, whilst  the bottom shows $(\bar{\gamma}-\gamma_{\text{true}})$, the bias of the dispersion posterior mean relative to the truth.}
    \label{fig:posterior_conc_sim}
\end{figure}

If the data were generated by an SIS model at known parameters, then ideally the posterior \eqref{eq:SIS_posterior} should concentrate about these as the sample size grows. The goal of the first two simulations is to confirm empirically that such behaviour is observed and to explore what factors might come into play. 
The high-level approach is the following. Given true mode $\iS^m_{\text{true}}$ and dispersion $\gamma_{\text{true}}$, we draw a sample $\data$ where $$\iS^{(i)} \sim \text{SIS}(\iS^m_{\text{true}}, \gamma_{\text{true}})$$
before obtaining samples $\{(\iS_i^m,\gamma_i)\}_{i=1}^m$ from the posterior $p(\iS^m,\gamma|\data)$. We then assess the behaviour of these samples via the following summary measures 
\begin{equation*}
    \begin{aligned}
        \bar{d}:=\frac{1}{m}\sum_{i=1}^m d_S(\iS^m_i, \iS^m_{\text{true}}) && && &&  \bar{\gamma} := \frac{1}{m}\sum_{i=1}^m \gamma_i
    \end{aligned}
\end{equation*}
where ideally $\bar{d}$ should be close to zero and $\bar{\gamma} \approx \gamma_{\text{true}}$. By repeating this a number of times for different $n$ and evaluating these summaries we can thus get a sense of how the posterior is concentrating about the true parameters.

\label{sec:sim_post_conc_gamma}

Now, recall the dispersion works inversely to the variance, in that lower values lead to more variable data (\Cref{fig:posterior_conc_sim}, top). Intuitively, when the data is more variable it will be harder to discern the true mode $\iS^m_{\text{true}}$, and thus we expect $\bar{d}$ to decrease more slowly for lower values of $\gamma_{\text{true}}$. Alternatively, as can be seen in \Cref{fig:posterior_conc_sim}, when $\gamma_{\text{true}}$ is smaller the difference of their parameterised distributions (as described by the distribution of distances to the mode) becomes more marked relative to neighbouring values. As such, we might also expect smaller values for the dispersion to be easier to recover.

To explore for such properties, we varied $\gamma_{\text{true}}$ and $n$ whilst keeping $\iS^m_{\text{true}}$ fixed. In particular, we considered $\gamma_{\text{true}}=3.5,3.7,4.0,4.3,4.6,4.9$ (highlighted in \Cref{fig:posterior_conc_sim}, top) and $n=25,50,75,100$. The distance we took to be $d_S=d_{\text{Edit}}$ with $d_I=d_{\text{LCS}}$ between paths. We fixed $V=20$, and constrained the sample space as discussed in \Cref{sec_sup:sample_spaces_bounded}, assuming at most $L=20$ paths in any observation, with each path being at most length $K=10$. 

The mode $\iS^m_{\text{true}}$ of length $N=10$ we fixed throughout, sampled from the Hollywood model of \cite{Crane2018}. In particular, we drew $\iS^m_{\text{true}} \sim \text{Hollywood}(\alpha, \theta, \nu)$ where
\begin{align*}
    \alpha = -0.3 && \theta = 0.3V && \nu = \text{TrPoisson}(3, 1, K),
\end{align*}
where $\text{TrPoisson}(\lambda,a,b)$ denotes a truncated Poisson distribution with $\lambda>0$ the parameter of a standard Poisson, whilst $0 \leq a < b \leq \infty$ are the lower and upper bounds. This set-up for the Hollywood model, with $\alpha < 0$ and $\theta = -V\alpha$, corresponds to the finite setting, implying the sampled interaction sequences will have at most $V$ vertices. 

Regarding priors, we considered an uninformative set-up with $(\iS_0,\gamma_0) = (\hat{\iS}, 0.1)$ where
$$\hat{\iS} := \argmin_{\iS \in \data} \sum_{i=1}^n d^2_S(\iS^{(i)}, \iS)$$
denotes the sample Fr\'{e}chet mean,
whilst we took $\gamma \sim \text{Uniform}(0.5,7.0)$. Here we note the sample $\data$ used to obtain $\hat{\iS}$ will be different in each repetition of the simulation, and consequently so will $\hat{\iS}$.

Now, for each pair $(\gamma_{\mathrm{true}}, n)$ we (i) sampled $n$ observations from an $\text{SIS}(\iS^m_{\mathrm{true}}, \gamma_{\mathrm{true}})$ model, using the iMCMC scheme outlined in \Cref{sec:inference_model_sampling}, with a burn-in period of 50,000 and taking a lag of 500 between samples (ii) obtained $m=250$ samples from the posterior using the component-wise MCMC scheme of \Cref{sec:inference_joint}, with a burn-in period of 25,000 and taking a lag of 100 between samples\footnote{One must also parameterise the MCMC algorithm used to sample the auxiliary data. These were tuned by considering acceptance probabilities observed when sampling from an $\text{SIS}(\iS^m_{\mathrm{true}},\gamma_{\mathrm{true}})$ distribution.}
 (iii) evaluated summary measures $\overline{d}$ and $\bar{\gamma}$.

We repeated (i)-(iii) 100 times in each case, the results of which are summarised in \Cref{fig:posterior_conc_sim}. Consulting the middle plot, we observe that $\bar{d}$ decreases with $n$ across all cases, indicating a concentration of the posterior about the true mode. Furthermore, this decrease is more gradual for lower values of $\gamma_{\text{true}}$, agreeing with intuition. Turning to the bottom plot, the  most obvious feature is bias in $\bar{\gamma}$ relative to the truth. Note this is expected, since we have used approximate MCMC samples within our component-wise scheme of \Cref{sec:inference_joint}. We do, however, see a reduction in this bias as the sample size grows. Furthermore, for the larger values of $n$ we begin to see a clearer difference in the variance of $\bar{\gamma}$ across different values of $\gamma_{\text{true}}$. In particular, the variance appears to be smaller for lower values of $\gamma_{\text{true}}$, agreeing with the intuition that these are easier to estimate.

\subsection{Effect of Mode Structure}
\label{sec:sim_hw}
Here we explored whether structural features of the mode might impact its inference. Adopting the same modelling set-up as the previous simulation, but in this case fixing the true dispersion to $\gamma_{\text{true}}=4.5$, we re-sampled the mode in each repetition via
$$\iS_{\text{true}} \sim \text{Hollywood}(\alpha, -\alpha V, \nu)$$
where we again take $V=20$ and $\nu=\text{TrPoisson}(3,1,K)$, whilst $\alpha<0$.
 
\begin{figure}
    \centering
    \includegraphics[width=0.9\linewidth]{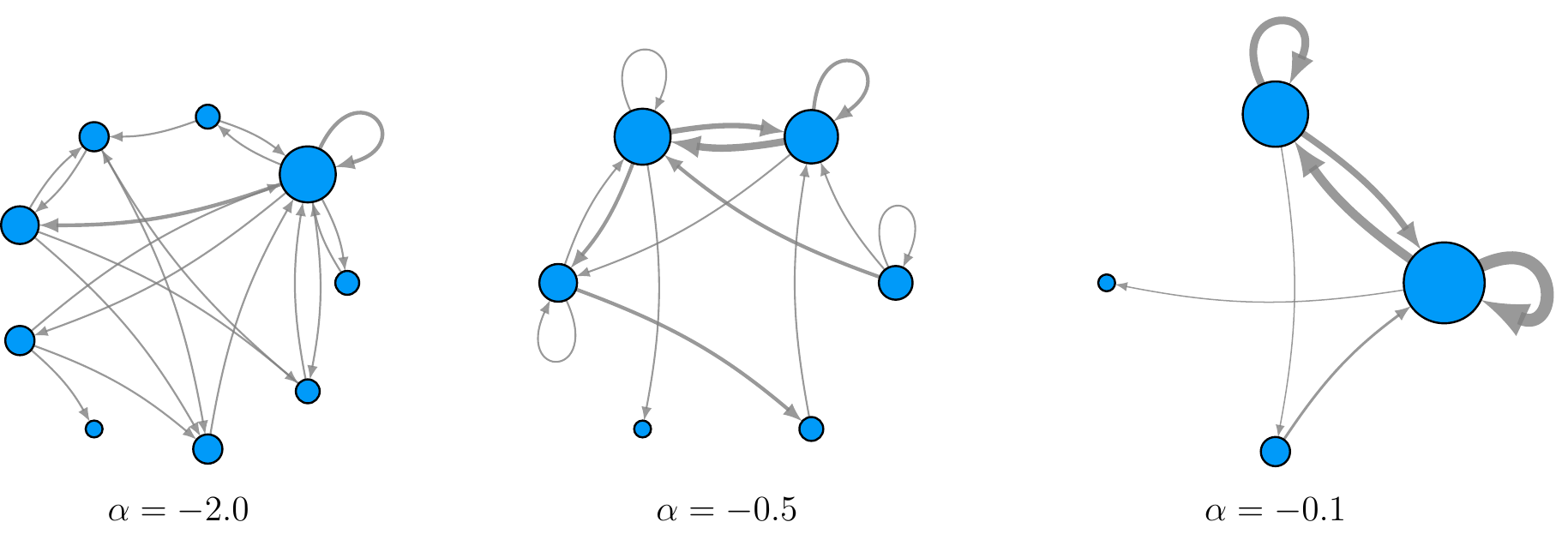}
    \caption{Visualising the role of $\alpha$ in the Hollywood model. Each plot shows an aggregate multigraph $\G_\iS$ where $\iS \sim \text{Hollywood}(\alpha, -\alpha V, \nu)$ with $V=10$, $\nu=\text{TrPoisson}(3,1,10)$ and $\alpha$ varying. Edge thickness reflects edge multiplicity, whilst vertex size is proportional to $k_\iS(v)$.  
    }
    \label{fig:hw_ex_samples}
    \includegraphics[width=0.95\linewidth]{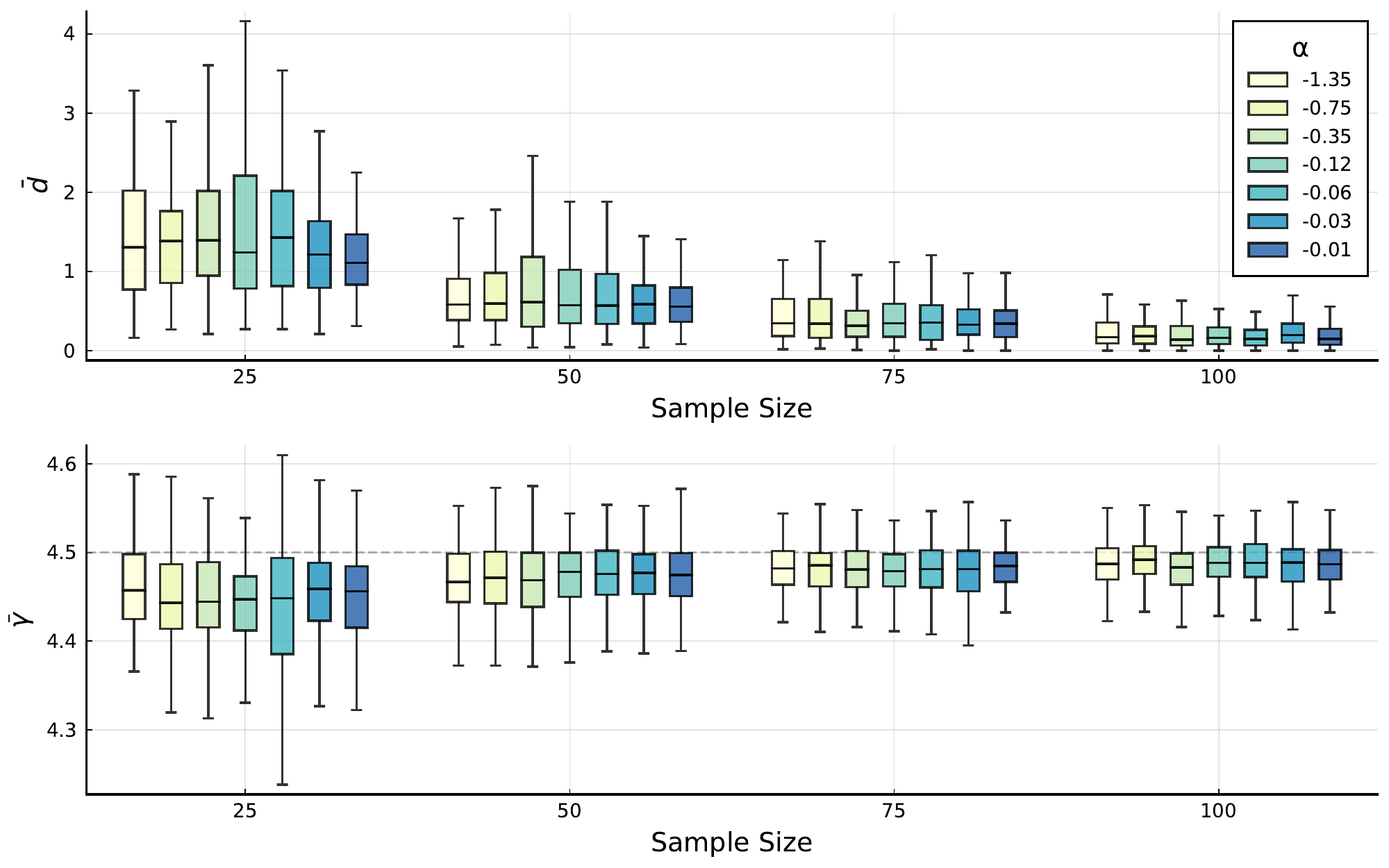}
    \caption{Summary of our second simulation (\Cref{sec:sim_hw}), where for each pair $(\alpha,n)$ the top subplot shows the distribution of $\bar{d}$, the average distance to the true mode, whilst the bottom shows the distribution of $\bar{\gamma}$, the posterior mean dispersion. }
    \label{fig:posterior_conc_sim_hw}
\end{figure}

The key idea underlying the Hollywood model is a `rich get richer' assumption made when sampling vertices. This results in $\alpha$ admitting an interpretation regarding the heavy-tailed nature of vertex counts. In particular, for a given interaction sequence $\iS$ and vertex $v \in \V$ one can define an analogue of the vertex degree (often defined for graphs) as follows $$k_{\iS}(v) := \text{\# times $v$ appears in $\iS$},$$
which thus implies, for each $\iS$, a sample $\{k_\iS(v) \,: v \in \V,\, k_\iS(v) >0\}$, similar in spirit to the degree distribution. Now, $\alpha$ can be seen to control the heavy-tailedness of this distribution (see \Cref{fig:hw_ex_samples}), whereby when $\alpha$ is low one tends to see vertices appearing a similar number of times, whilst when $\alpha$ is larger these counts become disproportionately focused on a smaller subset of vertices.
 
In this simulation, we considered $\alpha = -\tilde{\alpha}$ where $\tilde{\alpha} = 1.35, 0.75, 0.35, 0.12, 0.06, 0.03, 0.01$ (details on how these were chosen can be found in \Cref{sec_sup:hw_par_choices}) and $n=25,50,75,100$. For each pair $(\alpha, n)$ in a single repetition we (i) sampled $\iS_{\text{true}} \sim \text{Hollywood}(\alpha, -\alpha V, \nu)$, (ii) sampled $n$ observations from an $\text{SIS}(\iS_{\text{true}}, \gamma_{\text{true}})$ model (iii) obtained $m=250$ samples from the posterior, and (iv) evaluated summaries. For (ii) and (iii) we used exactly the same MCMC set-up as in the previous simulation.

\Cref{fig:posterior_conc_sim_hw} summarises the output of 100 repetitions for each pair $(\alpha, n)$. For each $\alpha$, we see values for $\bar{d}$ closer to zero as $n$ grows, indicating concentration about the true mode. Furthermore, $\alpha$ shows no clear sign of impacting this concentration. Regarding the dispersion posterior mean $\bar{\gamma}$, as in the previous simulation we observe bias relative to the truth, with this bias reducing as $n$ grows. Furthermore, this is the same across all $\alpha$, with no clear sign that $\alpha$ affects the inference of these values.

\subsection{Posterior Predictive Efficacy}

\label{sec:sim_predictive}

A desirable feature of the posterior predictive is a growing resemblance of the true data generating distribution as the sample size increases. In this simulation, we considered exploring such behaviour in the context of a missing data problem.

Suppose we have an observation $\iS$ in which a single entry is missing, for example
$$\iS = ((1,2,1,\bullet), (2,3,4,3), (1,2,2,1,2,3))$$
with $\bullet$ denoting the unknown entry. Towards predicting its value, let $\iS_x$ denote the observation obtained by taking this entry to be $x \in \V$, that is
$$\iS_x = ((1,2,1,x), (2,3,4,3), (1,2,2,1,2,3)),$$
and consider assigning a probability to each $x \in \V$ of being the true entry. If one knew $\iS \sim \text{SIS}(\iS^m, \gamma)$, then such a distribution could be obtained by comparing the relative probability of $\iS_x$ for each $x \in \V$, in particular we could consider
\begin{equation*}
    \begin{aligned}
        p(x|\iS^m,\gamma,\iS_{-x}) := \frac{1}{Z(\iS^m, \gamma, \iS_{-x})}\exp\{-\gamma \phi(d_S(\iS_x, \iS^m))\}
    \end{aligned}
\end{equation*}
with $Z(\iS^m, \gamma, \iS_{-x}) = \sum_{x\in \V} \exp\{-\gamma d_S(\iS_x, \iS^m)\}$ the normalising constant, where we introduce the notation $\iS_{-x}$ to indicate that we are conditioning on the other known entries (and implicitly also on the dimensions of the observation). We refer to this as the \textit{true predictive} for $x \in \V$.

In practice, with the true distribution unknown, one can instead leverage an observed sample $\data$ by averaging with respect to the posterior as follows 
\begin{equation*}
    \begin{aligned}
        p(x|\data,\iS_{-x}) &= \sum_{\iS^m \in \iSb} \int_{\mathbb{R}_+}p(x|\iS^m, \gamma,\iS_{-x})p(\iS^m, \gamma|\data)d\gamma, \\
    \end{aligned}
\end{equation*}
defining the \textit{posterior predictive} for $x \in \V$, which itself can be approximated using a sample $\{(\iS^m_i,\gamma_i)\}_{i=1}^m$ from the posterior via 
\begin{equation*}
    \begin{aligned}
        \hat{p}(x|\data,\iS_{-x}) &:= \frac{1}{m}\sum_{i=1}^m p(x|\iS^m_i,\gamma_i,\iS_{-x}),
    \end{aligned}
\end{equation*}
a derivation of which can be found in \Cref{sec_sup:simulation_predictive}. To now predict $x$, one can for example take the maximum \textit{a posteriori} (MAP) estimate 
$$\hat{x} = \argmax\limits_{x \in \V} \hat{p}(x | \data, \iS_{-x}).$$

In this simulation, we considered assessing the agreement of $\hat{p}(x|\data,\iS_{-x})$ with $p(x|\iS^m, \iS_{-x})$ as $n$ grows by comparing their predictive accuracy on a common test sample. We adopted the same modelling set-up as \Cref{sec:sim_post_conc_gamma}, jointly varying the dispersion and sample size, in this case considering $\gamma_{\text{true}}=3.7,4.2,4.5,4.9$ and $n=25,50,75,100$. However, in a slight deviation we here re-sampled the mode in each repetition from a fixed Hollywood model. 

Now, for a given pair $(\gamma_{\text{true}},n)$ and a pre-specified number of test samples $n_{\text{test}}$, in a single repetition we (i) sampled mode $\iS_{\text{true}} \sim \text{Hollywood}(\alpha, -\alpha V, \nu)$, with $\alpha=-0.35$ ($V$ and $\nu$ as in \Cref{sec:sim_post_conc_gamma,sec:sim_hw}) (ii) sampled training and testing data $\{\iS^{(i)}\}_{i=1}^{n+n_{\text{test}}}$ from an $\text{SIS}(\iS_{\text{true}}, \gamma_{\text{true}})$ model, (iii) obtained a sample $\{(\iS^m_i,\gamma_i)\}_{i=1}^m$ from the posterior $p(\iS^m,\gamma|\data)$, that is, using the $n$ training samples, (iv) for each $i=n+1,\dots,n+n_{\text{test}}$ and for each entry of $\iS^{(i)}$ (that is, each entry of each interaction) we assumed it to be missing and predicted its value via the MAP estimate of both $\hat{p}(x|\data,\iS_{-x})$ and $p(x|\iS^m, \iS_{-x})$, and finally (v) returned the proportion of times each prediction was correct. For (ii) and (iii) we used the same MCMC schemes as previous simulations.

\begin{figure}
    \centering
    \includegraphics[width=0.95\linewidth]{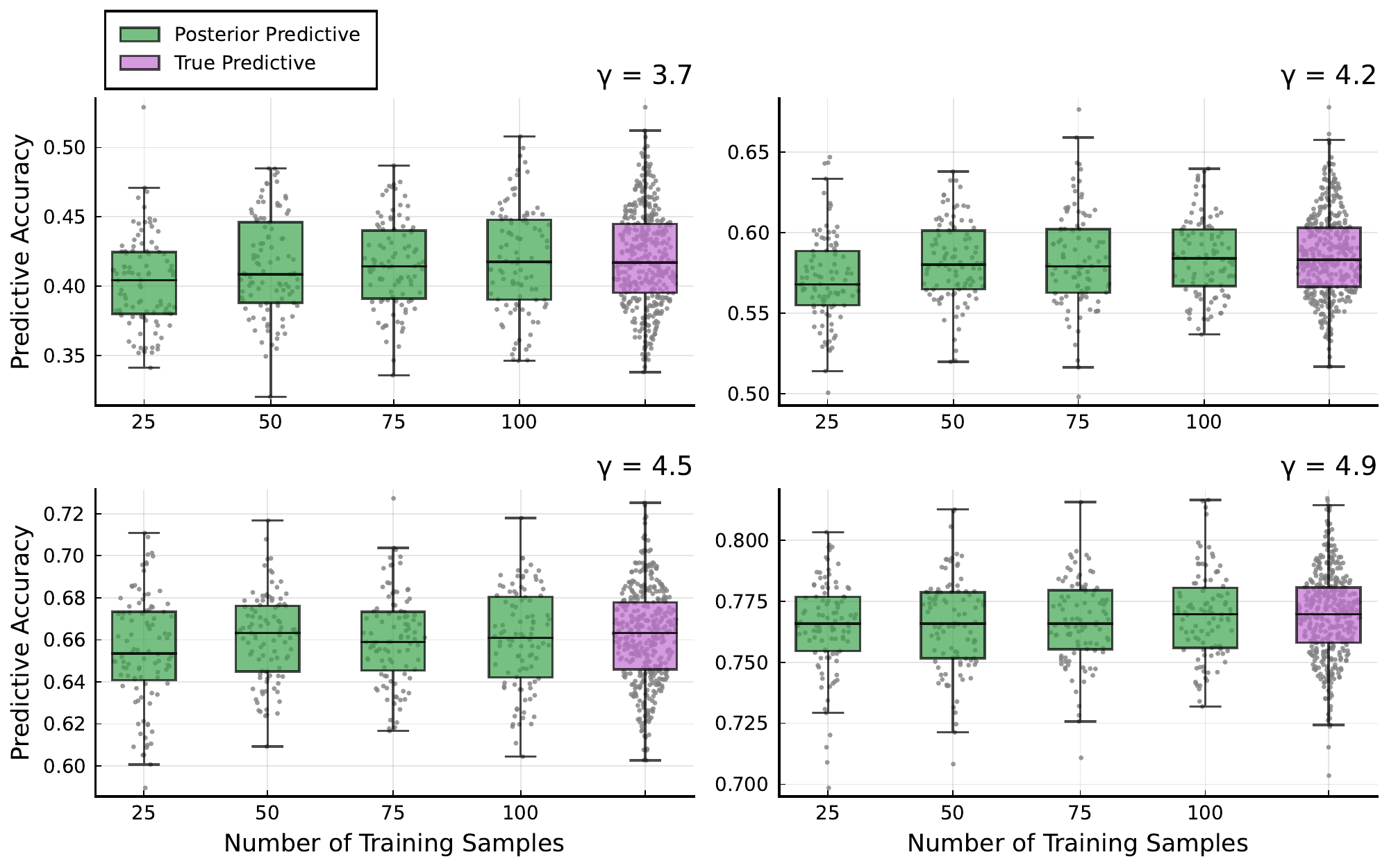}
    \caption{Summary of posterior predictive simulation (\Cref{sec:sim_predictive}). In each subplot, overlaid markers show the observed predictive accuracy being summarised by boxplots, with the four leftmost boxes concerning the posterior predictive at different sample sizes, whilst the rightmost box shows the accuracy of the true predictive across all sample sizes (collated together).}
    \label{fig:poster_pred_sim}
\end{figure}

\Cref{fig:poster_pred_sim} summarises the output of 100 repetitions for each pair $(\gamma_{\text{true}}, n)$, with $n_{\text{test}}=100$ in each repetition. For each $\gamma_{\text{true}}$, we see the accuracy of the posterior predictive is typically lower than the true predictive when the number of training samples is small ($n=25$), with this difference in accuracy diminishing as $n$ grows, so that by $n=100$ training samples the accuracy of the posterior predictive is almost indistinguishable from that of the true predictive. With similarity in predictive accuracy serving as a proxy for the coherence of the posterior predictive and the true data generating distribution, it thus appears the posterior predictive is more closely resembling the true data generating distribution as the number of training samples grows, as desired. One can also observe influence from variability as controlled by $\gamma_{\text{true}}$, where for larger $\gamma_{\text{true}}$ the predictive accuracy is often higher (for both the posterior and the true predictive), as one might expect since the test data in such cases will look more like the true mode and thus more entries thereof will be easily predicted.

\section{Real Data Analysis}
\label{sec:data_analysis}

In this section, we illustrate the applicability of our method by analysing the Foursquare data set of \cite{Yang2015}. As mentioned in \Cref{sec:intro}, an alternative approach to ours is to first aggregate observations to form graphs before applying a suitable graph-based method. As such, we compare our inference with some graph-based estimates. Note that in aggregating observations to form graphs one implicitly makes the assumption that the order of interaction arrival is irrelevant. Hence, for fairness we opt to make this comparison with our SIM model.

\subsection{Data Background and Processing}
\label{sec:data_analysis_processing}
For this analysis we looked at one month of check-in data (over the period from 12 April to 12 May 2012, from the New York and Tokyo data set\footnote{\url{https://sites.google.com/site/yangdingqi/home/foursquare-dataset}.}), focusing in particular on those in New York. Each check-in event consists of a (i) user id, identifying which user enacted the check-in (anonymised) (ii) venue id, unique to each venue, (iii) venue category, (iv) latitutde and longitude, and (v) timestamp.  

 As discussed in \Cref{sec:intro}, we view this as interaction network data by seeing a day of check-ins for a single user as a path through the venue categories. Note the venue category labels have a hierarchical structure, with those given by \cite{Yang2015} being the low-level. For example, the category ``Jazz Club'' is a subcategory of ``Music Venue'', which is itself a subcategory of ``Arts \& Entertainment''. In this analysis, we opted to use the highest-level venue categories, so for example here we would use ``Arts \& Entertainment''.  

Before proceeding with our analysis, we further filtered the data. Firstly, it is clearly possible a user might only check-in to a single venue on a given day. Since our analysis is based on interaction multisets and concerns the movements of users \textit{between} venue categories, such observations provide little information. Furthermore, such observations will be disregarded when aggregating to form graphs, and therefore would not feature in any of the graph-based approaches with which we intend to compare. As such, we considered only days where a user had checked-in to at least two venues. To further ensure each observation contained enough information, we considered only users with at least 10 observed days of check-ins. This left a total of 402 observations, from which we extracted a subset of 100 to analyse using a criterion based upon the distance metric used in our model fit (further details in \Cref{sec_sup:data_processing}).

\subsection{SIM Model Fit}
\label{sec:data_analysis_sim}

Following data processing, we were left with a sample of multisets $\edata$, where each $\E^{(i)} = \left\{\I^{(i)}_1,\dots,\I^{(i)}_{N^{(i)}}\right\}$ denotes the data of the $i$th user, with $\I^{(i)}_j$ denoting a single day of their check-ins. Recalling the inferential questions of interest outlined in \Cref{sec:intro}, we now use our methodology to obtain (a) an average multiset of paths, and (b) a measure of variability. 

In particular, using the Bayesian inference approach outlined in \Cref{sec_sup:inference_sim}, we fit our SIM model to these data. We made use of the matching distance $d_{\text{M}}$, with the LSP distance $d_{\text{LSP}}$ between paths. As discussed in \Cref{sec:modelling_choices}, a consequence of assuming this distance is that our inferred mode will contain paths often appearing as subpaths in the observed data. For our priors, we assumed $\E^m \sim \text{SIM}(\hat{\E}, 3.0)$, with $\hat{\E}$ denoting the sample Frech\'{e}t mean of the observed data $\edata$, whilst we assumed $\gamma \sim \text{Gamma}(5, 1.67)$. Via our MCMC scheme, we then obtained a sample $\{(\E_i^m,\gamma_i)\}_{i=1}^M$, from the posterior $p(\E^m,\gamma|\edata)$, obtaining a total of $M=500$ samples with a burn-in period of 25,000 and taking a lag of 50 between samples. 

Given the posterior sample $\{(\E_i^m,\gamma_i)\}_{i=1}^M$, we subsequently obtained point estimates $(\hat{\E}^m, \hat{\gamma})$, with the mode estimate $\hat{\E}^m$ functioning as our desired average, and $\hat{\gamma}$ a measure of data variability. In particular, we considered the following
\begin{align*}
 \hat{\E}^m = \argmin_{\E \in \{\E^m_i\}_{i=1}^M} \sum_{i=1}^M d^2_{\text{M}}(\E^m_i, \E) && \hat{\gamma} = \frac{1}{M} \sum_{i=1}^M\gamma_i
\end{align*}
 that is, the Frech\'{e}t mean for the mode and the arithmetic mean for the dispersion, both obtained from their respective posterior samples.
 
\begin{figure}
    \centering
    \includegraphics[width=0.8\linewidth, page=1]{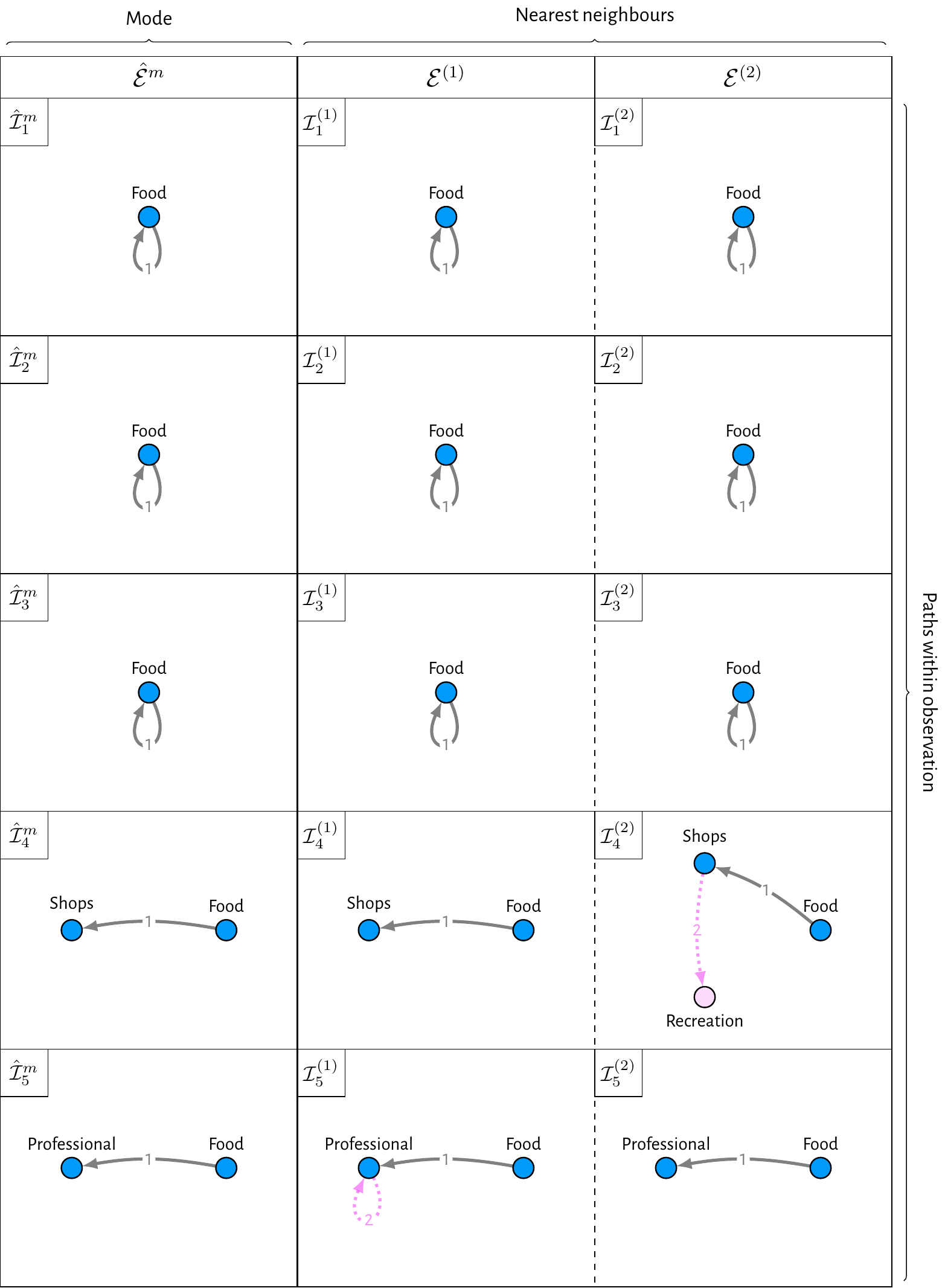}
    \caption{A subset of paths from our point estimate $\hat{\E}^m$ for the Foursquare data, alongside those of $\E^{(1)}$ and $\E^{(2)}$, its two nearest neighbours. Paths are aligned according to the optimal matching found when evaluating $d_{\text{M}}(\hat{\E}^m,\E^{(i)})$ for each neighbour $\E^{(i)}$. For each observed path $\I^{(i)}_j$, dashed pink edges and pink vertices indicate differences with $\hat{\I}^m_j$, with edges labels indicating the order of vertex visits. The remaining paths can be seen in \Cref{fig:fsq_sim_est2,fig:fsq_sim_est3}.}
    \label{fig:fsq_sim_est1}
\end{figure}

\begin{figure}
	\centering
	\includegraphics[page=2, width=0.85\linewidth]{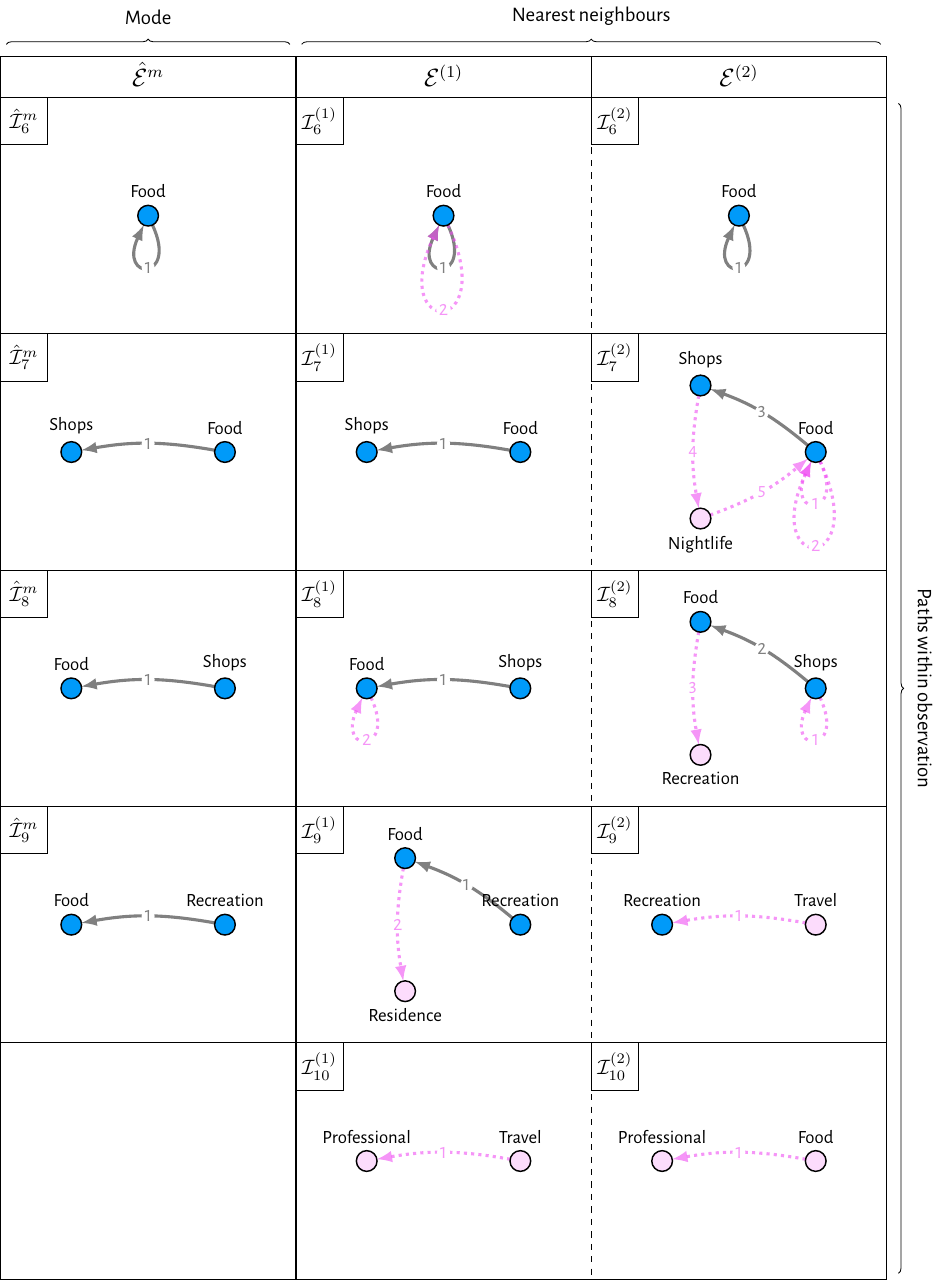}
	\caption{Paths of our point estimate $\hat{\E}^m$ for the Foursquare data, alongside those of $\E^{(1)}$ and $\E^{(2)}$, its two nearest neighbours. The remaining paths can be seen in \Cref{fig:fsq_sim_est1,fig:fsq_sim_est3}.}
	\label{fig:fsq_sim_est2}
\end{figure}

\begin{figure}
	\centering
	\includegraphics[page=3, width=0.85\linewidth]{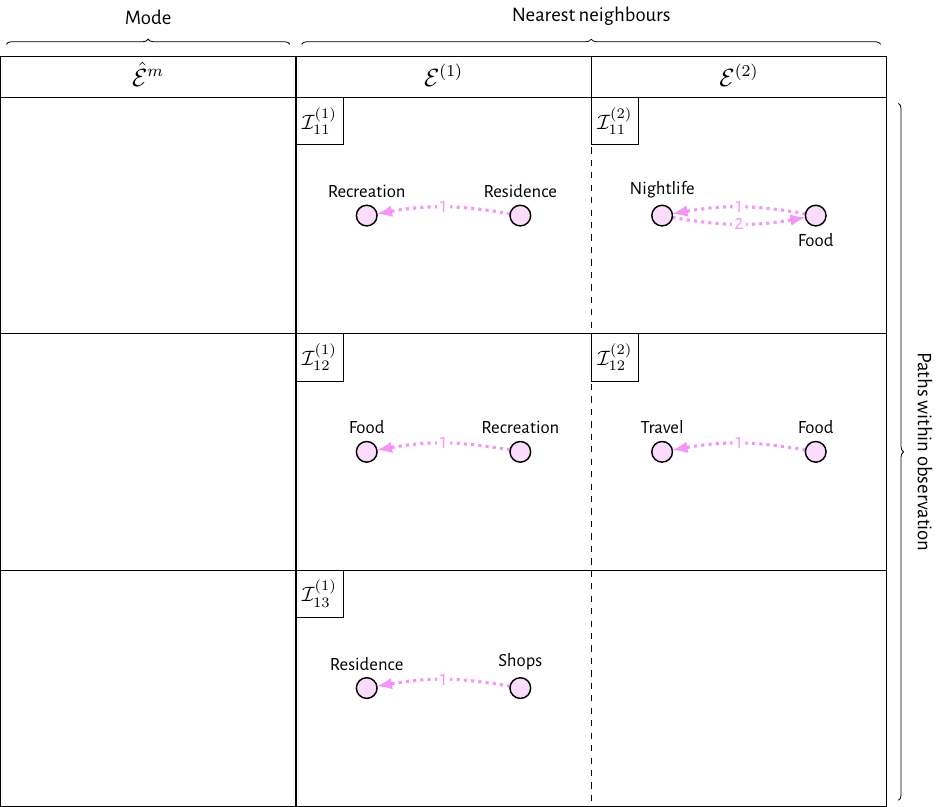} 
	\caption{Paths of our point estimate $\hat{\E}^m$ for the Foursquare data, alongside those of $\E^{(1)}$ and $\E^{(2)}$, its two nearest neighbours. The remaining paths can be seen in \Cref{fig:fsq_sim_est1,fig:fsq_sim_est2}.}
	\label{fig:fsq_sim_est3}
\end{figure}

As mentioned, due to our choice of distance, the inferred mode $\hat{\E}^m$ represents a collection of pathways frequently seen together in the observed data. To visualise this, we consider plotting the paths of $\hat{\E}^m$ alongside those of its two nearest observations. Supposing the data points have been labelled such that $\E^{(1)}$ and $\E^{(2)}$ denote the first and second nearest neighbours of $\hat{\E}^m$ with respect to $d_{\text{M}}$, writing these as follows
\begin{align*}
    \hat{\E}^m= \left\{\hat{\I}^m_1,\dots,\hat{\I}^m_{N^m}\right\} && \E^{(1)}=\left\{\I^{(1)}_1,\dots,\I^{(1)}_{N_1}\right\} && \E^{(2)} = \left\{\I^{(2)}_1,\dots,\I^{(2)}_{N_2}\right\},
\end{align*}
\Cref{fig:fsq_sim_est1,fig:fsq_sim_est2,fig:fsq_sim_est3} visualise the paths of $\hat{\E}^m$, $\E^{(1)}$ and $\E^{(2)}$ alongside one another. In each, the paths of $\E^{(1)}$ and $\E^{(2)}$ have been aligned in accordance with the optimal matching found when evaluating their distance from $\hat{\E}^m$ via $d_{\text{M}}$. In particular, in the $j$th row we plot $\hat{\I}^m_j$ alongside $\I^{(1)}_j$ and $\I^{(2)}_j$, denoting the paths matched to $\hat{\I}^m_j$ when evaluating $d_{\text{M}}(\hat{\E}^m, \E^{(1)})$ and $d_{\text{M}}(\hat{\E}^m, \E^{(2)})$, respectively. The paths of $\E^{(1)}$ and $\E^{(2)}$ not matched to any of $\hat{\E}^m$ are then shown in the remaining rows. 

Here one can observe paths of $\hat{\E}^m$ do indeed appear as subpaths within those of $\E^{(1)}$ and $\E^{(2)}$. In fact, in first three rows of \Cref{fig:fsq_sim_est1} they are equivalent, that is $\hat{\I}^m_j=\I^{(1)}_j=\I^{(2)}_j$, whilst for the remaining rows of \Cref{fig:fsq_sim_est1} and those of \Cref{fig:fsq_sim_est2,fig:fsq_sim_est3} we begin to see differences in the observed paths relative to those of the estimated mode, however, in almost all cases, the paths in the mode $\hat{\I}^m_j$ continue to feature as subpaths of both $\I^{(1)}_j$ and $\I^{(2)}_j$. Note also that no paths in $\hat{\E}^m$ are of length greater than two. At face value this might seem to imply use of this method gains nothing over a graph-based approach. However, as we illustrate in the next section, the subtle difference is that our inference is unambiguous. 

\begin{figure}
    \centering
    \includegraphics[width=\linewidth]{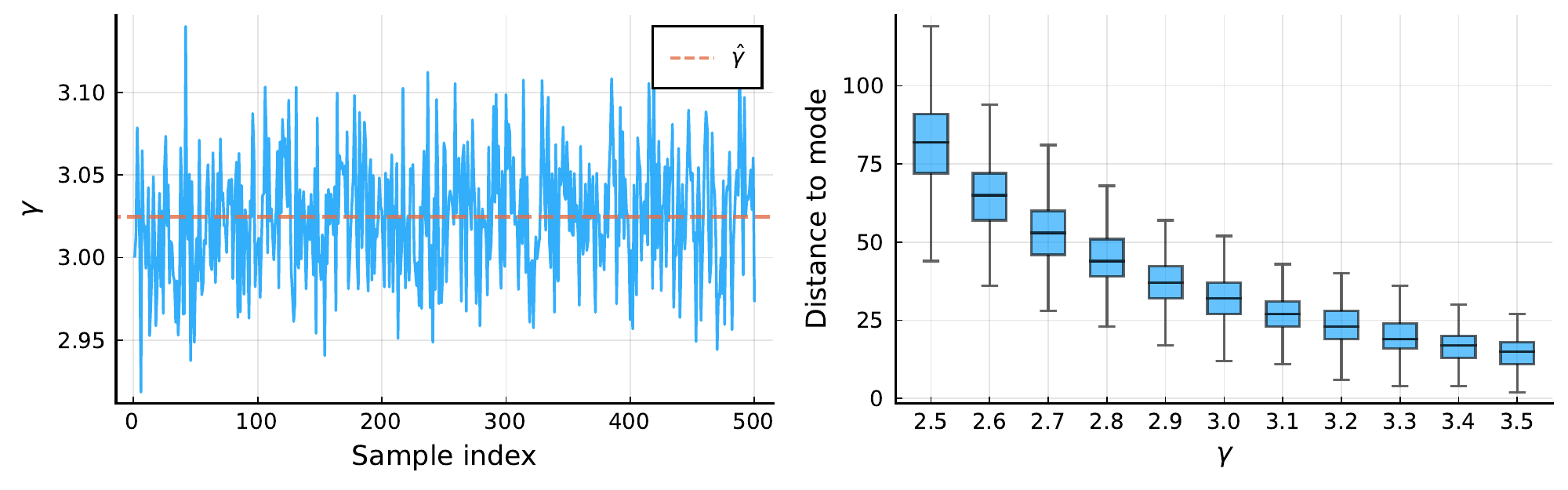}
    \caption{Summary of inference for the dispersion for the Foursquare data. Left shows a trace-plot of the posterior samples $\{\gamma_i\}_{i=1}^m$, whilst the right plot summarises the distribution of distances to the inferred mode for different values of $\gamma$, aiding interpretation of our estimate $\hat{\gamma}$.}
    \label{fig:fsq_sim_disp}
\end{figure}

For the dispersion, we have $\hat{\gamma}\approx 3.02$, with a trace-plot of the posterior samples $\{\gamma_i\}_{i=1}^M$ from which this estimate was obtained shown in the left-hand plot of \Cref{fig:fsq_sim_disp}. To aide interpretation of $\hat{\gamma}$, the right hand plot of \Cref{fig:fsq_sim_disp} visualises the distribution of $d_{\text{M}}(\E, \hat{\E}^m)$ where $\E \sim \text{SIM}(\hat{\E}^m, \gamma)$ for different values of $\gamma$, each boxplot summarising 1,000 samples drawn from the respective multiset model via our iMCMC algorithm (\Cref{sec_sup:inference_sim_model_sampling}). 
A comparison with our estimate $\hat{\gamma}$ shows that we expect the distance of samples to the mode to be around 25 (from $\gamma=3.0$), which, since we used the matching distance, can be seen as the average number of edit operations required to transform the mode into an observation. 
Considering the mode has 18 entries in total (9 paths of length two), this implies a reasonable amount of variability in the observed data.



\subsection{Comparison with Graph-Based Inferences}

\label{sec:data_analysis_graphs}


An alternative to applying our methodology is to convert observations to graphs in a pre-processing step, before applying a suitable graph-based method. As such, we consider striking a comparison between this approach and ours. The intention here is twofold. On one hand, to show the graph-based inferences are not too dissimilar from that obtained via our approach. Whilst on the other, that our approach goes beyond the graph-based methods, in so far as producing an inference which is unambiguous regarding the presence of higher-order information in the observed data. 

Given the observed sample $\edata$, one can obtain a sample of graphs $\gdata$ via aggregation, namely, by letting $\G^{(i)} = \G_{\E^{(i)}}$, the graph obtained by aggregating the paths of $\E^{(i)}$, as outlined in \Cref{sec:background}. In the same way that our estimate $\hat{\E}^m$ summarises the sample $\edata$, one can now consider obtaining a graph $\hat{\G}$ which summarises the sample $\gdata$. This can be achieved through a variety of different approaches, the choice of which will depend on whether the $\G^{(i)}$ are graphs or multigraphs. We will consider both cases here. In each instance, the ouput summary $\hat{\G}$ will be either a graph or multigraph, which we then compare with $\G_{\hat{\E}^m}$, the aggregation of our estimate $\hat{\E}^m$. 

To aide this exposition, we will make use of the graph \textit{adjacency matrix}. For a graph $\G = (\E,\V)$, where $\V = \{1,\dots,V\}$, its adjacency matrix $A^\G \in \{0,1\}^{V \times V}$ is the binary matrix with
\begin{equation*}
    \begin{aligned}
    A^\G_{ij} = \begin{cases}
    1 & \text{if } (i,j) \in \E \\ 
    0 & \text{else},
    \end{cases}
    \end{aligned}
\end{equation*}
whilst if $\G=(\E,\V)$ is a multigraph its adjacency matrix $A^\G \in \mathbb{Z}_{\geq 0}^{V \times V}$, is defined by letting $A^\G_{ij}$ equal the number of times $(i,j)$ appears in $\E$. Note there is a one-to-one correspondence between graphs and adjacency matrices, and as such they can be used interchangeably for convenience.

\begin{figure}
    \centering
    \begin{subfigure}{0.48\linewidth}
        \centering
        \includegraphics[page=3,width=0.7\textwidth]{img/data analysis/graph_model_mode_plots.pdf}
        \caption{$\hat{\G}_{\text{CER}}$}
        \label{fig:cer_est}
    \end{subfigure}
    \begin{subfigure}{0.48\linewidth}
        \centering
        \includegraphics[page=1,width=0.7\textwidth]{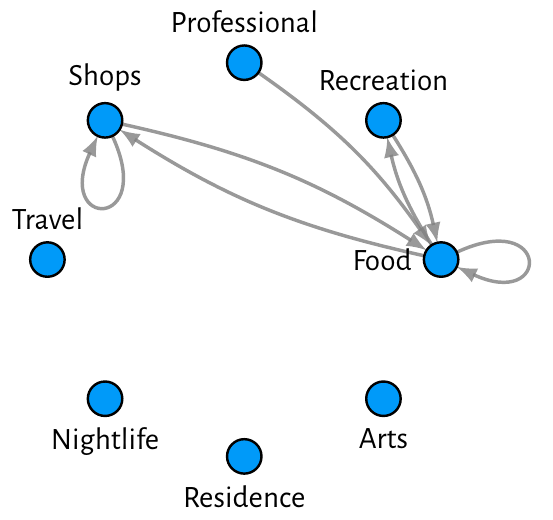}
        \caption{$\hat{\G}_{\text{MV}}$}
        \label{fig:mv_est}
    \end{subfigure}
    \begin{subfigure}{0.48\linewidth}
        \centering
        \vspace{0.5cm}
        \includegraphics[page=4,width=0.7\textwidth]{img/data analysis/graph_model_mode_plots.pdf}
        \caption{$\hat{\G}_{\text{SNF}}$}
        \label{fig:snf_est}
    \end{subfigure}
    \begin{subfigure}{0.48\linewidth}
        \centering
        \vspace{0.5cm}
        \includegraphics[page=2,width=0.7\textwidth]{img/data analysis/graph_model_mode_plots.pdf}
        \caption{$\hat{\G}_{\text{RM}}$}
        \label{fig:rm_est}
    \end{subfigure}
    \begin{subfigure}{0.48\linewidth}
        \centering
        \vspace{0.5cm}
        \includegraphics[page=5,width=0.7\textwidth]{img/data analysis/graph_model_mode_plots.pdf}
        \caption{$\G_{\hat{\E}^m}$}
        \label{fig:sim_est_agg}
    \end{subfigure}
    \caption{A comparison with graph-based inferences. Here (e) shows $\G_{\hat{\E}^m}$, the aggregate multigraph of our point estimate $\hat{\E}^m$ of \Cref{sec:data_analysis_sim}, whilst (a)-(d) show alternative inferences obtained via graph-based approaches outlined in \Cref{sec:data_analysis_graphs}. Note that (a) and (b) are graphs, whilst (c)-(e) are multigraphs, with edge thickness proportional to their weight. }
\end{figure}

In the case where each $\G^{(i)}$ is a graph, and thus each $A^{\G^{(i)}}$ is a binary matrix, a simple model-free summary is the majority vote, which we denote $\hat{\G}_{\text{MV}}$, where we include an edge if it was observed in at least one half of the observations. More formally, $\hat{\G}_{\text{MV}}$ can be defined in terms of its adjacency matrix as follows 
$$A^{\hat{\G}_{\text{MV}}}_{ij} = \mathds{1}(\bar{A}_{ij} \ge 0.5),$$
where $\bar{A}$ is the real-valued matrix with entries $\bar{A}_{ij} = \frac{1}{n}\sum_{k=1}^n A_{ij}^{\G^{(k)}}$, that is, the entry-wise average of the observed adjacency matrices.
As a model-based alternative, we turn to the centered Erd\"{o}s-R\'{e}nyi (CER) model of \cite{Lunagomez2021}. Using the notation $\G \sim \text{CER}(\G^m, \alpha)$ when a graph $\G$ was drawn from the CER model with mode $\G^m$ (a graph), and noise parameter $0 \leq \alpha \leq 0.5$, we assumed the following hierachical model
\begin{equation*}
    \begin{aligned}
    \G^{(i)} \, | \, \G^m, \alpha &\sim \text{CER}(\G^m, \alpha) \quad \text{(for $i=1,\dots,n$)}\\
    \G^m &\sim \text{CER}(\G_0, \alpha_0)\\
    \alpha &\sim 0.5 \cdot \text{Beta}(\beta_1, \beta_2)\\ 
    \end{aligned}
\end{equation*}
where $\G_0$ (a graph), $0 \leq \alpha_0 \leq 0.5$, $\beta_1>0$ and $\beta_2>0$ denote hyperparameters. For this analysis, we assumed an uniformative set-up, letting $\G_0=\hat{\G}_{\text{MV}}$ and $\alpha_0=0.5$, leading to a uniform distribution over the space of graphs for the prior on $\G^m$, whilst we took $\beta_1=\beta_2=1$, similarly leading to the uniform distribution over the interval $(0,0.5)$ for the prior on $\alpha$. Following the scheme of \cite{Lunagomez2021}, we drew a sample $\{(\G^m_i,\alpha_i)\}_{i=1}^M$ from the posterior $p(\G^m, \alpha | \gdata)$ via MCMC, obtaining the desired summary via the sample Frech\'{e}t mean $$\hat{\G}_{\text{CER}} = \argmin_{\G \in \{\G^m_i\}} \sum_{i=1}^n d^2_{\text{H}}(\G, \G^m_i)$$
where $d_{\text{H}}$ denotes the Hamming distance between graphs \citep{Lunagomez2021,Donnat2018b}. \Cref{fig:cer_est,fig:mv_est} show these two un-weighted summaries, $\hat{\G}_{\text{CER}}$ and $\hat{\G}_{\text{MV}}$, respectively, for the Foursquare data, where it transpires that  $\hat{\G}_{\text{CER}}=\hat{\G}_{\text{MV}}$. 

In the case where each $\G^{(i)}$ is a multigraph, and thus each $A^{\G^{(i)}}$ is a matrix of non-negative integers, an analogous model-free summary can be obtained by rounding the entries of $\bar{A}$ to the nearest integer. Referring to this as the rounded mean estimate and denoting it $\hat{\G}_{\text{RM}}$, it can be defined formally via its adjacency matrix as follows
$$A_{ij}^{\hat{\G}_{\text{RM}}} = \lfloor \bar{A}_{ij} \rfloor + \mathds{1}(\bar{A}_{ij} - \lfloor \bar{A}_{ij} \rfloor \geq 0.5),$$
where the notation $\lfloor x \rfloor$ for $x \in \mathbb{R}$ denotes the floor function. As a model-based approach, we consider using the SNF models proposed by \cite{Lunagomez2021}. Though originally proposed to model graphs, they can be readily extended to handle multigraphs (see \Cref{sec_sup:data_analysis_snf}). Use of the SNF, like our models, requires specification of a distance metric between graphs. We considered taking the absolute difference of edge multiplicities, or alternatively, the adjacency matrix entries, that is
$$d_1(\G, \G^\prime) = \sum_{i,j} \left|A^{\G}_{ij} - A^{\G^\prime}_{ij}\right|, $$
which can be seen as the generalisation of the Hamming distance to multigraphs. Adopting the notation $\G \sim \text{SNF}(\G^m, \gamma)$ when a graph $\G$ is drawn from the SNF model with mode $\G^m$ (a multigraph) and dispersion $\gamma>0$, we assumed the following hierarchical model
\begin{equation*}
    \begin{aligned}
    \G^{(i)} \, | \, \G^m, \gamma  &\sim \text{SNF}(\G^m, \gamma) \quad \text{(for $i=1,\dots,n$)}\\
    \G^m &\sim \text{SNF}(\G_0, \gamma_0)\\
    \gamma &\sim \text{Gamma}(\alpha, \beta)
    \end{aligned}
\end{equation*}
where $\G_0$ (a multigraph), $\gamma_0>0$, $\alpha>0$ and $\beta>0$ are hyperparameters. For this analysis, we took $\G_0$ to be the sample Frech\'{e}t mean of the observed multigraphs $\{\G^{(i)}\}_{i=1}^n$ with respect to the distance $d_1$, whilst we let $\gamma_0=0.1$, $\alpha=3$ and $\beta=1$. Again, we obtained a sample $\{(\G^m_i,\gamma_i)\}_{i=1}^M$ from the posterior $p(\G^m, \gamma | \gdata)$ via MCMC, before invoking the sample Frech\'{e}t mean to obtain the desired summary
$$\hat{\G}_{\text{SNF}} = \argmin_{\G \in \gdata} \sum_{i=1}^n d_1^2(\G, \G^m_i).$$
Note that the posterior here will be doubly-intractable, necessitating use of a specialised MCMC algorithm. \cite{Lunagomez2021} adopted the algorithm of \cite{moller2006}, however, since here we consider multigraphs, we cannot apply their scheme directly. Instead, we took an alternative approach via the exchange algorithm \citep{Murray2006}, details of which can be found in \Cref{sec_sup:data_analysis_snf}. Visualisations of these two multigraph summaries, $\hat{\G}_{\text{SNF}}$ and $\hat{\G}_{\text{RM}}$, can be seen in \Cref{fig:snf_est,fig:rm_est}, respectively.

Comparing the graph-based methods amongst themselves, we see a slight variation in the signal they uncover. For example, in taking edge multiplicities into account, the multigraph-based estimate $\hat{\G}_{\text{RM}}$ introduces edges which did not appear in either of the graphs $\hat{\G}_{\text{CER}}$ and $\hat{\G}_{\text{CER}}$, generally involving the node corresponding to food venues. Conversely, the SNF-based estimate $\hat{\G}_{\text{SNF}}$ appears to instead disregard edges which appear in $\hat{\G}_{\text{CER}}$ and $\hat{\G}_{\text{CER}}$. Sitting somewhere in between is the aggregate of our estimate $\G_{\hat{\E}^m}$, which appears to have some degree of similarity with each of the graph-based summaries, as we intended to confirm. Moreover, a common theme seems to appear (for all summaries). Namely, that visits to food venues feature strongly, often followed or preceded by a visit to another food venue or some other venue category, with shopping venues being a prevalent choice. 

Naturally, one is inclined to ask if the aggregation of our estimate $\G_{\hat{\E}^m}$ is not too dissimilar to those obtained via analysis of the data processed into graphs or multigraphs, then what does one gain by taking our approach? Recalling that we estimated $\hat{\E}^m$, a multiset of paths, we argue this contains more information regarding the signals present in the observed data than any graph-based summary, assuming the data were truly path-observed. This comes back to a point made in \Cref{sec:background}, namely that when one aggregates paths a loss of information is incurred, which will in turn limit the conclusions one can draw concerning the original data. For example, consider the CER-based summary $\hat{\G}_{\text{CER}}$ of \Cref{fig:cer_est}, where we see the following two edges 
\begin{align*}
     e_1 = (\text{recreation},\text{food}) && e_2 = (\text{food}, \text{shops}).
\end{align*}
This could imply at least two things. Perhaps many users went from a recreation venue to a food venue, and separately, that is, on a different day, from a food venue to a shopping venue. Alternatively, maybe many users traced the path recreation $\to$ food $\to$ shops in a single day. Both are possibilities, and from a graph-based summary there is no way of knowing which is the case. However, in directly estimating a collection of paths, we can make such distinctions. For example, considering our estimate $\hat{\E}^m$ for the Foursquare data (\Cref{fig:fsq_sim_est1,fig:fsq_sim_est2,fig:fsq_sim_est3}), it appears we are in the former case, since no paths therein are of length greater than two.

\section{Discussion}

\label{sec:discussion}

In this paper, we have motivated and instantiated the study of multiple interaction-network data. We have proposed a flexible Bayesian modelling framework capable of analysing such data without the need to perform any aggregation of observations. This has been supplemented with specialised MCMC schemes, facilitating inference for our proposed models. Through simulation studies, we have confirmed the efficacy of our approach and inference scheme, whilst the applicability of our methodology has been illustrated by an analysis of Foursquare check-in data, where we illustrated how our methodology can be used to answer inferential questions (a) and (b) posed in \Cref{sec:intro}. Moreover, in comparing with graph-based methods we highlighted the extra information one subtly gains by taking our approach.

Regarding future work, there are a few ways one might consider building upon what has been proposed here. Firstly, a natural extension of our models is to consider a mixture model, with our SIS or SIM models functioning as mixture components, which would allow one to capture heterogeneity in the observations, opening the door to answering question (c) of \Cref{sec:intro}. Secondly, on a more pragmatic note, one could also take steps to scale-up our approach computationally. For example, one might be able to circumvent the need to use the exchange algorithm if the normalising constant for a particular distance metric was derived, as was the case for the CER model in \cite{Lunagomez2021}. Finally, if one is able to make an exchangeability assumption for each observation, that is, the order in which paths arrive is not of interest, then a slightly modified model structure could be considered, reminiscent of the latent Dirichlet allocation (LDA) model \citep{Blei2003}. Namely, one could assume each observation was drawn from some mixture distribution over paths, with mixture components being shared between observations but mixture proportions differing. This would also have a natural non-parametric extension via the hierarchical Dirichlet process (HDP) \citep{Teh2006}. It would be interesting to see how the inferences from such an approach compare with ours, at least qualitatively, and whether any computational benefit would be achieved.


More tangentially, one could also follow the path laid in the wider literature on multiple networks and consider extending models designed to analyse a single interaction network, for example, the models of \cite{Crane2018} or \cite{Williamson2016}. 

Finally, it could be interesting to consider the situation where one has access to covariate information at the level of observations. For example, considering the Foursquare data, one might have additional information for each user, such as their line of work or country of residence. Interest might then be in defining a modelling framework which could be invoked to examine for a relationship between covariates and observed data. Such developments would mirror those in the wider literature on multiple-network data, such as work on hypothesis testing \citep{Ginestet2014,durante2018testing,ghoshdastidar2020,chen2021hypothesis}.

\begin{appendices}
	
	\appendix
	
	\section{Sample Spaces}

Here we provide extra details regarding the sample spaces for our SIS and SIM models of \Cref{def:SIS,def:SIM}. Recall that distributions within the defined families constitute distributions over sequences and multisets (of paths) respectively. Thus, their respective sample spaces consist of \textit{all} such objects. 

Assuming the vertex set $\V$ is fixed, we first define the space of all paths over $\V$ as follows
\begin{equation*}
    \begin{aligned}
        \Ib := \{(x_1,\dots,x_n) \, : \, x_i \in \V, n\geq1 \},
    \end{aligned}
\end{equation*}
subsequently defining the space of interaction sequences $\iSb$ via
\begin{equation*}
    \begin{aligned}
        \iSb := \{ (\I_1,\dots,\I_N) \, : \, \I_i \in \Ib, N \geq 1 \},
    \end{aligned}
\end{equation*}
whilst, with $\E_\iS$ denoting the multiset obtained from the sequence $\iS$ by disregarding the order of paths therein, the space of interaction multisets $\Eb$ can be defined via
\begin{equation*}
    \begin{aligned}
        \Eb := \{\E_\iS \, : \, \iS \in \iSb\}
    \end{aligned}
\end{equation*}
where here we are abusing notation slightly, since we can have $\E_\iS=\E_{\iS^\prime}$ for $\iS \not=\iS^\prime$ (when equal up to a permutation of interactions), but we just assume such values have been included once and so $\Eb$ is indeed a set. 

We note that $\Eb$ also admits another interpretation as a partitioning of $\iSb$ into equivalence classes. To see this, first define an equivalence relation on $\iSb$ via permutations, in particular we write $\iS \overset{p}{\sim} \iS^\prime$ if there is some permutation $\sigma$ such that $\iS^\prime = \iS^\sigma$, where $\iS^\sigma = (\I_{\sigma (1)},\dots,\I_{\sigma(N)})$ is the interaction sequence obtained by permuting the interactions of $\iS$ via $\sigma$. Now, observe that each $\E \in \Eb$ can be seen as an equivalence class of interaction sequences obtained via $\overset{p}{\sim}$, that is 
$$\E = \{\iS \in \iSb \, : \, \iS \overset{p}{\sim} \tilde{\iS}\}$$
where $\tilde{\iS}$ denotes some arbitrary ordering of the interactions of $\E$. Thus, $\Eb$ is in a sense the union of such sets and partitions $\iSb$.

\label{sec_sup:sample_spaces}

\subsection{Bounding Dimensions}

\label{sec_sup:sample_spaces_bounded}

As highlighted \Cref{sec:modelling}, it can in practice be somewhat easier to work with bounded sample spaces, since in the unbounded case our models are not guaranteed to be proper, which can lead to a divergence of dimension when doing MCMC sampling for some parameterisations (typically when $\gamma$ is low). This can further cause computational instabilities if one uses such MCMC sampling within other algorithms, as we do within our scheme to sample from the posterior. In this section, we state our notation for the bounded sample spaces.

With regards to the objects we consider, there are two things we can bound: (i) the size of paths and (ii) the number of paths. Referring to these as the inner and outer dimensions respectively, we specify two integers $K$ and $L$ bounding their values and define our sample spaces accordingly. Assuming that the vertex set $\V$ is fixed, and $K \in \mathbb{Z}_{\geq 1}$ we let 
\begin{equation*}
    \begin{aligned}
        \Ib_K := \{(x_1,\dots,x_n) \, : \, x_i \in \V, 1  \leq n\leq K \}
    \end{aligned}
\end{equation*}
denote the space of paths up to length $K$, and then with $L \in \mathbb{Z}_{\geq 1}$ we let 
\begin{equation*}
    \begin{aligned}
        \iSb_{K,L} := \{ (\I_1,\dots,\I_N) \, : \, \I_i \in \Ib_K, 1 \leq N \leq L \},
    \end{aligned}
\end{equation*}
denote the space of interaction sequences with at most $L$ paths of length at most $K$. The analogous bounded space of interaction multisets in then given by
\begin{equation*}
    \begin{aligned}
        \Eb_{K,L} := \{\E_\iS \, : \, \iS \in \iSb_{K,L}\}.
    \end{aligned}
\end{equation*}

\section{Simulation Studies}

This section contains supporting details for the simulation studies of \Cref{sec:simulations}. In particular, we discuss how parameters were chosen for the simulation of \Cref{sec:sim_hw}, and provide a derivation for the posterior predictive approximation used in \Cref{sec:sim_predictive}.

\subsection{Posterior Concentration Parameter Choices}

\label{sec_sup:hw_par_choices}

Recall that in the simulation of \Cref{sec:sim_hw} we re-sampled the true mode via
$$\iS_{\text{true}} \sim \text{Hollywood}(\alpha, -\alpha V, \nu)$$
where $V=20$ and $\nu = \text{TrPoisson}(3,1,10)$, whilst $\alpha<0$ we varied. As mentioned in \Cref{sec:sim_hw}, the parameter $\alpha$ can be seen to control the tail of the vertex count distribution. As such, rather than choosing $\alpha$ on an even grid we instead consult a summary measure quantifying the `heavy-tailedness' of the degree distribution, before choosing values so as to evenly represent different structures for $\iS_{\text{true}}$ (as quantified by this degree distribution).

For a given observation $\iS$, recall the following definition
$$k_{\iS}(v) := \text{\# times $v$ appears in $\iS$},$$
which for each $\iS$ implies a sample $\{k_\iS(v) \,: v \in \V,\, k_\iS(v) >0\}$, similar to the degree distribution. Now, the summary measure we considered was the 95\% quantile of this sample. 

Through simulation, we examined how $\alpha$ controls the expected value of this 95\% quantile (expected since $\iS$ is sample randomly from a Hollywood model). In particular, for a range of $\alpha$ values, we (i) drew a sample $\{\iS^{(i)}\}_{i=1}^n$, from a $\text{Hollywood}(\alpha, -\alpha V, \nu)$ model, taking $\nu$ and $V$ as above, drawing a total of $N=10$ paths in each case, then (ii) for $i=1,\dots,n$ we evaluated the 95\% quantile of the sample $\{k_{\iS^{(i)}}(v) \,: v \in \V,\, k_{\iS^{(i)}}(v) >0\}$, before returning the mean value of these quantiles. 

\begin{figure}
    \centering
    \includegraphics[width=0.95\linewidth]{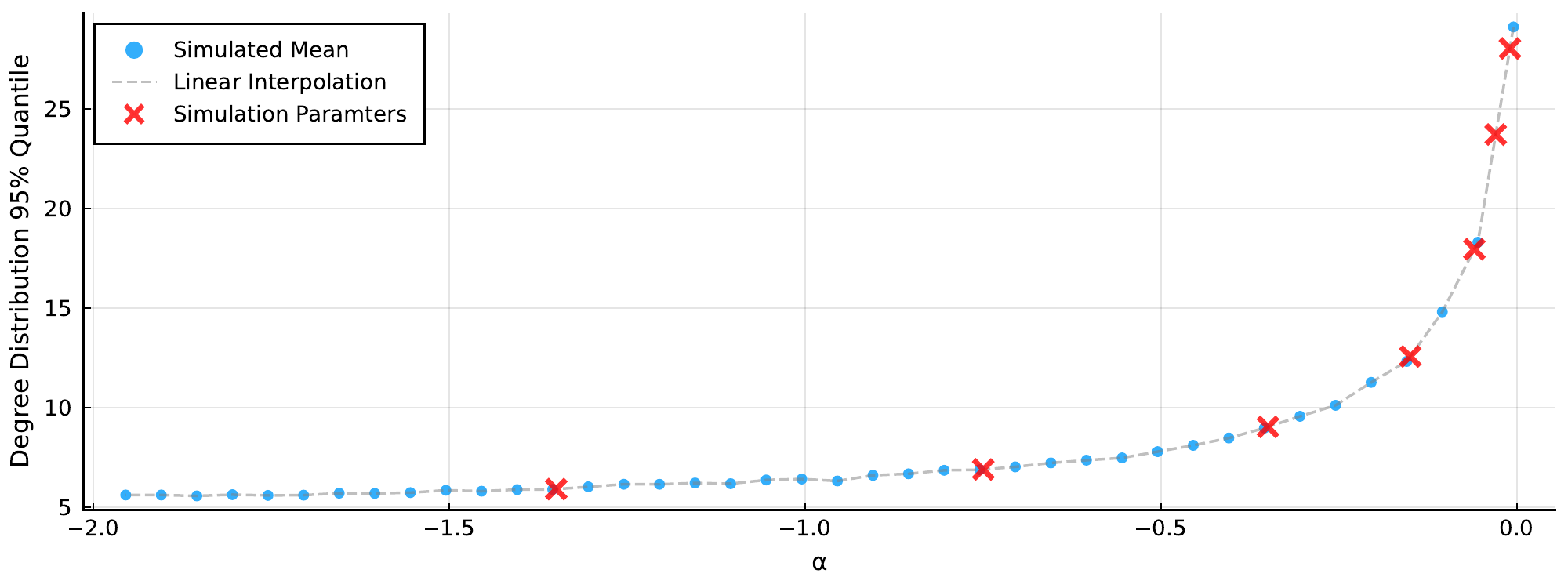}
    \caption{Summary of Hollywood model simulation used to select parameters for simulation of \Cref{sec:sim_hw}. Plot shows simulated mean degree distribution 95\% quantiles for $\text{Hollywood}(\alpha, -\alpha V, \nu)$ model, where $V=20$, $\nu=\text{TrPoisson}(3,1,10)$ and $\alpha$ varies. Via linear interpolation (dashed line), we choose $\alpha$ values (crosses) to get an even spread over the expected degree distribution quantiles.}
    \label{fig_sup:sim_hw_par_choices}
\end{figure}

\Cref{fig_sup:sim_hw_par_choices} summarises the output with $n=1000$ samples, where circular markers show the mean quantiles. Towards choosing simulation parameters, we next constructed a function mapping all $\alpha<0$ to an expected quantile via a linear interpolation, as shown in \Cref{fig_sup:sim_hw_par_choices} by the dashed line, which allowed us to select $\alpha$ (red crosses) providing an even spread of expected degree-distribution 95\% qauntiles.

\subsection{Posterior Predictive for Missing Entries}

\label{sec_sup:simulation_predictive}

Here we show how one can obtain an approximation for the missing-entry posterior predictive using a sample from the posterior, as used in \Cref{sec:sim_predictive}. First, observe that any sample $\{(\iS^m_i,\gamma_i)\}_{i=1}^m$ from the posterior implies the following atomic approximation thereof
\begin{equation}
    \begin{aligned}
        \hat{p}(\iS^m,\gamma | \data) &= \frac{1}{m} \sum_{i=1}^m \mathds{1}(\iS^m=\iS^m_i) \cdot \delta(\gamma-\gamma_i)
    \end{aligned}
    \label{eq_sup:SIS_predictive_approx}
\end{equation}
where $\delta(\cdot)$ is the Dirac delta function. 

As in \Cref{sec:sim_predictive}, with $\iS_x$ denoting the observation with missing entry filled in to be $x$, then given some parameters $(\iS^m,\gamma)$ we have the true predictive for $x$ given by
\begin{equation*}
    \begin{aligned}
        p(x|\iS^m,\gamma,\iS_{-x}) := \frac{1}{Z(\iS^m, \gamma, \iS_{-x})}\exp\{-\gamma d_S(\iS_x, \iS^m)\}
    \end{aligned}
\end{equation*}
with 
$$Z(\iS^m, \gamma, \iS_{-x}) = \sum_{x\in \V} \exp\{-\gamma d_S(\iS_x, \iS^m)\}.$$
The posterior predictive is now obtained by averaging with respect to the posterior 
\begin{equation*}
    \begin{aligned}
        p(x|\data,\iS_{-x}) &= \sum_{\iS^m \in \iSb} \int_{\mathbb{R}_+}p(x|\iS^m, \gamma,\iS_{-x})p(\iS^m, \gamma|\data)d\gamma, \\
    \end{aligned}
\end{equation*}
which we can now approximate by substituting in \eqref{eq_sup:SIS_predictive_approx} as follows
\begin{equation*}
    \begin{aligned}
        \hat{p}(x|\data,\iS_{-x}) &:=   \sum_{\iS^m \in \iSb} \int_{\mathbb{R}_+}p(x|\iS^m, \gamma,\iS_{-x})\hat{p}(\iS^m, \gamma|\data)d\gamma \\
        &=  \sum_{\iS^m \in \iSb} \int_{\mathbb{R}_+}p(x|\iS^m, \gamma)\left( \frac{1}{m}\sum_{i=1}^m \mathds{1}\left(\iS^m = \iS^m_i\right) \delta(\gamma-\gamma_i) \right)d\gamma \\
        &= \frac{1}{m}\sum_{i=1}^m p(x|\iS^m_i,\gamma_i),
    \end{aligned}
\end{equation*}
which is exactly as stated in \Cref{sec:sim_predictive}.

A pragmatic note here is that as the posterior concentrates the number of unique values in the sample $\{\iS^m_i\}_{i=1}^m$ will typically not be too large. Since we need only evaluate the distance metric (which is typically quite costly) at these values, this predictive is feasible to evaluate.

\end{appendices}


\newpage
\begin{appendices}
	\setcounter{section}{0}
	\renewcommand{\thesection}{S\arabic{section}}    
	
	\crefalias{section}{supp}
	\crefalias{subsection}{supp}
	\crefalias{subsubsection}{supp}
	
	\section*{\Large Supplementary Material}
	
	\vspace{0.5cm}
	
	\section{Montonicity of the Entropy}
\label{sec_sup:monotonicity}
Here we examine the entropy for the SIS and SIM model families (\Cref{def:SIS,def:SIM} respectively), in particular, we confirm it is monotonic with respect to the dispersion. A similar result was shown by \cite{Lunagomez2021} (Proposition 3.3, proved pages 41-43 of Supplementary Material), and as such we follow the reasoning used therein.

For the SIS model, recall the entropy  is given by
\begin{equation*}
    \begin{aligned}
    H(\iS^m, \gamma) &= - \mathbb{E}[\log p(\iS| \iS^m, \gamma)]\\
    &= - \sum_{\iS \in \iSb} \log \left(\frac{\exp\{ - \gamma d_S(\iS, \iS^m)\}}{Z(\iS^m, \gamma)} \right) \frac{\exp\{ - \gamma d_S(\iS, \iS^m)\}}{Z(\iS^m, \gamma)} \\
    &=  - \bigg( \sum_{\iS \in \iSb} - \gamma d_S(\iS, \iS^m) \frac{\exp\{ - \gamma d_S(\iS, \iS^m)\}}{Z(\iS^m, \gamma)}\\
    & \qquad \qquad -  \log Z(\iS^m, \gamma) \sum_{\iS \in \iSb} \frac{\exp\{ - \gamma d_S(\iS, \iS^m)\}}{Z(\iS^m, \gamma)} \bigg)\\
    &= \gamma \left(\sum_{\iS \in \iSb} d_S(\iS, \iS^m)\frac{\exp\{ - \gamma d_S(\iS, \iS^m)\}}{Z(\iS^m, \gamma)}\right)+ \log Z(\iS^m, \gamma)\\
    &= \gamma \times \mathbb{E}[d_S(\iS, \iS^m)] + \log Z(\iS^m,\gamma).
    \end{aligned}
\end{equation*}

Unfortunately, as was the case for the normalising constant $Z(\iS^m, \gamma)$ (\Cref{sec:models_properties}), since $\iSb$ is infinite we have no guarantee that $H(\iS^m, \gamma)$ will exist. However, what we can say is that, when $H(\iS^m,\gamma)$ exists, it is monotonic in $\gamma$. To show this, we first differentiate $H(\iS^m, \gamma)$ with respect to $\gamma$
\begin{equation*}
    \begin{aligned}
     \frac{\partial}{\partial \gamma}H(\iS^m, \gamma) = \frac{\partial}{\partial \gamma} \mathbb{E}[d_S(\iS, \iS^m)] + \mathbb{E}[d_S(\iS, \iS^m)] + \frac{\partial}{\partial \gamma} \log Z(\iS^m,\gamma) 
    \end{aligned}
\end{equation*}
where one has
\begin{equation}
    \label{eq_sup:sim_monotone_log_normconst_derivative}
    \begin{aligned}
    \frac{\partial}{\partial \gamma} \log Z(\iS^m, \gamma) &= \frac{\frac{\partial}{\partial \gamma} Z(\iS^m, \gamma)}{Z(\iS^m, \gamma)} \\
    &= \frac{1}{Z(\iS^m, \gamma)}\frac{\partial}{\partial \gamma} \left(\sum_{\iS \in \iSb} \exp \left\{- \gamma d_S(\iS, \iS^m) \right\} \right) \\
    &= \frac{1}{Z(\iS^m, \gamma)} \sum_{\iS \in \iSb} \frac{\partial}{\partial \gamma}\exp \left\{- \gamma d_S(\iS, \iS^m) \right\} \\
    &= \frac{1}{Z(\iS^m, \gamma)} \sum_{\iS \in \iSb} (-d_S(\iS, \iS^m)) \exp \left\{- \gamma d_S(\iS, \iS^m) \right\} \\
    &= - \sum_{\iS \in \iSb} d_S(\iS, \iS^m)\frac{1}{Z(\iS^m, \gamma)} \exp \left\{- \gamma d_S(\iS, \iS^m) \right\}  \\
    &= - \mathbb{E}[d_S(\iS, \iS^m)],
    \end{aligned}
\end{equation}
thus implying
\begin{equation*}
    \begin{aligned}
    \frac{\partial}{\partial \gamma}H(\iS^m, \gamma) = \frac{\partial}{\partial \gamma} \mathbb{E}[d_S(\iS, \iS^m)].
    \end{aligned}
\end{equation*}
Now, we have 

\begingroup
\allowdisplaybreaks
\begin{align}
    \frac{\partial}{\partial \gamma} \mathbb{E}[d_S(\iS, \iS^m)] &= \frac{\partial}{\partial \gamma} \left(\frac{1}{Z(\iS^m,\gamma)}\sum_{\iS \in \iSb} d_S(\iS, \iS^m)\exp\{ - \gamma d_S(\iS, \iS^m)\}\right) \nonumber\\
    &=-\frac{\frac{\partial}{\partial \gamma} Z(\iS^m, \gamma)}{Z(\iS^m, \gamma)^2}\left(\sum_{\iS \in \iSb} d_S(\iS, \iS^m)\exp\{ - \gamma d_S(\iS, \iS^m)\}\right) \nonumber \\
    &\qquad \qquad - \frac{1}{Z(\iS^m,\gamma)} \left(\sum_{\iS \in \iSb} d_S(\iS, \iS^m)^2\exp\{ - \gamma d_S(\iS, \iS^m)\}\right)\nonumber\\
    &=-\frac{\frac{\partial}{\partial \gamma} Z(\iS^m, \gamma)}{Z(\iS^m, \gamma)^2}\left(\sum_{\iS \in \iSb} d_S(\iS, \iS^m)\exp\{ - \gamma d_S(\iS, \iS^m)\}\right)\nonumber \\
    &\qquad \qquad - \frac{1}{Z(\iS^m,\gamma)} \left(\sum_{\iS \in \iSb} d_S(\iS, \iS^m)^2\exp\{ - \gamma d_S(\iS, \iS^m)\}\right) \nonumber\\
    \label{eq_sup:sim_monotone_step1}
    &= -\frac{\frac{\partial}{\partial \gamma} Z(\iS^m, \gamma)}{Z(\iS^m, \gamma)}\mathbb{E}[d_S(\iS, \iS^m)] - \mathbb{E}[d_S(\iS, \iS^m)^2]  \\[0.2cm] \label{eq_sup:sim_monotone_step2}
    &= \mathbb{E}[d_S(\iS, \iS^m)]^2 - \mathbb{E}[d_S(\iS, \iS^m)^2] \\[0.2cm]
    &= \mathbb{V}\text{ar}[d_S(\iS, \iS^m)] \nonumber,
\end{align}
\endgroup
where \eqref{eq_sup:sim_monotone_step2} follows from \eqref{eq_sup:sim_monotone_step1} by applying \eqref{eq_sup:sim_monotone_log_normconst_derivative}. Now, observe that if $\frac{\partial}{\partial \gamma}H(\iS^m,\gamma) >0$ this implies $H(\iS^m,\gamma)$ is monotonic in $\gamma$, as desired. By the derivations above, this is equivalent to saying we have montonicity provided $\mathbb{V}\text{ar}[d_S(\iS, \iS^m)]> 0$. This result, identical to that of \cite{Lunagomez2021}, essentially says we have monotonicity of the entropy with respect to $\gamma$ provided our distribution is not a point mass. 

Similar derivations can be obtained for the multiset models (\Cref{def:SIM}) by a simple change of notation. For brevity, we do not repeat this here. 


	
	\section{The iExchange Algorithm}

\label{sec_sup:iexchange_alogrithm}

In this section, we outline the \textit{iExchange} algorithm (\Cref{alg_sup:iexchange}), a generalisation of exchange algorithm \citep{Murray2006} obtained by incorporating the proposal generating mechanism of the iMCMC algorithm \citep{Neklyudov2020}. As we show, the iExchange algorithm is itself an iMCMC algorithm (with a particular form of involution), providing the necessary theoretical justification. For completeness, we give background details regarding both the exchange and iMCMC algorithms, before showing how they can be combined.

\begin{algorithm}
	\SetAlgoLined
	\KwIn{target density $p(\theta | \mathbf{x}) \propto  p(\theta)\gamma(\mathbf{x}|\theta)/Z(\theta)$}
	\KwIn{auxiliary density $q(u|\theta)$}
	\KwIn{involution $f(\theta,u)$, i.e. $f^{-1}(\theta,u) = f(\theta,u)$}	
	initialise $\theta$\\
	\For{$i=1,\dots,n$}{
		sample $u \sim q(u|\theta)$ \\
		invoke involution $(\theta^\prime, u^\prime) = f(x, u)$ \\
		sample $\mathbf{y} \sim p(\mathbf{y} | \theta^\prime)$ \\
		evaluate $\alpha(\theta, \theta^\prime) = \min\left\{1, \frac{p(\theta^\prime) \gamma(\mathbf{x} | \theta^\prime) \gamma(\mathbf{y}| \theta) q(u^\prime| \theta^\prime) }{p(\theta) \gamma(\mathbf{x} | \theta) \gamma(\mathbf{y}| \theta^\prime)  q(u| \theta)}\left| \frac{\partial f(\theta, u)}{\partial (\theta, u)}\right|\right\}$ \\
		$\theta_i = \begin{cases}
			\theta^\prime & \text{with probability } \alpha(\theta,\theta^\prime) \\
			\theta & \text{with probability } 1-\alpha(\theta,\theta^\prime)
		\end{cases}$ \\
		$\theta \leftarrow \theta_i$\\
	}
	\caption{Involutive exchange (iExchnage) algorithm}
	\label{alg_sup:iexchange}
\end{algorithm}

Let us first set the context. We have some data $\mathbf{x}$ which is assumed to have been drawn via a model $p(\mathbf{x} | \theta)$, where $\theta$ denote parameters, taking the following form
\begin{equation}
    \begin{aligned}
        p(\mathbf{x} |\theta) = \frac{\gamma(\mathbf{x} | \theta) }{Z(\theta)}
    \end{aligned}
    \label{eq_sup:general_model}
\end{equation}
where $Z(\theta) = \int\gamma(\mathbf{x} |\theta) d\mathbf{x}$ denotes its normalising constant, assumed to be \textit{intractable}. If one is taking a Bayesian approach to inference and has specified a prior $p(\theta)$, this leads to the following posterior
\begin{equation}
    \begin{aligned}
        p(\theta | \mathbf{x}) = \frac{p(\mathbf{x} | \theta) p(\theta)}{p(\mathbf{x})}
    \end{aligned}
    \label{eq_sup:general_posterior}
\end{equation}
where $p(\mathbf{x}) = \int p(\mathbf{x}|\theta)p(\theta)d\theta$ is the marginal probability of the data, which in most cases is also intractable. Due to these two elements of intractability, such posteriors are often referred to as \textit{doubly}-intractable \citep{Murray2006}. For example, the posteriors resulting from both our SIS and SIM models are doubly-intractable.

A typical approach to circumvent the intractability present in Bayesian posterior distributions is to use MCMC algorithms to sample from them, with the Metropolis-Hastings (MH) algorithm being a prevalent choice. However, for doubly-intractable posteriors, many standard MCMC algorithms are not feasible. To illustrate, consider using the MH algorithm. Here, with $\theta$ the current state and $q(\theta^\prime|\theta)$ some proposal density, in a single iteration one would sample proposal $\theta^\prime$ from $q(\theta^\prime|\theta)$ and accept this with the following probability 
\begin{equation}
    \begin{aligned}
        \alpha(\theta, \theta^\prime)&= \min\left\{ 1, 
        \frac{p(\theta^\prime | \mathbf{x})q(\theta|\theta^\prime)}{p(\theta|\mathbf{x})q(\theta^\prime|\theta)}
        \right\} \\
        &= \min\left\{ 1, 
        \frac{\gamma(\mathbf{x}|\theta^\prime)/Z(\theta^\prime)p(\theta^\prime)q(\theta|\theta^\prime)}{\gamma(\mathbf{x}|\theta)/Z(\theta)p(\theta)q(\theta^\prime|\theta)}\right\},
    \end{aligned}
    \label{eq:mh_acc_prob}
\end{equation}
so that, starting from some initial state $\theta_0$ one obtains a sample $\{\theta_i\}_{i=1}^m$ which is (approximately) distributed according to $p(\theta|\mathbf{x})$. However, though the marginal probability $p(\mathbf{x})$ cancels out in \eqref{eq:mh_acc_prob}, the normalising constants $Z(\theta)$ and $Z(\theta^\prime)$ do not. Moreover, since these are by assumption intractable, $\alpha(\theta,\theta^\prime)$ cannot be evaluated, ruling out use of the MH algorithm. 

This necessitates the proposal of specialised MCMC algorithms to sample from doubly-intractable posterior distributions, and herein lies the motivation for the exchange and iExchange algorithms.

\subsection{Exchange Algorithm}

\label{sec_sup:exchange_algorithm_background}

In this section, we give a high-level overview of the exchange algorithm (\Cref{alg_sup:exchange}), proposed by \cite{Murray2006}. This is similar in structure to MH algorithm, but with some extra sampling in each iteration. Namely, one samples so-called \textit{auxiliary data}, which subsequently appears in the acceptance probability, inducing cancellation of intractable normalising constants. Effectively, it targets an augmented distribution which admits the posterior of interest as its marginal \citep{Murray2006}.

As in the MH algorithm, we have some proposal distribution $q(\theta^\prime \, | \, \theta)$ which is pre-specified. We also introduce an auxiliary data set $\mathbf{y}$ which lies in the same space as the observed data $\mathbf{x}$. Now, given current state $\theta$ a single iteration consists of the following
\begin{enumerate}
    \item Sample proposal $\theta^\prime$ via $q(\theta^\prime \, | \, \theta)$
    \item Sample auxiliary data $\mathbf{y} \, | \, \theta^\prime$ via $p(\mathbf{y} \, | \, \theta^\prime)$ of \eqref{eq_sup:general_model} (sample from the model)
    \item Evaluate acceptance probability
    \begin{equation}
        \label{eq:exchange_acc_prob}
        \begin{aligned}
            \alpha(\theta, \theta^\prime) 
            &= \min\left\{1, \frac{p(\theta^\prime| \mathbf{x}) q(\theta| \theta^\prime) p(\mathbf{y}| \theta) }{p(\theta| \mathbf{x}) q(\theta^\prime | \theta) p(\mathbf{y}| \theta^\prime) }\right\} \\
            &= \min\left\{1, \frac{p(\theta^\prime) \gamma(\mathbf{x} | \theta^\prime) \gamma(\mathbf{y}| \theta) q(\theta| \theta^\prime) }{p(\theta) \gamma(\mathbf{x} | \theta) \gamma(\mathbf{y}| \theta^\prime)  q(\theta^\prime | \theta)}\right\}
        \end{aligned}
    \end{equation}
    \item With probability $\alpha(\theta, \theta^\prime)$ we move to state $\theta^\prime$, otherwise we stay at $\theta$.
\end{enumerate}

Observe the absence of normalising constants here makes $\alpha(\theta, \theta^\prime)$ tractable. Repeating this a number of times, as summarised in \Cref{alg_sup:exchange}, produces a Markov chain admitting $p(\theta \, | \, \mathbf{x})$ as its stationary distribution \citep{Murray2006}. An alternative justification to that given by \cite{Murray2006} comes by viewing this as an instance of iMCMC, which we detail in the next section.

\begin{algorithm}
	\SetAlgoLined
	\KwIn{target density $p(\theta | \mathbf{x}) \propto  p(\theta)\gamma(\mathbf{x}|\theta)/Z(\theta)$}
	\KwIn{proposal distribution $q(\theta^\prime|\theta)$}
	initialise $\theta$\;
	\For{$i=1,\dots,n$}{
		sample $\theta^\prime$ via $q(\theta^\prime|\theta)$ \\
		sample $\mathbf{y}$ via $p(\mathbf{y} | \theta^\prime)$ (from the model) \\
		evaluate $\alpha(\theta, \theta^\prime) = \min\left\{1, \frac{p(\theta^\prime) \gamma(\mathbf{x} | \theta^\prime) \gamma(\mathbf{y}| \theta) q(\theta| \theta^\prime) }{p(\theta) \gamma(\mathbf{x} | \theta) \gamma(\mathbf{y}| \theta^\prime)  q(\theta^\prime| \theta)}\right\}$ \\
		$\theta_i = \begin{cases}
			\theta^\prime & \text{with probability } \alpha(\theta,\theta^\prime) \\
			\theta & \text{with probability } 1-\alpha(\theta,\theta^\prime)
		\end{cases}$ \\
		$\theta \leftarrow \theta_i$
		
	}
	\KwOut{sample $\{\theta_i\}_{i=1}^n$}
	\caption{Exchange algorithm}
	\label{alg_sup:exchange}
\end{algorithm}

\subsection{Involutive MCMC (iMCMC)}
\label{sec_sup:comp_imcmc_background}
The iMCMC algorithm of \cite{Neklyudov2020} considers the problem of sampling from a general target distribution $p(x)$ over some space $\X$, for example, this might be our posterior from \eqref{eq_sup:general_posterior} (replacing $x$ with $\theta$). Like all MCMC algorithms, it does so by sampling a Markov chain admitting $p(x)$ as its stationary distribution, using in particular a combination of random sampling and involutive deterministic maps. The result is a very general framework which includes many well-known MCMC algorithms as special cases. 

As the name suggests, iMCMC uses a particular type of deterministic function know as an \textit{involution}. This is a function which serves as its own inverse, that is, if $f \, : \X \to \X$ then one has $f^{-1}(x) = f(x)$. Equivalently, a composition $f$ with itself leads to the identity
$$f(f(x)) = x.$$
Towards targeting $p(x)$ one introduces auxiliary variables $u \in \mathcal{U}$ with conditional density $q(u|x)$ over an auxiliary space $\U$ (which need not be equal to $\X$), augmenting the target as follows 
$$p(x,u) = p(x) q(u | x)$$
which is now a distribution over $\X \times \U$. Observe this admits $p(x)$ as its marginal and hence one can obtain samples thereof by targeting $p(x,u)$ and disregarding the $u$ samples. To do so, suppose an involution $f \, : \, \X \times \mathcal{U} \to \X \times \mathcal{U}$ has been specified along with the auxiliary distribution $q(u|x)$. In structure reminiscent of the MH algorithm, a single iteration consists of the following. With current state $(x,u)$,  an auxiliary variable $u \in \U$ is first drawn from $q(u|x)$, before the involution $f$ is invoked to get a proposal $(x^\prime, u^\prime) = f(x,u)$, which is subsequently accepted with the following probability
\begin{equation*}
    \begin{aligned}
        \alpha\left((x,u),(x^\prime,u^\prime)\right) &= \min\left\{1, \frac{p(f(x,u))}{p(x,u)} \left|\frac{\partial f(x,u) }{\partial (x,u)}\right| \right\} \\
        &= \min\left\{1, \frac{p(x^\prime)q(u^\prime|x^\prime)}{p(x)q(u|x)} \left|\frac{\partial f(x,u) }{\partial (x,u)}\right| \right\},
    \end{aligned}
\end{equation*}
leading to a Markov chain admitting $p(x,u)$ as its stationary distribution \cite[Proposition 2]{Neklyudov2020}. 

Observe that since auxiliary variables $u$ are re-sampled in each iteration they do not need to be stored, and can instead be discarded as the algorithm proceeds. In this way, one may also drop their reference in the acceptance probability denoting this simply $\alpha(x,x^\prime)$. This leads to the algorithm to target $p(x)$ as outlined in \Cref{alg_sup:iMCMC}.

\begin{algorithm}
	\SetAlgoLined
	\KwIn{target density $p(x)$}
	\KwIn{auxiliary density $q(u|x)$}
	\KwIn{involution $f(x,u)$}	
	initialise $x$\;
	\For{$i=1,\dots,n$}{
		sample $u \sim q(u|x)$ \\
		invoke involution $(x^\prime, u^\prime) = f(x, u)$ \\
		evaluate 
		$\alpha(x,x^\prime) = \min\left\{1, \frac{p(x^\prime)q(u^\prime|x^\prime)}{p(x)q(u|x)} \left|\frac{\partial f(x,u) }{\partial (x,u)}\right| \right\}$ \\
		$x_i = \begin{cases}
			x^\prime & \text{with probability } \alpha(x,x^\prime) \\
			x & \text{with probability } 1-\alpha(x,x^\prime)
		\end{cases}$ \\
		$x \leftarrow x_i$
	}
	\KwOut{sample $\{x_i\}_{i=1}^n$}
	\caption{Involutive MCMC (iMCMC)}
	\label{alg_sup:iMCMC}
\end{algorithm}

As mentioned, many known MCMC algorithms can be written in this form. For example, if one assumes $\mathcal{U}=\X$, with $q(x^\prime|x)$ the auxiliary distribution and $f(x,x^\prime) = (x^\prime, x)$ the involution defined by swapping entries, then one obtains the Metropolis-Hastings algorithm with proposal distribution $q(x^\prime|x)$. Further examples of MCMC algorithms which can be cast in the iMCMC framework are given in \cite{Neklyudov2020}, Appendix B. 

Another iMCMC special case which is of relevance to us is the exchange algorithm. To see this, we let $u=(\theta^\prime, \mathbf{y})$, where $\mathbf{y}$ denotes the auxiliary data, as seen in \Cref{sec_sup:exchange_algorithm_background}. Moreover, we define our involution as follows 
$$f(\theta, u) = (\theta^\prime, (\theta, \mathbf{y})),$$ 
that is, we simply swap $\theta \leftrightarrow \theta^\prime$. Observe we have
\begin{equation*}
    \begin{aligned}
        f(f(\theta, u))&=f(f(\theta,(\theta^\prime,\mathbf{y}))\\
        &=f(\theta^\prime,(\theta, \mathbf{y}))\\
        &= (\theta, (\theta^\prime, \mathbf{y}))\\
        &=(\theta,u)
    \end{aligned}
\end{equation*}
so that $f$ is indeed an involution. We now derive the Jacobian term. For convenience, drop the inner parenthesis and write $(\theta, u) = (\theta, \theta^\prime, \mathbf{y})$, for which we have $f(\theta, \theta^\prime, \mathbf{y}) = (\theta^\prime, \theta, \mathbf{y})$. Now, we have
\begin{equation*}
    \begin{aligned}
        \frac{\partial f(\theta, \theta^\prime, \mathbf{y})}{\partial (\theta, \theta^\prime, \mathbf{y})} &= \begin{bmatrix}
        \frac{\partial f_1}{\partial \theta} & \frac{\partial f_1}{\partial \theta^\prime}&\frac{\partial f_1}{\partial \mathbf{y}} \\[0.2cm]
        \frac{\partial f_2}{\partial \theta} & \frac{\partial f_2}{\partial \theta^\prime}&\frac{\partial f_2}{\partial \mathbf{y}}\\[0.2cm]
        \frac{\partial f_3}{\partial \theta} & \frac{\partial f_3}{\partial \theta^\prime}&\frac{\partial f_3}{\partial \mathbf{y}}
        \end{bmatrix} 
        =\begin{bmatrix}
        \frac{\partial \theta^\prime}{\partial \theta} & \frac{\partial \theta^\prime}{\partial \theta^\prime}&\frac{\partial \theta^\prime}{\partial \mathbf{y}} \\[0.2cm]
        \frac{\partial \theta}{\partial \theta} & \frac{\partial \theta}{\partial \theta^\prime}&\frac{\partial \theta}{\partial \mathbf{y}}\\[0.2cm]
        \frac{\partial \mathbf{y}}{\partial \theta} & \frac{\partial \mathbf{y}}{\partial \theta^\prime}&\frac{\partial \mathbf{y}}{\partial \mathbf{y}}
        \end{bmatrix} = \begin{bmatrix}
        0 & 1 & 0 \\
        1 & 0 & 0 \\
        0 & 0 & 1
        \end{bmatrix}
    \end{aligned}
\end{equation*}
and taking determinants
\begin{equation*}
    \left| \frac{\partial f(\theta, \theta^\prime, \mathbf{y})}{\partial (\theta, \theta^\prime, \mathbf{y})}\right| = 1 \cdot \left| 
    \begin{matrix}
    0 & 1 \\
    1 & 0 
    \end{matrix}\right| + 0 + 0 = 1.
\end{equation*}
Finally, with $q(\theta^\prime|\theta)$ denoting the proposal density of the exchange algorithm, define the auxiliary distribution as follows
\begin{equation*}
    \begin{aligned}
        q(u|\theta) = q(\theta^\prime|\theta) p(\mathbf{y}|\theta^\prime)
    \end{aligned}
\end{equation*}
where $p(\mathbf{y}|\theta^\prime)$ is the likelihood of auxiliary data $\mathbf{y}$ under the assumed model \eqref{eq_sup:general_model}. With these elements, an iMCMC algorithm targeting $p(\theta|\mathbf{x})$ would (i) sample $u$ from $q(u|\theta)$, which amounts to first sampling $\theta^\prime$ from $q(\theta^\prime|\theta)$, before drawing $\mathbf{y}$ from $p(\mathbf{y}|\theta^\prime)$, and (ii) accept $\theta^\prime$ with probability 
\begin{equation*}
    \begin{aligned}
        \alpha(\theta,\theta^\prime) &= \min\left\{1,\frac{p(\theta^\prime|\mathbf{x})p(u^\prime|\theta^\prime)}{p(\theta|\mathbf{x})p(u|\theta)}\left|\frac{\partial f(\theta, u)}{\partial (\theta,u)}\right|\right\} \\
        &= \min\left\{1, \frac{p(\theta^\prime) \gamma(\mathbf{x} | \theta^\prime) \gamma(\mathbf{y}| \theta) q(\theta| \theta^\prime) }{p(\theta) \gamma(\mathbf{x} | \theta) \gamma(\mathbf{y}| \theta^\prime)  q(\theta^\prime | \theta)}\left|\frac{\partial f(\theta, \theta^\prime, \mathbf{y})}{\partial (\theta,\theta^\prime, \mathbf{y})}\right|\right\}\\
        &=\min\left\{1, \frac{p(\theta^\prime) \gamma(\mathbf{x} | \theta^\prime) \gamma(\mathbf{y}| \theta) q(\theta| \theta^\prime) }{p(\theta) \gamma(\mathbf{x} | \theta) \gamma(\mathbf{y}| \theta^\prime)  q(\theta^\prime | \theta)}\right\},
    \end{aligned}
\end{equation*}
which is nothing more than the exchange algorithm (\Cref{alg_sup:exchange}).



\subsection{Defining the iExchange Algorithm}

\label{sec_sup:comp_justify_iexchange}

We now define our extension of the exchange algorithm. We will assume that an iMCMC scheme to target $p(\theta|\mathbf{x})$ has been defined, that is, auxiliary variables $u$, involution $f(\theta,u) = (\theta^\prime, u^\prime)$ and conditional distribution $q(u|\theta)$ have all been specified. When the posterior is doubly-intractable, in general one will not be able to implement this algorithm due to the intractability of the acceptance probability. However, in spirit of the exchange algorithm, we can choose auxiliary variables and their conditional distribution to induce cancellation of normalising constants in the acceptance probability. 

In particular, we let $\tilde{u} = (u, \mathbf{y})$, where $\mathbf{y}$ denotes an auxiliary dataset lying in the same space as $\mathbf{x}$. Now, writing $f(\theta, u)=(f_1(\theta,u),f_2(\theta,u))=(\theta^\prime,u^\prime)$ we define an involution $g(\theta,\tilde{u})$ as follows
\begin{equation*}
    \begin{aligned}
        g(\theta, \tilde{u}) =g(\theta,(u,\mathbf{y})) &= (f_1(\theta, u), (f_2(\theta, u),\mathbf{y}))\\
        &= (\theta^\prime, (u^\prime, \mathbf{y}))
    \end{aligned}
\end{equation*}
for which we have 
\begin{equation*}
    \begin{aligned}
        g(g(\theta,\tilde{u}))&= g(\theta^\prime, (u^\prime, \mathbf{y}))\\
        &= (f_1(\theta^\prime,u^\prime), (f_2(\theta^\prime,u^\prime),\mathbf{y}))\\
        &=(\theta, (u,\mathbf{y}))\\
        &=(\theta,\tilde{u})
    \end{aligned}
\end{equation*}
that is, $g$ is indeed an involution. Now, as in Section \ref{sec_sup:comp_justify_iexchange}, drop the inner parenthesis and write $(\theta, \tilde{u})=(\theta, u, \mathbf{y})$. The Jacobian is now given by
\begin{equation*}
    \begin{aligned}
        \frac{\partial g(\theta, \tilde{u})}{\partial (\theta, \tilde{u})} =\frac{\partial g(\theta, u, \mathbf{y})}{\partial (\theta, u, \mathbf{y})} = \begin{bmatrix}
            \frac{\partial g_1}{\partial \theta} & \frac{\partial g_1}{\partial u} & \frac{\partial g_1}{\partial \mathbf{y}} \\[0.2cm]
            \frac{\partial g_2}{\partial \theta} & \frac{\partial g_2}{\partial u} & \frac{\partial g_2}{\partial \mathbf{y}}\\[0.2cm]
            \frac{\partial g_3}{\partial \theta} & \frac{\partial g_3}{\partial u} & \frac{\partial g_3}{\partial \mathbf{y}}
        \end{bmatrix}=\begin{bmatrix}
            \frac{\partial f_1}{\partial \theta} & \frac{\partial f_1}{\partial u} & \frac{\partial f_1}{\partial \mathbf{y}} \\[0.2cm]
            \frac{\partial f_2}{\partial \theta} & \frac{\partial f_2}{\partial u} & \frac{\partial f_2}{\partial \mathbf{y}}\\[0.2cm]
            \frac{\partial f_3}{\partial \theta} & \frac{\partial f_3}{\partial u} & \frac{\partial f_3}{\partial \mathbf{y}}
        \end{bmatrix}=\begin{bmatrix}
            \frac{\partial f_1}{\partial \theta} & \frac{\partial f_1}{\partial u} & 0 \\[0.2cm]
            \frac{\partial f_2}{\partial \theta} & \frac{\partial f_2}{\partial u} & 0\\[0.2cm]
            0 & 0 & 1
        \end{bmatrix}
    \end{aligned}
\end{equation*}
and taking determinants we get the following 
\begin{equation*}
    \begin{aligned}
        \left|\frac{\partial g(\theta, \tilde{u})}{\partial (\theta, \tilde{u})}\right| &= 1 \cdot \left| \begin{bmatrix}
            \frac{\partial f_1}{\partial \theta} & \frac{\partial f_1}{\partial u} \\[0.2cm]
            \frac{\partial f_2}{\partial \theta} & \frac{\partial f_2}{\partial u} 
        \end{bmatrix}\right| + 0 + 0 = \left| \frac{\partial f(\theta, u)}{\partial (\theta, u)}\right|.
    \end{aligned}
\end{equation*}

The final element to define is the auxiliary distribution. Given current state $\theta$ we consider sampling $\tilde{u} = (u, \mathbf{y})$ as follows: (i) sample $u$ from $q(u|\theta)$, then (ii) sample $\mathbf{y}$ from $p(\mathbf{y}|\theta^\prime)$ (the model) where $\theta^\prime = f_1(\theta, u)$. This leads to the following auxiliary density 
\begin{equation*}
    \begin{aligned}
        q(\tilde{u}|\theta) = q(u|\theta)p(\mathbf{y}|\theta^\prime).
    \end{aligned}
\end{equation*}
All the elements of an iMCMC algorithm have now been defined, a single iteration of which consists of the following. Given current state $\theta$, we first sample $\tilde{u}=(u, \mathbf{y})$ via $q(\tilde{u}|\theta)$ as above. We then invoke involution $g(\theta, \tilde{u}) = (\theta^\prime, \tilde{u}^\prime) = (\theta^\prime, (u^\prime,\mathbf{y}))$, generating a proposal $\theta^\prime$ which we accept with the following probability
\begin{equation*}
    \begin{aligned}
        \alpha(\theta, \theta^\prime) &= \min \left\{1,  \frac{p(g(\theta, \tilde{u}))}{p(\theta, \tilde{u})}\left| \frac{\partial g(\theta, \tilde{u})}{\partial (\theta, \tilde{u})}\right|\right\} \\
        &= \min \left\{1,  \frac{p(\theta^\prime|\mathbf{x}) q(\tilde{u}^\prime|\theta^\prime)}{p(\theta|\mathbf{x})q(\tilde{u}|\theta)}\left| \frac{\partial g(\theta, \tilde{u})}{\partial (\theta, \tilde{u})}\right|\right\}\\
        &= \min \left\{1,  \frac{p(\theta^\prime|\mathbf{x}) q(u^\prime|\theta^\prime) p(\mathbf{y}|\theta)}{p(\theta|\mathbf{x})q(u|\theta)p(\mathbf{y}|\theta^\prime)}\left| \frac{\partial f(\theta, u)}{\partial (\theta, u)}\right|\right\}\\
        &= \min \left\{1,  \frac{p(\theta^\prime)\gamma(\mathbf{x}|\theta^\prime) \gamma(\mathbf{y}|\theta)q(u^\prime|\theta^\prime)}{p(\theta)\gamma(\mathbf{x}|\theta)\gamma(\mathbf{y}|\theta^\prime)q(u|\theta)}\left| \frac{\partial f(\theta, u)}{\partial (\theta, u)}\right|\right\},
    \end{aligned}
\end{equation*}
where as in the exchange algorithm we observe cancellation of normalising constants thanks to the introduction of auxiliary data. Note the Jacobian term here concerns the involution of the original iMCMC scheme to sample from $p(\theta|\mathbf{x})$, and thus the key difference here is the introduction of auxiliary data. The result is what we call the iExchange algorithm (\Cref{alg_sup:iexchange}).

	\section{Bayesian Inference: Extra Details}

\label{sec_sup:mcmc_extra_details}

In this section we provide extra details concerning our MCMC scheme for the interaction-sequence models outlined in \Cref{sec:inference}, including explicit specification of proposal distributions, involutions and auxiliary distributions, derivations of closed-form acceptance probabilities and pseudocode.

\subsection{Dispersion Conditional}

\label{sec_sup:inference_dispersion_cond}

The dispersion conditional can be obtained directly from \eqref{eq:SIS_posterior} by conditioning on the mode $\iS^m$, in particular we have
\begin{equation}
    \label{eq_sup:posterior_cond_dispersion}
        \begin{aligned}
            p(\gamma | \iS^m, \data) &\propto Z(\iS^m, \gamma)^{-n}\exp\left\{ - \gamma \sum_{i=1}^n d_S(\iS^{(i)}, \iS^m)\right\} p(\gamma).
        \end{aligned}
\end{equation}
To target \eqref{eq_sup:posterior_cond_dispersion} we use the exchange algorithm of \cite{Murray2006} (see \Cref{sec_sup:exchange_algorithm_background} for background details). As a proposal $q(\gamma^\prime|\gamma)$ we consider sampling $\gamma^\prime$ uniformly over a $\varepsilon$-neighbourhood of $\gamma$ with reflection at zero, this is, we first sample $\gamma^* \sim \text{Uniform}(\gamma-\varepsilon, \gamma+\varepsilon)$ and then let $\gamma^\prime = \gamma^*$ if $\gamma^* > 0$ and let $\gamma^\prime = -\gamma^*$ otherwise. The density is thus given by the following (for $\gamma > 0$)
\begin{equation}
    \label{eq:gamma_proposal_density}
    \begin{aligned}
        q(\gamma^\prime|\gamma) = \begin{cases}
        \frac{1}{2\varepsilon} & \text{if } \gamma^\prime > 0 \text{ and }\gamma + \gamma^\prime > \varepsilon \\
        \frac{1}{\varepsilon} & \text{if } \gamma^\prime > 0 \text{ and }\gamma + \gamma^\prime < \varepsilon \\
        0 & \text{if } \gamma^\prime \leq 0.
        \end{cases}
    \end{aligned}
\end{equation}
whilst $q(\gamma^\prime  |\gamma)=0$ for $\gamma \leq 0$. Observe this proposal is symmetric, in that $q(\gamma^\prime | \gamma) = q(\gamma|\gamma)$.

Now, a single iteration consists of the following. Assuming $\gamma$ is our current state, we first sample proposal $\gamma^\prime$ from $q(\gamma^\prime | \gamma)$. Next, we sample auxiliary data $\adata$ i.i.d. from the appropriate model, namely 
$$\iS^*_i \sim \text{SIS}(\iS^m, \gamma^\prime) \quad (\text{for }i=1,\dots,n),$$
which we note implies 
$$p(\adata \, | \, \iS^m, \gamma^\prime) = Z(\iS^m, \gamma^\prime)^{-n}\exp\left\{ -\gamma^\prime \sum_{i=1}^n d_S(\iS^*_i,\iS^m)\right\}.$$
Finally, we accept this proposal with the following probability 
\begin{equation}
    \label{eq_sup:posterior_cond_dipsersion_acc_prob}
    \begin{aligned}
        \alpha(\gamma, \gamma^\prime) = \min\left\{1, H(\gamma, \gamma^\prime) \right\}
    \end{aligned}
\end{equation}
where
\begin{equation*}
    \begin{aligned}
        H(\gamma, \gamma^\prime) &= \frac{p(\gamma^\prime|\iS^m, \data)p(\adata|\iS^m, \gamma)}{p(\gamma|\iS^m, \data)p(\adata|\iS^m, \gamma^\prime)} \frac{q(\gamma|\gamma^\prime)}{q(\gamma^\prime| \gamma)} \\
        &=\exp\left\{-(\gamma^\prime-\gamma) \left(\sum_{i=1}^n d_S(\iS^{(i)},\iS^m) - \sum_{i=1}^n d_S(\iS^*_i,\iS^m)\right)\right\}\frac{p(\gamma^\prime)}{p(\gamma)},
    \end{aligned}
\end{equation*}
where we note normalising constants of the (conditional) posterior and auxiliary data cancel one another out, whilst the proposal density terms cancel due to its symmetry. This is summarised in \Cref{alg_sup:sis_dispersion_accept_reject}, which details a single accept-reject step for updating the dispersion.

\subsection{Mode Conditional}

\label{sec_sup:inference_mode_cond}

By conditioning on $\gamma$ in \eqref{eq:SIS_posterior} we get the following form for the mode conditional posterior 
\begin{equation}
    \label{eq_sup:posterior_cond_mode}
        \begin{aligned}
            p(\iS^m| \gamma, \data) &\propto Z(\iS^m, \gamma)^{-n}\exp\left\{ - \gamma \sum_{i=1}^n d_S(\iS^{(i)}, \iS^m) - \gamma_0d_S(\iS^m, \iS_0)\right\},
        \end{aligned}
\end{equation}
which as outlined in \Cref{sec:inference_mode_cond} we target via the iExchange algorithm (\Cref{alg_sup:iexchange}). For further details on the iExchange algorithm, including justification as an instance of iMCMC, please see \Cref{sec_sup:iexchange_alogrithm}. 

Supposing that auxiliary variables $u$, involution $f(\modecurr, u)$ and auxiliary distribution $q(u|\modecurr)$ have all be specified, a single iteration of the iExchange algorithm in this case consists of the following. With $\gamma$ fixed and $\iS^m$ denoting our current state we first sample auxiliary variable $u$ according to $q(u|\modecurr)$. We then invoke the involution $f(\iS^m, u) = (\modeprop, u^\prime)$, which generates our proposal $\modeprop$. Next, we sample auxiliary data $\adata$ i.i.d. where 
$$\iS_i^* \sim \text{SIS}(\modeprop, \gamma).$$
Finally, we accept $\modeprop$ with the following probability 
\begin{equation}
    \begin{aligned}
        \alpha\left(\modecurr, \modeprop\right) = \min\left\{1,H(\modecurr, \modeprop)\right\}
    \end{aligned}
    \label{eq_sup:posterior_cond_mode_acc_prob}
\end{equation}
where 
\begin{equation}
    \begin{aligned}
        H(\modecurr, \modeprop) &= \frac{p(\modeprop \, | \, \gamma, \data)}{p(\modecurr \, | \, \gamma, \data)}
        \frac{p(\adata \, | \, \modecurr, \gamma)}{p(\adata \, | \, \modeprop, \gamma)}
        \frac{q(u^\prime \, | \,\modeprop)}{q(u \, | \,\modecurr)} \\[0.25cm]
        &= \exp\Bigg\{-\gamma\left( \sum_{i=1}^n d_S(\iS^{(i)},\modeprop) - \sum_{i=1}^nd_S(\iS^{(i)},\modecurr)\right) \\ &\quad -\gamma\left(\sum_{i=1}^n d_S(\iS^*_i,\modecurr) - \sum_{i=1}^n d_S(\iS^*_i, \modeprop) \right) 
        \\ &\qquad -\gamma_0\left( d_S(\modeprop, \iS_0) - d_S(\modecurr,\iS_0)\right)\Bigg\}\frac{q(u^\prime \, | \,\modeprop)}{q(u \, | \,\modecurr)} 
    \end{aligned}
    \label{eq_sup:posterior_cond_mode_ratio_term}
\end{equation}
where the ratio $q(u^\prime \, | \,\modeprop)/q(u \, | \,\modecurr)$ is move-dependent. Again, we have the normalising constants of the conditional posterior and auxiliary data cancelling one another out.

\subsection{Edit Allocation Move}

\label{sec_sup:inference_edit_alloc}

Supposing that $\iS^m = (\I_1,\dots,\I_N)$ denotes the current state, recall that for this move we have an auxiliary variable given by 
$$u = (\delta, \bm{z}, u_1,\dots, u_N)$$
where (i) $\delta$ denotes the total number of edits (entry insertions and deletions), (ii) $\bm{z} =  (z_1,\dots,z_N)$ denotes the allocation of edits to paths, that is, $z_i \in \mathbb{Z}_{\geq 0}$ is the number of edits allocated to the $i$th path, where $\sum_{i=1}^N z_i = \delta$, and (iii) $u_i = (d_i, \bm{v}_i, \bm{v}_i^\prime, \bm{y}_i)$ describes the edits to the $i$th path, where $d_i$ is the number of deletions, $\bm{v}_i$ and $\bm{v}_i^\prime$ are subsequences indexing entry insertions and deletions and $\bm{y}_i$ denotes entries to be inserted.

We now define the involution of this iMCMC move. Writing the required involution as follows 
$$f(\iS^m, u) = (f_1(\iS^m, u), f_2(\iS^m, u)) = (\modeprop, u^\prime)$$
as outlined in \Cref{sec:inference_edit_alloc} in enacting the operations parameterised by $u$ we define the first component $f_1(\iS^m, u) = \modeprop = (\I^\prime_1,\dots,\I_N^\prime)$. The second component we define as follows
$$f_2(\iS^m, u) = (\delta, \bm{z}, u^\prime_1,\dots,u^\prime_N)$$
where 
\begin{equation}
    u_i^\prime = (z_i-d_i, \bm{v}^\prime_i, \bm{v}_i, (\I_i)_{\bm{v}_i})
    \label{eq_sup:edit_alloc_second_component_submap}
\end{equation}
where $(\I_i)_{\bm{v}_i} = (x_{iv_1},\dots,x_{iv_{d_i}})$ is the subsequence of $\I_i$ indexed by $\bm{v}_i = (v_1,\dots,v_{d_i})$. On an intuitive level, $u^\prime_i$ parameterises the edits to the $i$th path $\I_i$ which are exactly the opposite of those parameterised by $u_i$, namely we delete $z_i-d_i$ entries indexed by $\bm{v}^\prime_i$, then insert entries $(\I_i)_{\bm{v}_i}$ at locations indexed by $\bm{v}_i$. In this way, enacting the operations parameterised by $u^\prime$ will take us back to $\iS^m$, that is 
$$f_1(\modeprop,u^\prime) = \iS^m,$$
furthermore observe that reapplying the operations of \eqref{eq_sup:edit_alloc_second_component_submap} to $u_i^\prime$ itself takes us back to $u_i$ 
$$(z_i - (z_i-d_i),\bm{v}_i, \bm{v}^\prime_i, (\I_i^\prime)_{\bm{v}^\prime_i}) = (d_i, \bm{v}_i, \bm{v}^\prime_i, \bm{y}_i) = u_i$$
where $\bm{y}_i = (\I^\prime_i)_{\bm{v}^\prime_i}$ since $\bm{v}^\prime_i$ indexed where there entries $\bm{y}_i$ were inserted in $\I^\prime_i$. This implies
\begin{equation*}
    \begin{aligned}
        f_2(\modeprop, u^\prime) &= (\delta, \bm{z}, u_1,\dots,u_N)
    \end{aligned}
\end{equation*}
and hence
$$f(f(\iS^m, u))=f(\modeprop, u^\prime)=(\iS^m,u)$$
so that $f(\iS^m, u)$ is indeed an involution. 

Turning now to the auxiliary distribution $q(u|\iS^m)$, recall the following assumptions stated in \Cref{sec:inference_edit_alloc}
\begin{equation*}
    \begin{aligned}
        \delta &\sim \text{Uniform}\{1,\dots, \nu_{\text{ed}}\} \\
        \bm{z} \, | \, \delta &\sim \text{Multinomial}(\delta \, ; \, 1/N, \dots, 1/N)\\
        d_i \, | \,  z_i &\sim \text{Uniform}\{0,\dots,\min(z_i, n_i)\} & (\text{for } i=1,\dots,N),\\
    \end{aligned}
\end{equation*}
whilst we sample indexing subsequences $\bm{v}_i$ and $\bm{v}_i^\prime$ uniformly. Regarding this latter assumption, recall that $\bm{v}_i$ is a length $d_i$ (number of deletions) subsequence of $[n_i]$ ($n_i$ is the length of $\I_i$), whilst $\bm{v}^\prime_i$ is a length $a_i:=z_i-d_i$ (number of insertions) subsequence of $[m_i]$, where $m_i = n_i -d_i + a_i$ (length of the $i$th proposed path $\I^\prime_i$). Thus sampling these uniformly implies 
\begin{equation*}
    \begin{aligned}
        q(\bm{v}_i | d_i) = \binom{n_i}{d_i}^{-1} && q(\bm{v}^\prime_i| d_i,z_i) = \binom{m_i}{a_i}^{-1}.
    \end{aligned}
\end{equation*}
Finally, regarding sampling entry insertions we for now assume these are drawn via some general distribution which may be dependent on the current state, namely we assume each $\bm{y}_i$ was drawn via $q(\bm{y}|\I_i)$. Together this implies the following closed form for the auxiliary distribution
\begin{equation}
    \begin{aligned}
        q(u | \iS^m) &= q(\delta) q(\bm{z} | \delta) \prod_{i=1}^N q(d_i) q(\bm{v}_i|d_i)q(\bm{v}^\prime_i|d_i,z_i) q(\bm{y}_i|\I_i) \\
        &=\frac{1}{\nu_{\text{ed}}} \left(\frac{1}{N}\right)^\delta \prod_{i=1}^N \frac{1}{\min(n_i,z_i)+1} \binom{n_i}{d_i}^{-1}\binom{m_i}{a_i}^{-1}q(\bm{y}_i|\I_i).
        \end{aligned}
    \label{eq_sup:aux_dist_edit_alloc_curr}
\end{equation}
whilst if $(\modeprop,u^\prime) = f(\iS^m, u)$ has been obtained by the involution above we have
\begin{equation}
    \begin{aligned}
        q(u^\prime |\modeprop) &= q(\delta) q(\bm{z} | \delta) \prod_{i=1}^N q(a_i) q(\bm{v}_i^\prime|a_i)q(\bm{v}_i|a_i,z_i) q((\I_i)_{\bm{v}_i}|\I^\prime_i)\\
        &=\frac{1}{\nu_{\text{ed}}} \left(\frac{1}{N}\right)^\delta \prod_{i=1}^N \frac{1}{\min(m_i,z_i)+1} \binom{m_i}{a_i}^{-1} \binom{n_i}{d_i}^{-1}q((\I_i)_{\bm{v}_i} |\I^\prime_i).
    \end{aligned}
    \label{eq_sup:aux_dist_edit_alloc_prop}
\end{equation}

Taking the ratio of \eqref{eq_sup:aux_dist_edit_alloc_prop} and \eqref{eq_sup:aux_dist_edit_alloc_curr} leads to the following
\begin{equation}
    \begin{aligned}
        \frac{q(u^\prime | \modeprop)}{q(u |\iS^m)} &= \prod_{i=1}^N \frac{\min(n_i,z_i)+1}{\min(m_i,z_i)+1} \frac{q((\I_i)_{\bm{v}_i} |\I^\prime_i)}{q(\bm{y}_i |\I_i)}.
    \end{aligned}
    \label{eq_sup:edit_alloc_ratio}
\end{equation}
which is the key term appearing in the acceptance probability of this move, as seen in \eqref{eq_sup:posterior_cond_mode_ratio_term}.

We finalise these details on the edit allocation move with a discussion on entry insertion distributions. The simplest option here is to sample entries uniformly over the vertex set $\V$. In this case, with $V = |\V|$, we have
\begin{equation}
    \begin{aligned}
        q(\bm{y}_i |\I_i) = \left(\frac{1}{V}\right)^{a_i}
    \end{aligned}
    \label{eq_sup:edit_alloc_entry_insertion_uniform}
\end{equation}
which implies
$$\frac{q((\I_i)_{\bm{v}_i} |\I^\prime_i)}{q(\bm{y}_i |\I_i)}= \left(\frac{1}{V}\right)^{d_i-a_i} = \left(\frac{1}{V}\right)^{2d_i-z_i} = \left(\frac{1}{V}\right)^{n_i - m_i}$$
any of which can be plugged into \eqref{eq_sup:edit_alloc_ratio}. 

As an alternative choice, one can consider informing the entry insertions from observed data. This approach is based on the following assumption: If two vertices have been observed in the same path across many observations then the probability of proposing one given the other is already present should be higher within the MCMC algorithm.

To reflect this assumption in a proposal, we first extract the necessary information from the observed data. Letting 
$$\iS^{(1)},\dots,\iS^{(n)} $$
denote the observed sample we construct a \textit{co-occurrence matrix} $\mathbf{A} \in \mathbb{Z}_{\geq 0}^{V\times V}$, defined as follows
\begin{equation*}
    \begin{aligned}
        \mathbf{A}_{vv^\prime} &= \# \text{observations with an interaction containing both $v$ and $v^\prime$}\\
        &= |\{k  \, : \, \exists\,\, \I \in \iS^{(k)} \text{ with } v, v^\prime  \in \I  \}|\\
    \end{aligned}
\end{equation*}
where $v \ne v^\prime$, whilst for $v = v^\prime$ we let
\begin{equation*}
    \begin{aligned}
        \mathbf{A}_{vv} &= \# \text{observations with an interaction containing $v$ at least twice}\\
        &= |\{k  \, : \, \exists\,\, \I \in \iS^{(k)} \text{ with } v \in \I \text{ at least twice} \}|,\\
    \end{aligned}
\end{equation*}
which can be seen as the adjacency matrix of a weighted graph describing the co-occurrence structure observed in the data. Now, given $\mathbf{A}$ we construct a probability matrix $\mathbf{P} \in \mathbb{R}^{V \times V}$ by normalising the rows, that is
\begin{equation*}
    \begin{aligned}
        \mathbf{P}_{v v^\prime} &= \mathbf{A}_{vv^\prime}/Z_v 
    \end{aligned}
\end{equation*}
where $Z_v = \sum_{v^\prime \in \V} \mathbf{A}_{v v^\prime}$ is the normalising constant of the $v$th row. Intuitively, the entry $\mathbf{P}_{vv^\prime}$ can be seen as the probability of observing $v^\prime$ in an interaction given $v$ is known to already be present. We consider using $\Pb$ to inform entry insertions as follows. Suppose that $\I_i = (x_{i1},\dots,x_{in_i})$ denotes the path being edited, with $\bm{v}_i$ denoting the subsequence of $[n_i]$ indexing which entries are to be deleted. Introduce the notation $\bm{v}_i^c$ for the complement of $\bm{v}_i$, which is the subsequence of $[n_i]$ containing the entries not in $\bm{v}_i$. For example, with $\bm{v} = (1,2,5) \in [5]$ we would have $\bm{v}^c=(3,4)$. Now, observed that $(\I_i)_{\bm{v}_i^c}$ denotes the entries of $\I_i$ \textit{not} being deleted, that is, those being preserved. Our approach is to now propose entries which have often been observed in the data alongside those being preserved. Since each unique preserved entry has an associated distribution over $\V$ given by the respective row of $\mathbf{P}$, we can consider mixing these distributions together with equal weight to form an entry proposal distribution. In particular, we sample entry insertions for the $i$th path i.i.d. via the following 
\begin{equation*}
    \label{eq:informed_prop_simple}
    \begin{aligned}
        q(y|\I_i) \propto \sum_{v \in (\I_i)_{\bm{v}_i^c}} \mathbf{P}_{v y}.
    \end{aligned}
\end{equation*}

One can also introduce a tuning parameter to control the extent to which proposals are informed by the data. In particular, with $\alpha>0$ first alter the probability matrix as follows
\begin{equation*}
    \begin{aligned}
        \mathbf{P}^\alpha_{v v^\prime} \propto \mathbf{P}_{v v^\prime} + \alpha
    \end{aligned}
\end{equation*}
which normalises to 
\begin{equation*}
    \begin{aligned}
        \mathbf{P}^\alpha_{v v^\prime} = \frac{\mathbf{P}_{v v^\prime} + \alpha}{1 + V\alpha},
    \end{aligned}
\end{equation*}
for which $\mathbf{P}^\alpha_{v v^\prime} \to 1/V$ as $\alpha \to \infty$, that is, the rows converge to the uniform distribution over $\V$. We can now define an analogous insertion distribution
\begin{equation*}
    \begin{aligned}
         q_\alpha(y|\I_i) \propto \sum_{v\in (\I_i)_{\bm{v}_i^c}} \mathbf{P}^\alpha_{v y}
    \end{aligned}
\end{equation*}
where as $\alpha \to \infty$ this will converge to a mixture of uniform distributions over $\V$, that is, also a uniform distribution. In this way, one has a proposal which is informed by the data, but becomes less informed as the tuning parameter $\alpha \to \infty$. 

We finish with a note regarding evaluation of \eqref{eq_sup:edit_alloc_ratio} for this informed proposal. Supposing that $\I^\prime_i$ is $i$th path in the proposal $\modeprop$ (obtained by deleting $d_i$ entries of $\I_i$ indexed by $\bm{v}_i$, and inserting entries $\bm{y}_i$ at locations indexed by $\bm{v}^\prime_i$), then observe we have $(\I_i)_{\bm{v}_i^c} = (\I^\prime_i)_{(\bm{v}^\prime_i)^c}$ (preserved entries)
which thus implies $q_\alpha(y|\I_i) = q_\alpha(y|\I^\prime_i)$. Consequently we can write the following
$$\frac{q_\alpha((\I_i)_{\bm{v}_i} |\I^\prime_i)}{q_\alpha(\bm{y}_i |\I_i)}=\frac{q_\alpha((\I_i)_{\bm{v}_i} |\I_i)}{q_\alpha(\bm{y}_i |\I_i)}$$
and hence only the single mixed distribution $q_\alpha(y |\I_i)$ needs to be constructed. This is helpful to bare in mind when evaluating \eqref{eq_sup:edit_alloc_ratio}.

\subsection{Path Insertion and Deletion Move}

\label{sec_sup:inference_trans_dim}
Supposing that $\iS^m = (\I_1,\dots,\I_N)$ denotes the current state, recall that for this move we have an auxiliary variable given by 
$$u = (\varepsilon, d, \bm{v}, \bm{v}^\prime, \I^*_1,\dots,\I^*_a)$$ 
where (i) $\varepsilon$ denotes the total number of paths to be inserted or deleted, (ii) $d$ denotes the number of paths to be deleted, implying $a = \varepsilon-d$ insertions, (iii) $\bm{v}$ and $\bm{v}^\prime$ denote subsequences indexing path deletions and insertions respectively, and (iv) $(\I^*_1,\dots,\I^*_a)$ denote the paths to be inserted. 

We now define the involution of this iMCMC move. As outlined in \Cref{sec:inference_trans_dim}, if we decompose the the involution as follows 
$$f(\iS^m, u) = (f_1(\iS^m, u), f_2(\iS^m, u)) = (\modeprop, u^\prime)$$
then enacting the path insertions and deletions parameterised by $u$ defines the first component $f_2(\iS^m, u) =  \modeprop$. The second component we define a follows 
\begin{equation*}
    \begin{aligned}
        f_2(\iS^m, u) = (\varepsilon, \varepsilon-d, \bm{v}^\prime, \bm{v}, \I_{v_1},\dots,\I_{v_{d}})
    \end{aligned}
\end{equation*}
which intuitively parameterises the exact opposite set of operations to $u$, namely where we make $\varepsilon$ total insertions and deletions but instead delete $\varepsilon-d = a$ paths indexed by $\bm{v}^\prime$, before inserting the paths $(\I_{v_1}, \dots,\I_{v_d})$ (of $\iS^m$) into locations indexed by $\bm{v}$. As such, we have the following 
$$f_1(\modeprop, u^\prime) = \modecurr$$
furthermore, reapplying the second component just defined leads to 
\begin{equation*}
    \begin{aligned}
        f_2(\modeprop, u^\prime) &= \left(\varepsilon,  \varepsilon - (\varepsilon-d), \bm{v}, \bm{v}^\prime, \I^\prime_{v^\prime_1},\dots,\I^\prime_{v^\prime_{\varepsilon-d}}\right)\\
        &=(\varepsilon, d, \bm{v}, \bm{v}^\prime, \I^*_1,\dots,\I^*_a)
    \end{aligned}
\end{equation*}
using the fact that $\I^\prime_{v^\prime_i}=\I^*_i$, since by definition $\I^*_i$ was inserted to the $(v^\prime_i)$th entry of $\modeprop$. Altogether this implies 
$$f(f(\modecurr, u)) = f(\modeprop, u^\prime) = (\modecurr, u)$$
that is, $f(\modecurr, u)$ is an involution. 

Regarding the auxiliary distribution $q(u |\modecurr)$, recall the following assumptions stated in \Cref{sec:inference_trans_dim}
\begin{equation*}
    \begin{aligned}
        \varepsilon & \sim \text{Uniform}\{1,\dots,\nu_{\text{td}}\} \\
        d \, | \, \varepsilon &\sim \text{Uniform}\{0,\dots,\min(N, \varepsilon)\} \\
    \end{aligned}
\end{equation*}
whilst we sample indexing subsequences $\bm{v}$ and $\bm{v}^\prime$ uniformly and assume path insertions a drawn via some general distribution $q(\I | \modecurr)$. In this instance, recall that $\bm{v}$ is a subsequence of $[N]$ of size $d$, whilst $\bm{v}^\prime$ is a subsequence of $[M]$ of size $a$, where $M=N-d+a$ is the length of $\modeprop$. Sampling these uniformly thus implies 
\begin{equation*}
    \begin{aligned}
        q(\bm{v}|d) = \binom{N}{d}^{-1} && q(\bm{v}^\prime | \varepsilon, d) = \binom{M}{a}^{-1} 
    \end{aligned}
\end{equation*}
leading to the following closed form
\begin{equation*}
    \begin{aligned}
        q(u |\modecurr)&= q(\varepsilon)q(d|\varepsilon)q(\bm{v}|d)q(\bm{v}^\prime|\varepsilon,d)\prod_{i=1}^a q(\I^*_i|\modecurr) \\
        &= \frac{1}{\nu_{\text{td}}} \frac{1}{\min(N,\varepsilon)+1} \binom{N}{d}^{-1}\binom{M}{a}^{-1}\prod_{i=1}^a q(\I^*_i |\modecurr)
    \end{aligned}
\end{equation*}
whilst, if $(\modeprop, u^\prime) = f(\modecurr, u)$ has been obtained by the involution above, we have 
\begin{equation*}
    \begin{aligned}
        q(u^\prime |\modeprop)&= q(\varepsilon)q(a|\varepsilon)q(\bm{v}^\prime|a)q(\bm{v}|\varepsilon,a)\prod_{i=1}^d q(\I_{v_i}|\modeprop) \\
        &= \frac{1}{\nu_{\text{td}}} \frac{1}{\min(M,\varepsilon)+1} \binom{M}{a}^{-1} \binom{N}{d}^{-1}\prod_{i=1}^d q(\I_{v_i} |\modeprop).
    \end{aligned}
\end{equation*}

Taking the ratio of these leads to the following 
\begin{equation}
    \begin{aligned}
        \frac{q(u^\prime |\modeprop)}{q(u|\modecurr)} = \frac{\min(N,\varepsilon)+1}{\min(M,\varepsilon)+1} \frac{\prod_{i=1}^d q(\I_{v_i} | \modeprop)}{\prod_{i=1}^a q(\I^\prime_{v^\prime_i} | \modecurr)},
    \end{aligned}
    \label{eq_sup:trans_dim_ratio}
\end{equation}
which can be substituted into  \eqref{eq_sup:posterior_cond_mode_ratio_term} to evaluate the acceptance probability of this move (here we again use the fact $\I^\prime_{v^\prime_i}=\I^*_i$). 

We finalise by discussing possible choices for the path insertion distribution. The simplest approach is to combine a distribution on path length with uniform sampling of entries. In particular, to sample some path $\I=(x_1,\dots,x_m)$ we (i) sample its length $m$ via some distribution $q(m)$ (ii) sample entries $x_i$ uniformly from $\V$. This implies 
$$q(\I | \modecurr) = q(\I) = q(m)\left(\frac{1}{V}\right)^m$$
where $V = |\V|$, which can be substituted into \eqref{eq_sup:trans_dim_ratio}. 

One can also consider informing entry insertions from observed data. With 
$$\iS^{(1)},\dots,\iS^{(n)}$$
a sample, for each $v \in \V$ we let
$$c_v = |\{k \, : \, \exists \, \I \in \iS^{(k)} \text{ with } v \in \I\}|$$
denote the number of observations with at least one path containing the vertex $v$. Normalising this leads to
$$p_v = \frac{c_v}{\sum{v \in \V}c_v}$$
which can be seen as the probability a randomly selected observation contains $v$. Introducing the parameter $\alpha>0$ we let 
$$q_\alpha(v) \propto p_v + \alpha$$
which normalises to 
$$q_\alpha(v) = \frac{p_v + \alpha}{1 + \alpha V}.$$
One can now use this to sample path entries, namely to sample $\I=(x_1,\dots,x_m)$ we (i) sample length $m$ via some $q(m)$, (ii) sample entries $x_i$ via $q_\alpha(x_i)$. Observe that if $\alpha=0$ we have $q_\alpha(v)=p_v$, and the entry insertion distribution is fully informed by the data, whilst as $\alpha \to \infty$ we have $q_\alpha(v) \to 1/V$, and we recover uniform entry insertions. 

\subsection{Model Sampling}
\label{sec_sup:inference_model_sampling}
In this section we provide supporting details regarding our iMCMC algorithm to sample from the SIS models outlined in \Cref{sec:inference_model_sampling}. Recall that for the SIS model (\Cref{def:SIS}) the (normalised) probability of observing $\iS$ is given by
\begin{equation*}
    \begin{aligned}
        p(\iS|\iS^m, \gamma) = \frac{\exp\left\{ - \gamma d_S(\iS,\iS^m)\right\}}{Z(\iS^m, \gamma)}.
    \end{aligned}
    \label{eq_sup:sis_normalised_probability}
\end{equation*}
implying the following closed form for the acceptance probability \eqref{eq:sis_model_sampling_imcmc_acc_prob} 
\begin{equation}
    \begin{aligned}
        \alpha(\iS, \iS^\prime) = \min\left\{ 1, H(\iS, \iS^\prime)\right\}
    \end{aligned}
    \label{eq_sup:sis_model_sampling_imcmc_acc_prob}
\end{equation}
where 
\begin{equation*}
    \begin{aligned}
        H(\iS, \iS^\prime) &= \frac{p(\iS^\prime | \iS^m, \gamma)}{p(\iS|\iS^m, \gamma)} \frac{q(u^\prime|\iS^\prime)}{q(u|\iS)}\\
        &= \exp\left\{-\gamma \bigg(d_S(\iS^\prime,\iS^m)-d_S(\iS,\iS^m) \bigg) \right\}\frac{q(u^\prime|\iS^\prime)}{q(u|\iS)},
    \end{aligned}
\end{equation*}
where the value of $q(u^\prime|\iS^\prime)/q(u|\iS)$ will depend on the iMCMC specification. 

As mentioned in \Cref{sec:inference_model_sampling}, we consider re-using the iMCMC moves of our iExchange scheme used to sample from the mode conditional (\Cref{sec_sup:inference_edit_alloc,sec_sup:inference_trans_dim}). For ease of reference, we summarise the corresponding ratios for each move:
\begin{itemize}
    \item \textbf{Edit allocation} - suppose that $u$, $f(u,\iS)$ and $q(u|\iS)$ are defined as in \Cref{sec_sup:inference_edit_alloc} (replacing $\iS^m$ with $\iS$ and $\modeprop$ with $\iS^\prime$) with a uniform entry insertion distribution \eqref{eq_sup:edit_alloc_entry_insertion_uniform}. With $\iS=(\I_1,\dots,\I_N)$ the current state, supposing $u=(\delta, \bm{z}, u_1,\dots,u_N)$ has been sampled via $q(u|\iS)$ mapping to $(\iS^\prime,u^\prime)=f(\iS,u)$ with $\iS^\prime=(\I^\prime_1,\dots,\I^\prime_N)$ we will have 
    \begin{align}
        \frac{q(u^\prime | \iS^\prime)}{q(u |\iS)} &= \prod_{i=1}^N \frac{\min(n_i,z_i)+1}{\min(m_i,z_i)+1} \left(\frac{1}{V}\right)^{n_i - m_i}
        \label{eq_sup:sis_model_sampling_edit_alloc_ratio}
    \end{align}
    where $n_i$ and $m_i$ denote the lengths of the $i$th path in $\iS^m$ and $\modeprop$ respectively;
    \item \textbf{Path insertion and deletion} - suppose that $u$, $f(u,\iS)$ and $q(u|\iS)$ are defined as in \Cref{sec_sup:inference_trans_dim} (again using $\iS$ and $\iS^\prime$ instead of $\iS^m$ and $\modeprop$). With $\iS=(\I_1,\dots,\I_N)$ the current state, supposing $u=(\varepsilon, d, \bm{v}, \bm{v}^\prime, \I^*_1,\dots,\I^*_a)$ (where $a=\varepsilon-d$) has been sampled via $q(u|\iS)$ mapping to $(\iS^\prime,u^\prime)=f(\iS,u)$ with $\iS^\prime=(\I^\prime_1,\dots,\I^\prime_M)$ we will have 
    \begin{align}
        \frac{q(u^\prime |\iS^\prime)}{q(u|\iS)} = \frac{\min(N,\varepsilon)+1}{\min(M,\varepsilon)+1} \frac{\prod_{i=1}^d q(\I_{v_i} | \iS^\prime)}{\prod_{i=1}^a q(\I^\prime_{v^\prime_i} | \iS)}.
        \label{eq_sup:sis_model_sampling_trans_dim_ratio}
    \end{align}
\end{itemize}

As mentioned in \Cref{sec:inference_model_sampling}, we follow the approach used for the posterior mode conditional and consider mixing together these two iMCMC moves with some proportion $\beta \in (0,1)$, left as a tuning parameter. Pseudocode of the resultant algorithm can be found in \Cref{alg_sup:SIS_model_sampling}.

	\section{Bayesian Inference for Multiset Models}
\label{sec_sup:inference_sim}

Here we detail the approach to inference for the interaction-multiset models (\Cref{def:SIM}). This is very similar to the interaction-sequence models outlined in \Cref{sec:inference}, with priors, hierarchical model and posterior are all being essentially the same (albeit with different notation). Computationally, we again use MCMC to sample from the posterior, adapting the scheme proposed for the interaction-sequence models.

\subsection{Priors, Hierarchical Model and Posterior}

To specify priors, we follow \Cref{sec:inference_priors} and assume the mode was itself sampled from an SIM model, namely 
$$\E^m \sim \text{SIM}(\E_0, \gamma_0)$$
where $(\E_0, \gamma)$ are hyperparameters, whilst we assume the dispersion was drawn from some distribution $p(\gamma)$ whose support is a subset of the non-negative reals. Given these specifications, an observed sample $\edata$ is assumed to be drawn via 
\begin{equation*}
    \begin{aligned}
    \E^{(i)} \, | \, \E^m, \gamma &\sim \text{SIM}(\E^m, \gamma) & (\text{for }i=1,\dots,n) \\
    \E^m &\sim \text{SIM}(\E_0,\gamma_0) & \\
    \gamma &\sim p(\gamma). &\\
    \end{aligned}
\end{equation*}

The likelihood of $\edata$ is given by 
\begin{equation*}
    \begin{aligned}
    p(\edata \, | \, \E^m, \gamma) &= \prod_{i=1}^n p(\E^{(i)} \, | \, \E^m, \gamma) \\
    &=Z(\E^m, \gamma)^{-n} \exp\left\{-\gamma \sum_{i=1}^n d_E(\E^{(i)}, \E^m) \right\}
    \end{aligned}
\end{equation*}
which implies a posterior given by
\begin{equation}
    \begin{aligned}
    p(\E^m, \gamma \, | \, \edata) &\propto p(\edata \, | \, \E^m, \gamma) p(\E^m \,) p(\gamma)\\
    &= Z(\E^m, \gamma)^{-n} \exp\left\{-\gamma \sum_{i=1}^n d_E(\E^{(i)}, \E^m) \right\} \\
    &\qquad \exp\{-\gamma_0 d_E(\E^m,\E_0)\}p(\gamma).
    \end{aligned}
    \label{eq_sup:SIM_posterior}
\end{equation}

\subsection{Posterior Sampling}

As for the interaction-sequence models, we consider sampling from the posterior \eqref{eq_sup:SIM_posterior} via component-wise MCMC algorithm, alternating between sampling from the two conditionals 
\begin{equation*}
    \begin{aligned}
        p(\E^m \, | \, \gamma , \edata) && \text{and} && p(\gamma \, | \, \E^m, \edata)
    \end{aligned}
\end{equation*}
in both of which the normalising constant of \eqref{eq_sup:SIM_posterior} will persist, making them doubly intractable \citep{Murray2006} and motivating the use of the exchange and iExchange algorithms. 

There are two key differences here compared with the setting of \Cref{sec:inference}. Firstly, the mode in this instance is a multiset, implying the mode conditional is a distribution over multisets rather than sequences. Secondly, to induce the required cancellation of normalising constants, sampling of auxiliary data in the exchange (or iExchange) algorithms must be from the multiset models. 

In both cases, the challenge lies in sampling from distributions over multisets (of paths). As will be seen in subsequent sections, a solution can be found by first extending these to distributions over sequences, before using the iMCMC-based algorithms proposed for the interaction-sequence models (\Cref{sec:inference,sec_sup:mcmc_extra_details}) to target them.

\subsection{Dispersion Conditional}

\label{sec_sup:inference_sim_dispersion}

Conditioning on $\E^m$ in \eqref{eq_sup:SIM_posterior} we have the following
\begin{equation*}
    \begin{aligned}
        p(\gamma \, | \, \E^m, \edata) \propto Z(\E^m,\gamma)^{-n}\exp\left\{-\gamma \sum_{i=1}^n d_E(\E^{(i)}, \E^m) \right\} p(\gamma)
    \end{aligned}
\end{equation*}
which to target we follow \Cref{sec:inference_dispersion_cond,sec_sup:inference_dispersion_cond} and use the exchange algorithm \citep{Murray2006}. For the proposal $q(\gamma^\prime | \gamma)$ we again consider sampling $\gamma^\prime$ uniformly over a $\varepsilon$-neighbourhood of $\gamma$ with reflection at zero (see \Cref{sec_sup:inference_dispersion_cond}). With this choice of proposal, a single iteration consists of the following. Assuming $\gamma$ is the current state, we first sample proposal $\gamma^\prime$ via $q(\gamma^\prime | \gamma)$. Next, we sample auxiliary data $\eadata$ i.i.d. from the appropriate multiset model, namely
$$\E^*_i \sim \text{SIM}(\E^m, \gamma^\prime) \quad (\text{for }i=1,\dots,n),$$
for which we have 
$$p(\eadata \, | \, \E^m, \gamma^\prime) = Z(\E^m, \gamma^\prime)^{-n}\exp\left\{ -\gamma^\prime \sum_{i=1}^n d_E(\E^*_i,\E^m)\right\}.$$
Finally, we accept this proposal with the following probability 
\begin{equation}
    \begin{aligned}
        \alpha(\gamma, \gamma^\prime) = \min\{1,H(\gamma, \gamma^\prime)\}
    \end{aligned}
    \label{eq_sup:sim_posterior_cond_dispersion_acc_prob}
\end{equation}
where 
\begin{equation*}
    \begin{aligned}
        H(\gamma, \gamma^\prime) &=  \frac{p(\gamma^\prime \,|\,\E^m, \edata)p(\eadata \,| \,\E^m, \gamma)}{p(\gamma\, | \,\E^m, \edata)p(\eadata\,|\,\E^m, \gamma^\prime)} \frac{q(\gamma|\gamma^\prime)}{q(\gamma^\prime| \gamma)} \\[0.25cm]
        &=\exp\left\{-(\gamma^\prime-\gamma) \left(\sum_{i=1}^n d_E(\E^{(i)},\E^m) - \sum_{i=1}^n d_E(\E^*_i,\E^m)\right)\right\}\frac{p(\gamma^\prime)}{p(\gamma)},
    \end{aligned}
\end{equation*}
where, as in \Cref{sec_sup:inference_dispersion_cond}, normalising constants of the (conditional) posterior and auxiliary data cancel one another out, whilst the proposal density terms cancel due to its symmetry. 

\subsection{Mode Conditional}

\label{sec_sup:inference_sim_mode}

Conditioning on $\gamma$ in \eqref{eq_sup:SIM_posterior} we have the following 
\begin{equation}
    \begin{aligned}
        p(\E^m \, | \, \gamma, \edata) \propto Z(\E^m,\gamma)^{-n}\exp\left\{-\gamma \sum_{i=1}^n d_E(\E^{(i)}, \E^m) - \gamma_0d_E(\E^m, \E_0)\right\},
    \end{aligned}
    \label{eq_sup:sim_posterior_mode_cond}
\end{equation}
which is a distribution over $\Eb$, that is, the space of multisets. To re-use the iExchange scheme of \Cref{sec:inference_mode_cond} we instead need a distribution over the space of interaction sequences $\iSb$. To this end, we extend \eqref{eq_sup:sim_posterior_mode_cond} to a distribution over interaction sequences. 

Consider the general problem of extending some distribution $\pi(\E)$ over $\Eb$ to one over $\iSb$. Firstly, observe each $\E$ is associated with a set of sequences, obtained by placing the interactions of $\E$ in different orders. More formally, $\E$ can be seen as equivalence class of sequences (see \Cref{sec_sup:sample_spaces}). 
As such, one can consider assigning equal probability to each unique ordering of $\E$. In particular, for $\iS \in \iSb$ we let
\begin{equation*}
\label{eq_sup:multiset_extended_target}
    \begin{aligned}
        \tilde{\pi}(\iS) = \frac{1}{A(\E)}\pi(\E)
    \end{aligned}
\end{equation*}
where $\E$ is the multiset obtained from $\iS$ by disregarding the order of interactions, and $A(\E)$ denotes the number of unique orderings of the paths in $\E$. 

The form of $A(\E)$ can be obtained as follows. Suppose that $\E$ consists of $N$ paths, with $M\leq N$ \textit{unique} paths. Without loss of generality label the unique paths $1$ to $M$ and let $w_i$ denote the multiplicity of the $i$th path. Now, if each path of $\E$ is different there are $N!$ possible ways to order them. However, if there are repeated paths this will include double counting. Therefore, in general we must further divide by $(w_i)!$ leading to the familiar multinomial term 
\begin{equation}
    \label{eq_sup:multiset_extension_combi_term}
    A(\E) := \binom{N}{w_1,\dots,w_N} = \frac{N!}{w_1!\cdots w_N!}.
\end{equation}

Through this reasoning we can extend \eqref{eq_sup:sim_posterior_mode_cond} as follows
\begin{equation}
    \begin{aligned}
        \tilde{p}(\iS^m \, | \, \gamma, \edata) = \frac{1}{A(\E^m)}p(\E^m \, | \, \gamma, \edata)
    \end{aligned}
    \label{eq_sup:sim_posterior_mode_cond_ext}
\end{equation}
where now $\iS^m \in \iSb$ and $\E^m$ is the multiset obtained from $\iS^m$ by disregarding the order of paths. 

We can now re-use the iExchange algorithm of \Cref{sec:inference_mode_cond,sec_sup:inference_mode_cond} to target \eqref{eq_sup:sim_posterior_mode_cond_ext}. However, note the normalising constant appearing in \eqref{eq_sup:sim_posterior_mode_cond}, and hence also in \eqref{eq_sup:sim_posterior_mode_cond_ext}, is that of an SIM model. Thus, for the iExchange algorithm to induce the necessary cancellation auxiliary data must be sampled from an SIM model. 

A single iteration of the resultant algorithm consists of the following. Suppose that $\E^m$ denotes our current state and $\gamma$ is fixed. We first construct an interaction sequence $\iS^m$ by placing the interactions of $\E^m$ in an arbitrary order. Now, assuming $u$, $q(u |\iS^m)$ and $f(\iS^m, u)$ is some iMCMC specification as used in \Cref{sec:inference}, we sample auxiliary variables $u$ via $q(u|\iS^m)$, before invoking the involution to obtain $(\modeprop, u^\prime)=f(\modecurr, u)$, where $\modeprop$ denotes our proposal. By now disregarding the order of interactions in $\modeprop$, we obtain a proposal $\emodeprop$. We then sample auxiliary data $\eadata$ i.i.d. where 
$$\E^*_i \sim \text{SIM}(\emodeprop, \gamma)$$
which implies 
$$p(\eadata \, | \, \emodeprop, \gamma) = Z(\emodeprop, \gamma)^{-n}\exp\left\{ -\gamma \sum_{i=1}^n d_E(\E^*_i,\emodeprop)\right\},$$
before accepting $\emodeprop$ with the following probability
\begin{equation}
    \begin{aligned}
        \alpha(\E^m, \emodeprop) = \min\left\{ 1, H(\E^m, \emodeprop) \right\}
    \end{aligned}
    \label{eq_sup:sim_posterior_cond_mode_acc_prob}
\end{equation}
where 
\begin{equation}
    \begin{aligned}
        H(\E^m, \emodeprop) &=\frac{\tilde{p}(\modeprop \, | \, \gamma, \edata)}{\tilde{p}(\iS^m \, | \, \gamma, \edata)}\frac{p(\eadata \, | \, \E^m, \gamma)}{p(\eadata \, | \, \emodeprop, \gamma)}\frac{q(u^\prime \, |\,\modeprop)}{q(u\,|\,\iS^m)} \\[0.25cm] 
        &=\frac{\frac{1}{A(\emodeprop)}p(\emodeprop \, | \, \gamma, \edata)}{\frac{1}{A(\E^m)}p(\E^m \, | \, \gamma, \edata)}\frac{p(\eadata \, | \, \E^m, \gamma)}{p(\eadata \, | \, \emodeprop, \gamma)}\frac{q(u^\prime \, |\,\modeprop)}{q(u\,|\,\iS^m)} \\[0.25cm]
        &=\frac{A(\E^m)}{A(\emodeprop)} \exp\Bigg\{-\gamma\left( \sum_{i=1}^n d_E(\E^{(i)},\emodeprop) - \sum_{i=1}^n d_E(\E^{(i)},\E^m)\right) \\ &\quad -\gamma\left(\sum_{i=1}^n d_E(\E^*_i,\E^m) - \sum_{i=1}^n d_E(\E^*_i, \emodeprop) \right) 
        \\ &\qquad -\gamma_0\left( d_E(\emodeprop, \E_0) - d_E(\emodecurr,\E_0)\right)\Bigg\} \frac{q(u^\prime \, |\,\modeprop)}{q(u\,|\,\iS^m)}
    \end{aligned}
\end{equation}
where here $\iS^m$ and $\modeprop$ correspond to those used above to generate the proposal $\emodeprop$. Again, we observe cancellation of normalising constants due to the introduction of auxiliary data. We also see the introduction of a combinatorial term, namely
\begin{equation}
    \label{eq:combinatorial_term_acc_prob}
    \begin{aligned}
        \frac{A(\E^m)}{A(\emodeprop)} = \frac{N!(w^\prime_1!\cdots w^\prime_{M^\prime}!)}{M!(w_1!\cdots w_M!)}
    \end{aligned}
\end{equation}
where $N$ and $M$ are the cardinalities of $\E^m$ and $\emodeprop$ respectively, $w_i$ is the multiplicity of the $i$th unique path in $\E^m$ and $w^\prime_i$ is the multiplicity of the $i$th unique path in $\emodeprop$. 

Clearly, this all depends on a particular iMCMC specification (auxiliary variables, involution and auxiliary distribution). For this we can use the edit allocation (\Cref{sec:inference_edit_alloc,sec_sup:inference_edit_alloc}) and interaction insertion and deletion (\Cref{sec_sup:inference_trans_dim,sec_sup:inference_trans_dim}) moves, which we again mix together with proportion $\beta \in (0,1)$, left as a tuning parameter. A pseudocode summary of the resulting algorithm to update the mode can be seen in \Cref{alg_sup:sim_mode_accept_reject}. 

One pragmatic note to be made here is that computationally it is often easier to work with sequences than multisets, since the former can be stored as a vector. To this end, one can store observations as sequences of paths but interpret them as multisets of paths. Furthermore, we can take the order in which they are stored as the `arbitrary order' referred to in \Cref{alg_sup:sim_mode_accept_reject}, and in this way the whole algorithm can be enacted on vectors of paths, simply interpreting the output samples as multisets of paths.

\subsection{Model Sampling}

\label{sec_sup:inference_sim_model_sampling}

The exchange-based algorithms to update $\E^m$ and $\gamma$ both require exact sampling of auxiliary data from the SIM models. As for the interaction-sequence models (\Cref{sec:inference_model_sampling}), this is not possible in general. As such, we replace this with approximate samples obtained via an MCMC algorithm. 

Towards proposing a suitable MCMC algorithm, we follow the reasoning of \Cref{sec_sup:inference_sim_mode} and extend the target distribution (over multisets of paths) to one over sequences of paths, before appealing to the iMCMC scheme proposed to sample from the SIS models (\Cref{sec:inference_model_sampling,sec_sup:inference_model_sampling}). Recalling that for the SIM model (\Cref{def:SIM}) the (normalised) probability of observing $\E \in \Eb$ is given by 
$$p(\E | \E^m, \gamma) = \frac{1}{Z(\E^m, \gamma)} \exp\{ -\gamma d_E(\E, \E^m)\},$$
we can assign any $\iS \in \iSb$ the following probability
\begin{equation}
    \label{eq_sup:sim_model_ext}
    \begin{aligned}
        \tilde{p}(\iS| \E^m, \gamma) = \frac{1}{A(\E)} p(\E|\E^m, \gamma)
    \end{aligned}
\end{equation}
where $\E$ is multiset obtain from $\iS$ by disregarding order, and $A(\E)$ is as defined in \eqref{eq_sup:multiset_extension_combi_term}, thus defining an extended distribution over $\iSb$.

We can now target \eqref{eq_sup:sim_model_ext} via iMCMC as in \Cref{sec:inference_model_sampling}. In particular, suppose that one would like to sample from an $\text{SIM}(\E^m, \gamma)$ model. With $u$, $q(u|\iS)$ and $f(\iS,u)$ some iMCMC specification as used therein, and $\E$ the current state, a single iteration of will consist of the following
\begin{enumerate}
    \item Construct interaction sequence $\iS$ by placing the paths of $\E$ in an arbitrary order
    \item Sample $u \sim q(u|\iS)$
    \item Invoke involution $f(\iS, u)=(\iS^\prime, u^\prime)$
    \item Disregard order in $\iS^\prime$ to obtain proposed multiset $\E^\prime$
    \item Evaluate the following probability 
    \begin{equation}
        \label{eq_sup:sim_model_sampling_acc_prob}
        \begin{aligned}
            \alpha(\E, \E^\prime) = \min\left\{1, \frac{\tilde{p}(\iS^\prime|\E^m, \gamma)}{\tilde{p}(\iS|\E^m, \gamma)}\frac{q(u^\prime |\iS^\prime)}{q(u|\iS)} \right\}
        \end{aligned}
    \end{equation}
    \item Move to state $\E^\prime$ with probability $\alpha(\E, \E^\prime)$, staying at $\E$ otherwise.
\end{enumerate}

Clearly, this is conditional upon the choice of iMCMC specification. Here, we follow \Cref{sec:inference_model_sampling} and recycle the edit allocation (\Cref{sec:inference_edit_alloc} and \Cref{sec_sup:inference_edit_alloc}) and path insertion/deletion moves (\Cref{sec:inference_trans_dim} and \Cref{sec_sup:inference_trans_dim}), again mixing them together with proportion $\beta \in (0,1)$, left as a tuning parameter.

A closed form for \eqref{eq_sup:sim_model_sampling_acc_prob} can be derived as follows. Writing $\alpha(\E,\E^\prime) = \min\{1,H(\E,\E^\prime)\}$ we have 
\begin{equation*}   
    \label{eq_sup:sim_model_sampling_acc_prop_closed}
    \begin{aligned}
        H(\E, \E^\prime) &= \frac{\tilde{p}(\iS^\prime|\E^m, \gamma)}{\tilde{p}(\iS|\E^m,\gamma)} \frac{q(u^\prime|\iS^\prime)}{q(u|\iS)}\\
        &=\frac{ \frac{1}{A(\E^\prime)} p(\E^\prime|\E^m, \gamma)}{ \frac{1}{A(\E)} p(\E|\E^m, \gamma)}\frac{q(u^\prime|\iS^\prime)}{q(u|\iS)}\\
        &= \frac{A(\E)}{A(\E^\prime)} \exp \left\{ - \gamma\bigg( d_E(\E^\prime,\E^m) - d_E(\E,\E^m) \bigg)\right\}\frac{q(u^\prime|\iS^\prime)}{q(u|\iS)}
    \end{aligned}
\end{equation*}
where 
\begin{equation*}
    \begin{aligned}
        \frac{A(\E)}{A(\E^\prime)} = \frac{N!(w^\prime_1!\cdots w^\prime_{M^\prime}!)}{M!(w_1!\cdots w_M!)}
    \end{aligned}
\end{equation*}
with $N$ and $M$ the cardinalities of $\E$ and $\E^\prime$ respectively, $w_i$ the multiplicity of the $i$th unique path in $\E$ and $w^\prime_i$ the multiplicity of the $i$th unique path in $\E^\prime$. As when sampling from the interaction-sequence models (\Cref{sec_sup:inference_model_sampling}), the ratio $q(u^\prime|\iS^\prime)/q(u|\iS)$ will be move dependent and identical to those appearing in \Cref{sec_sup:inference_model_sampling}, namely \eqref{eq_sup:sis_model_sampling_edit_alloc_ratio} for the edit allocation move and \eqref{eq_sup:sis_model_sampling_trans_dim_ratio} for the path insertion/deletion move. The whole procedure to sample from the SIM models is summarised in the pseudocode of \Cref{alg_sup:SIM_model_sampling}.

Finally we note that, as for the interaction-sequence models, by using approximate as opposed to exact sampling in the exchange-based algorithms of \Cref{sec_sup:inference_sim_dispersion} and \Cref{sec_sup:inference_sim_mode} we will no longer target the true posterior, but instead an approximation thereof. This approximation can be improved, however, by obtaining samples which look `more exact', often achievable by increasing the burn-in period and/or introducing a lag between samples ($b$ and $l$ of \Cref{alg_sup:SIM_model_sampling}).
	
	\section{Real Data Analysis}

In this section, we provide details supporting the data analysis of \Cref{sec:data_analysis}. This includes further details on the data and how it was processed, and extra information regarding the integer-weighted extension of the SNF model \citep{Lunagomez2021} used in \Cref{sec:data_analysis_graphs}. 

\subsection{Foursquare Data Processing}

\label{sec_sup:data_processing}

The data analysed in \Cref{sec:data_analysis} was obtained from the New York and Tokyo data set of \cite{Yang2015}\footnote{\url{https://sites.google.com/site/yangdingqi/home/foursquare-dataset\#h.p_ID_46}}, which contains a total of 10 months of check-in activity (from 12 April 2012 to 16 February 2013). Each check-in has an associated time stamp, GPS location and venue category information. In particular, for each city, there is a tsv file containing the following columns
\begin{enumerate}
    \item User ID - unique identifier for the user, e.g. \texttt{479}
    \item Venue ID - unique identifier for the venue, e.g. \texttt{49bbd6c0f964a520f4531fe3}
    \item Venue category ID - unique identifier for the venue category, e.g. \\ \texttt{4bf58dd8d48988d127951735}
    \item Venue category name - name for venue category, e.g. \texttt{Arts \& Crafts}
    \item Latitude \& longitude - geographical location for venue, e.g. \texttt{(40.41,-74.00)}
    \item UTC time - time of check-in, to the second, e.g. \texttt{Tue Apr 03 18:00:09 +0000 2012} 
    \item Time zone offset - the offset of local time from UTC for venue (in minutes), e.g. \texttt{-240}
\end{enumerate}

As outlined in \Cref{sec:intro}, we converted this raw data to a sequence or multiset of paths. In particular, we let the vertices $\V$ denote venue categories with a path then representing a day of check-ins for a given user. Notice not all of the information above is required to enact this operation. In particular, all one requires are user IDs, venue category names (or IDs) and local time (a function of UTC and time zone offset).

\subsubsection{Venue Category Hierarchy}

As discussed in \Cref{sec:data_analysis}, the venue categories have a hierarchical structure. For example a venue of category ``Tram Station" is a sub-category of ``Train Station", which is itself a sub-category of ``Travel \& Transport", implying a hierarchical label given by ``Travel \& Transport $>$ Train Station $>$ Tram Station''. As it comes, the data set of \cite{Yang2015} uses low-level category names (``Tram Station"), whilst we consider the highest-level (``Travel \& Transport"). However, we do note that \cite{Yang2015} do not appear to have used the \textit{lowest} level in all cases. 

To get the hierarchical category names we made use of information on the Foursquare site (\href{https://developer.foursquare.com/docs/legacy-foursquare-category-mapping}{see here}). Note that since the release of this data set it appears that Foursquare have changed how they label venues, thus there is another set of venue category names (\href{https://developer.foursquare.com/docs/categories}{see here}). However, the data set of \cite{Yang2015} appears to be congruent with the former. Using this information we were able to essentially `fill-in' the higher-level category labels for each category name appearing in the data set of \cite{Yang2015}, mapping their low-level labels to top-level ones.

\subsubsection{Data Filtering}

As mentioned in \Cref{sec:data_analysis}, we analysed only a subset of 100 data points. This was due to issues causes by the presence of outliers. In this subsection, we outline exactly how this subset of data points was chosen.  

Following processing of the raw data we were left with a sample of interaction multisets $\edata$ with $n=928$. As we discussed in \Cref{sec:data_analysis_processing}, after some initial filtering, including the removal of all length one paths and observations with less than 10 paths, we were left with $n=402$ observations. To get the final $n=100$ data points we further subset these data by making use of a distance metric between observations.

Suppose that $d_E$ is some distance between interaction multisets, then one can choose a subset of size $m$ as follows: find the data point which has the smallest total distance to its $m$ nearest neighbours, taking this neighbourhood as the subset. More formally, introducing the notation $\mathcal{N}_m(\E)$ for the indices of the $m$ nearest neighbours of $\E$ with respect to $d_E$ in the sample, we let
\begin{equation*}
    \begin{aligned}
         \E^* = \argmin_{\E \in \edata} \left[\sum_{i \in \mathcal{N}_m(\E)} d_E(\E, \E^{(i)})\right],
    \end{aligned}
\end{equation*}
with the desired subset then being given by $\{\E^{(i)}\}_{i \in \mathcal{N}_m(\E^*)}$.

Regarding the choice of distance $d_E$, we opted for that used in the model-fit, namely the matching distance with an LSP distance between paths. Moreover, since the observations were of quite different sizes, it made sense to also normalise this distance. In particular, we used the so-called \textit{Steinhaus transform} \cite[Sec. 2]{Bolt2022} which defines a new distance $\bar{d}_{\mathrm{M}}$ as follows 
\begin{equation*}
    \begin{aligned}
        \bar{d}_{\mathrm{M}}(\E, \E') = \frac{2 d_{\mathrm{M}}(\E, \E') }{d_{\mathrm{M}}(\E,\emptyset) + d_{\mathrm{M}}(\E',\emptyset) + d_{\mathrm{M}}(\E,\E')}
    \end{aligned}
\end{equation*}
where here $\emptyset$ denotes the empty multiset of paths, which functions as our reference element in the space of interaction multisets. Note that $d_{\mathrm{M}}(\E,\emptyset)$ is equivalent to the sum of path lengths in $\E$, since each path $\I$ in $\E$ is un-matched and hence penalised by $d_{\mathrm{LSP}}(\I,\Lambda)$. It can be shown that this new distance takes values in the unit interval, that is $\bar{d}_{\mathrm{M}}(\E,\E') \in [0,1]$ \citep{Bolt2022}, where if $\bar{d}_{\mathrm{M}}(\E,\E')=1$ this implies $\E$ and $\E$ are more-or-less completely different. To see why using this normalised distance is sensible an example is helpful. Consider comparing $\E = \{(1,1,1)\}$ with the following two observations
\begin{align*}
    \E^{(1)} = \{(2,2,2)\} && \E^{(2)} = \{(1,1,1),(2,2,2),(2,2,2)\}.
\end{align*}
Observe that $\E^{(1)}$ shares nothing in common with $\E$ whilst $\E^{(2)}$ and $\E$ share a common path, namely $(1,1,1)$. As such, intuitively we might say $\E^{(2)}$ is more similar to $\E$ than $\E^{(1)}$ is, that is, its distance should be lower. However, in this case we will have 
\begin{align*}
    d_{\mathrm{M}}(\E,\E^{(1)}) = 6 && d_{\mathrm{M}}(\E,\E^{(2)}) = 6
\end{align*}
which appears to contradict this intuition. The problem here is the difference in the observation sizes; though $\E^{(2)}$ is more similar to $\E$ it is also larger, hence pushing up its distance. However, by taking sizes into account, the normalised distances evaluates to 
\begin{align*}
    \bar{d}_{\mathrm{M}}(\E,\E^{(1)}) &= \frac{2 \times 6}{3 + 3 + 6} & \bar{d}_{\mathrm{M}}(\E,\E^{(2)}) &= \frac{2 \times 6}{3 + 9 + 6} \\
    &= 1 &  &= \frac{2}{3} 
\end{align*}
which better agrees with the intuition that $\E^{(2)}$ is closer to $\E$. As such, if we use the normalised distance we are likely to select a sample of data points which share things in common, hence providing an underlying signal which our method can uncover. If we instead used the regular distance it is possible we may choose a sample of data which has no such common signal, causing our method to output inferences of little interest. 

\subsection{Multigraph SNF Model}

\label{sec_sup:data_analysis_snf}

Here we provide extra details regarding the generalisation of the SNF models \citep{Lunagomez2021} used in \Cref{sec:data_analysis_graphs}. In particular, we extend the SNF to model multigraphs. Let $\V = \{1,\dots,V\}$ denote the fixed set of vertices, and let $\G  = (\V, \E)$ denote a multigraph (directed or un-directed, and possibly with self-loops), where $\E$ is a \textit{multiset} of edges, so that an edge $(i,j)$ can appear more than once in $\E$. A multigraph $\G$ can also be represented uniquely by its adjacency matrix $A^\G \in \mathbb{Z}_{\geq 0}^{V \times V}$, where $A^\G_{ij} \in \mathbb{Z}_{\geq 0}$ denotes the multiplicity of the edge $(i,j)$ in $\E$. 

To define a model, we place a probability distribution over \textit{all} multigraphs (over the vertex set $\V$). This space, which we denote $\mathscr{G}$, can be defined via the one-to-one correspondence with adjacency matrices as follows
$$\mathscr{G} = \{ \G \, : \, A^\G \in \mathbb{Z}_{\geq 0}^{V \times V}\},$$
so that we seek to assign each $\G \in \mathscr{G}$ a probability. Following the same rationale as the SNF models (and the models of this paper), we construct this model via location and scale. Moreover, this is done with the use of distance metrics, this time between multigraphs. We have two parameters, the mode $\G^m \in \mathscr{G}$ (location) and the dispersion $\gamma>0$ (scale). We also assume that a distance metric has been pre-specified $d_G(\G, \G^\prime)$, quantifying the dissimilarity of any two multigraphs $\G$ and $\G^\prime$. Given this, we assume the probability of $\G \in \mathscr{G}$ is, up to proportionality, the following
\begin{equation}
    \begin{aligned}
    p(\G | \G^m,\gamma) \propto \exp\{ - \gamma \phi(d_G(\G, \G^m))\}
    \end{aligned}
    \label{eq:multigraph_snf}
\end{equation}
where $\phi(\cdot)$ is a non-negative strictly increasing function with $\phi(0)=0$. The notation $\G \sim \text{SNF}(\G^m, \gamma)$ is used when $\G$ is assumed to have been sampled from this probability distribution. The normalising constant of \eqref{eq:multigraph_snf} is given by the following 
\begin{equation*}
    \begin{aligned}
    Z(\G^m, \gamma) = \sum_{\G \in \mathscr{G}} \exp\{ - \gamma \phi(d_G(\G, \G^m))\},
    \end{aligned}
\end{equation*}
which, with $\mathscr{G}$ being an infinite space, will in general be intractable.

Note this is more-or-less identical the SNF models seen in \cite{Lunagomez2021}, Definition 3.4. The only differences being (i) the sample space $\mathscr{G}$ is now all muligraphs over $\V$, and (ii) the distance metrics $d_G(\cdot,\cdot)$ are between multigraphs.

Supposing that a sample of multigraphs $\gdata$ has been observed, as discussed in \Cref{sec:data_analysis_graphs}, we can use this mutligraph-based SNF to construct the following hierarchical model
\begin{equation*}
    \begin{aligned}
    \G^{(i)} &\sim \text{SNF}(\G^m, \gamma) \quad \text{(for $i=1,\dots,n$)}\\
    \G^m &\sim \text{SNF}(\G_0, \gamma_0)\\
    \gamma &\sim p(\gamma)
    \end{aligned}
\end{equation*}
where $\G_0 \in \mathscr{G}$ and $\gamma_0>0$ are hyperparameters, and $p(\gamma)$ denotes a prior for the dispersion. The goal of inference is to now estimate $\G^m$ and $\gamma$, representing notations of average and variance, respectively, and can be achieved by sampling from the posterior via MCMC. The posterior in this case is given by the following 
\begin{equation*}
    \begin{aligned}
        p(\G^m, \gamma | \gdata) &\propto \left(\prod_{i=1}^n p(\G^{(i)}|\G^m, \gamma)\right) p(\G^m)p(\gamma) \\
        &= Z(\G^m, \gamma)^{-n}\exp \left\{ - \gamma \sum_{i=1}^n \phi(d_G(\G^{(i)}, \G^m))\right\} \\
        & \quad\quad \times \exp\left\{ - \gamma_0 \phi(d_G(\G^m, \G_0))\right\} p(\gamma),
    \end{aligned}
\end{equation*}
which, since $Z(\G^m,\gamma)$ is intractable and depends on the parameters being sampled, is doubly-intractable \citep{Murray2006}. As such, to sample from it one must use a specialised MCMC algorithm. Since we are dealing with multigraphs, we cannot apply the scheme proposed by \cite{Lunagomez2021} directly, and instead propose an alternative approach via the exchange algorithm \citep{Murray2006}. In particular, we considered a component-wise MCMC algorithm which alternates between sampling from the two conditionals (i) $p(\gamma | \G^m, \gdata)$, and (ii) $p(\G^m| \gamma, \gdata)$. For (i) we apply the exchange algorithm directly, whilst for (ii) do an exchange-within-Gibbs step, updating each edge in turn in a single repetition. 

We first outline the procedure to update the dispersion. Assume that $q(\gamma'|\gamma)$ denotes a suitable proposal density. With $\G^m$ fixed and current state $\gamma$, first sample proposal $\gamma'$ from $q(\gamma'|\gamma)$. Next, sample auxiliary data $\gadata$ i.i.d. where $\G^*_i \sim \text{SNF}(\G^m, \gamma')$ and then accept $\gamma'$ with the following probability
\begin{equation*}
    \begin{aligned}
        \alpha(\gamma',\gamma) &= \min\left\{1, \frac{p(\gamma'| \G^m, \gdata) \prod_{i=1}^n p(\G^*_i|\G^m, \gamma) q(\gamma|\gamma^\prime)}{p(\gamma| \G^m, \gdata) \prod_{i=1}^n p(\G^*_i|\G^m, \gamma') q(\gamma'|\gamma)} \right\} \\
        &=\min\left\{1, H(\gamma',\gamma) \right\}
    \end{aligned}
\end{equation*}
where 
\begin{equation}
    \begin{aligned}
        H(\gamma',\gamma) &= \exp \Bigg\{-(\gamma'-\gamma) \left( \sum_{i=1}^n \phi(d_G(\G^{(i)},\G^m)) - \sum_{i=1}^n \phi(d_G(\G^*_i, \G^m)) \right) \Bigg\}\\
        &\qquad \times \frac{p(\gamma')}{p(\gamma)} \frac{q(\gamma|\gamma')}{q(\gamma'|\gamma)}.
    \end{aligned}
    \label{eq:snf_disp_update_acc_prob}
\end{equation}
For the proposal $q(\gamma'|\gamma)$ we consider sampling uniformly over a $\varepsilon$-neighbourhood of $\gamma$ with reflection at zero (see \Cref{sec_sup:inference_dispersion_cond}, Eq. \ref{eq:gamma_proposal_density}), for which one has $q(\gamma'|\gamma)=q(\gamma|\gamma')$.

To update the mode, we consider a exchange-within-Gibbs scheme, whereby we scan through all edges, propose new multiplicities and accept these with some probability. Assume one has defined a proposal $q(x^\prime|x)$, which proposes a new edge multiplicity $x^\prime \in \mathbb{Z}_{\geq 0}$ given current value $x \in \mathbb{Z}_{\geq 0}$. With $\gamma$ fixed and current state $\G^m$, with $A^m$ its adjacency matrix (abbreviating notation for readability), we first generate proposal $\G^{m'}$ by proposing a new multiplicity for $(i,j)$. More precisely, letting $x=A^m_{ij}$ denote the current multiplicity, we sample $x'$ from $q(x'| x)$, then construct proposal $\G^{m^\prime}$ via its adjacency matrix $A^{m^\prime}$, defined to be
\begin{equation*}
    \begin{aligned}
    A^{m^\prime}_{kl} = \begin{cases}
        x^\prime & \text{if }(k,l)=(i,j) \\
        A^m_{kl} & \text{else}
    \end{cases}
    \end{aligned}
\end{equation*}
that is, $A^{m^\prime}$ is equal to $A^m$ with the $(ij)$th entry altered from $x$ to $x'$. Note this step will alter if we are considering un-directed graphs, where we must let $A^{m'}_{ij}=A^{m'}_{ij}=x^\prime$, since the adjacency matrices must be symmetric. Here we will assumed graphs to be directed. Given proposal $\G^{m'}$, we next sample auxiliary data $\gadata$ i.i.d. where $\G^*_i \sim \text{SNF}(\G^{m^\prime}, \gamma)$ and then accept $\G^{m'}$ with the following probability 
\begin{equation*}
    \begin{aligned}
        \alpha(\G^{m^\prime}, \G^m) &= \min\left\{1, \frac{p(\G^{m^\prime}| \gamma, \gdata) \prod_{i=1}^n p(\G^*_i|\G^m, \gamma) q(x|x^\prime)}{p(\G^{m}| \gamma, \gdata) \prod_{i=1}^n p(\G^*_i|\G^{m^\prime}, \gamma) q(x^\prime|x)} \right\} \\
        &=\min\left\{1, H(\G^{m^\prime},\G^m) \right\}
    \end{aligned}
\end{equation*}
where 
\begin{equation}
    \begin{aligned}
        H(\G^{m^\prime},\G^m) &= \exp \Bigg\{-\gamma \left( \sum_{i=1}^n \phi(d_G(\G^{(i)},\G^{m'})) - \sum_{i=1}^n \phi(d_G(\G^{(i)}, \G^m)) \right)  \\
        &\qquad -\gamma \left( \sum_{i=1}^n \phi(d_G(\G^*_i,\G^m)) - \sum_{i=1}^n \phi(d_G(\G^*_i, \G^{m'})) \right) \\
        &\qquad \qquad - \gamma_0 \left( \phi(d_G(\G^{m'},\G_0)) - \phi(d_G(\G^m, \G_0))\right)
        \Bigg\} \frac{q(x|x')}{q(x'|x)}.
    \end{aligned}
    \label{eq:snf_mode_update_acc_prob}
\end{equation}
The steps above update the multiplicity of a single edge $(i,j)$. In a single iteration of updating the mode $\G^m$, we consider looping over each $(i,j) \in \V \times \V$, updating their multiplicity in this manner, leading to what can be seen as an exchange-within-Gibbs step for sampling from $p(\G^m|\gamma, \gdata)$. 

For the proposal $q(x'|x)$, we consider uniform sampling over a $\nu$-neighbourhood of $x$ with reflection as zero. More precisely, given current state $x \in \mathbb{Z}_{\geq 0}$, sample proposal $x^\prime$ via 
\begin{enumerate}
    \item Sample $x^* \sim \text{Uniform}(A)$ where 
        $$A = \{j \in \mathbb{Z} \, : \,x-\nu \leq j \leq x+\nu\} \setminus \{x\}$$
        is the $\nu$-neighbourhood of $x$ in $\mathbb{Z}$, excluding $x$, then
    \item If $x^* \geq 0$ let $x^\prime=x^*$, else let $x^\prime=-x^*$,
\end{enumerate}
for which one has
\begin{equation*}
    \begin{aligned}
        q(x^\prime | x) = \begin{cases}
        0  &  \text{if $x=x'$}\\
        \frac{1}{\nu} & \text{if $x+x' \leq \nu$} \\
        \frac{1}{2\nu} & \text{else}
        \end{cases}
    \end{aligned}
\end{equation*}
and hence $q(x'|x)=q(x|x')$, which will lead to cancellation of such terms in \eqref{eq:snf_mode_update_acc_prob}.

Finally, we note that both of these schemes to sample from $p(\G^m|\gamma, \gadata)$ and $p(\gamma|\G^m,\gdata)$ require the ability to obtain an i.i.d. sample $\{\G^*_i\}_{i=1}^n$ where $\G^*_i \sim \text{SNF}(\G^m,\gamma)$ for some given $(\G^m,\gamma)$. Unfortunately, this cannot be done in general. However, we can replace this with approximate MCMC-based samples, exactly as we did for our interaction-sequence and interaction-multiset models (\Cref{sec:inference_model_sampling}). To do so, we re-use the scheme above (without auxiliary sampling). 

In particular, with current state $\G$, we update edge $(i,j)$ as follows. Letting $x = A^\G_{ij}$, we sample $x'$ from $q(x'|x)$ (via $\nu$-neighbourhood as above), constructing proposal $\G'$ via its adjacency matrix 
\begin{equation*}
    \begin{aligned}
        A^{\G'}_{kl} = \begin{cases}
        x' & \text{if $(k,l)=(i,j)$}\\
        A^\G_{kl} & \text{else}
        \end{cases}
    \end{aligned}
\end{equation*}
that is, $\G'$ is equivalent to $\G$ with the multiplicity of edge $(i,j)$ flipped from $x$ to $x'$. We then accept $\G'$ with the following probability 
\begin{equation*}
    \begin{aligned}
        \alpha(\G, \G') &= \min\left\{1, \frac{p(\G'|\G^m,\gamma) q(x|x')}{p(\G|\G^m,\gamma)q(x'|x)} \right\} \\
        &=\min\left\{1, \exp\Big\{ -\gamma \left( \phi(d_G(\G',\G^m)) - \phi(d_G(\G,\G^m)) \right) \Big\}\frac{q(x|x')}{q(x'|x)} \right\}.
    \end{aligned}
\end{equation*}

Note this will update a single edge $(i,j)$. One could now follow the approach used to update the mode $\G^m$, looping over all edges in turn. However, in this case we opt to instead choose a single edge at random to update. That is, in a single iteration, we choose $(i,j)$ uniformly from $\V \times \V$, and update it as above. This can be seen as a Gibbs sampler with a randomised sweep strategy \citep{levine2006optimizing}.


	\newpage
	\begin{algorithm}[ht]
    \setstretch{1.1}
    \SetAlFnt{\sfdefault} 
	\SetAlgoLined
	\DontPrintSemicolon
	\KwIn{observed data $\data$}
	initialise $(\iS_0^m, \gamma_0)$ \;
	\For{$i=1,\dots,m$}{
	    \tcp{Update gamma}
	   $\gamma_i$ = \texttt{dispersion_update}$(\iS^m_{i-1}$, $\gamma_{i-1})$
	   \tcp{(\Cref{alg_sup:sis_dispersion_accept_reject})}
	    \tcp{Update mode}
	    $\iS^m_i$ = \texttt{mode_update($\iS^m_{i-1}, \gamma_{i}$)} \tcp{(\Cref{alg_sup:sis_mode_accept_reject})}
	}
	\KwOut{sample $\{(\iS^m_i, \gamma_i)\}_{i=1}^m$}
	\caption{SIS posterior component-wise MCMC}
	\label{alg_sup:sis_joint_posterior}
\end{algorithm}

\begin{algorithm}[ht]
    \setstretch{1.1}
	\SetAlgoLined
	\DontPrintSemicolon
	\KwIn{$(\iS^m_i,\gamma_i)$}
	\KwOut{$\gamma_{i+1}$}
	\SetKwFunction{FMain}{dispersion_update}
    \SetKwProg{Fn}{Function}{:}{}
	\Fn{\FMain{$\iS^m_i$, $\gamma_i$}}{
	    \vspace{.1cm}
	    let $(\iS^m,\gamma) = (\iS^m_i,\gamma_i)$\;
	    sample $\gamma^\prime$ via $q(\gamma^\prime|\gamma)$ of (\ref{eq:gamma_proposal_density}) \tcp*{Sample proposal}
	    sample $\adata$ i.i.d. from $\text{SIS}(\iS^m, \gamma^\prime)$ \tcp*{Sample auxiliary data}
	    evaluate $\alpha=\alpha(\gamma, \gamma^\prime)$ of (\ref{eq_sup:posterior_cond_dipsersion_acc_prob}) \tcp*{Acceptance probability}
		$\gamma_{i+1} = \begin{cases}
	        \gamma^\prime & \text{with probability } \alpha \\
	        \gamma & \text{with probability } (1-\alpha)
	        \end{cases}$ \;
	   \KwRet $\gamma_{i+1}$ \tcp*{Accept/reject proposal}
	   \vspace{.1cm}
	}
	\textbf{end}
	\caption{SIS posterior dispersion conditional accept-reject}
	\label{alg_sup:sis_dispersion_accept_reject}
\end{algorithm}

\begin{algorithm}[ht]
    \setstretch{1.1}
	\SetAlgoLined
	\DontPrintSemicolon
	\KwIn{$(\iS^m_i,\gamma_i)$}
	\KwOut{$\iS^m_{i+1}$}
	\SetKwFunction{FMain}{mode_update}
    \SetKwProg{Fn}{function}{:}{}
	\Fn{\FMain{$\iS^m_i$, $\gamma_i$}}{
	    \vspace{.1cm}
	    let $(\iS^m,\gamma) = (\iS^m_i,\gamma_i)$\;
	    sample $z \sim \text{Bernoulli}(\beta)$ \;
	    \eIf{z = 1}{
	        \tcp{Edit allocation move} 
	        let $u$, $f(u,\iS^m)$ and $p(u|\iS^m)$ be as in \Cref{sec_sup:inference_edit_alloc}\;
	        sample $u$ via $p(u | \iS^m)$ \tcp*{Sample auxiliary variable}
            $(\modeprop, u^\prime) = f(\iS^m, u)$ \tcp*{Invoke involution}
            sample $\adata$ i.i.d. from $\text{SIS}(\modeprop, \gamma)$ \tcp*{Sample auxiliary data}
            $\alpha=\alpha(\modecurr, \modeprop)$ of \eqref{eq_sup:posterior_cond_mode_acc_prob}, using ratio \eqref{eq_sup:edit_alloc_ratio} \tcp*{Acceptance probability}
	    }{  
	        \tcp{Path insertion \& deletion move}
	        let $u$, $f(u,\iS^m)$ and $p(u|\iS^m)$ be as in \Cref{sec_sup:inference_trans_dim}\;
	        sample $u$ via $p(u | \iS^m)$ \tcp*{Sample auxiliary variable}
            $(\modeprop, u^\prime) = f(\iS^m, u)$ \tcp*{Invoke involution}
    	    sample $\adata$ i.i.d. from $\text{SIS}(\modeprop, \gamma)$ \tcp*{Sample auxiliary data}
    	    $\alpha=\alpha(\modecurr, \modeprop)$ of \eqref{eq_sup:posterior_cond_mode_acc_prob}, using ratio \eqref{eq_sup:trans_dim_ratio}\tcp*{Acceptance probability}
	    }
	    \vspace{.1cm}
		$\iS^m_{i+1} = \begin{cases}
	        \modeprop & \text{with probability } \alpha \\
	        \modecurr & \text{with probability } (1-\alpha)
	        \end{cases}$ \tcp*{Accept/reject proposal}
	   \vspace{.1cm}
	   \KwRet $\iS^m_{i+1}$\;
	   \vspace{.1cm}
	}
	\textbf{end}
	\caption{SIS posterior mode conditional accept-reject}
	\label{alg_sup:sis_mode_accept_reject}
\end{algorithm}

\begin{algorithm}[ht]
    \SetAlFnt{\sfdefault} 
	\SetAlgoLined
	\DontPrintSemicolon
	\KwIn{$(\iS^m, \gamma)$ (model parameters)}
	\KwIn{$\nu_{\text{ed}}$, $\nu_{\text{td}}$, $p(\I|\iS)$, $\beta$ (MCMC tuning parameters)}
	\KwIn{$m$ (sample size), $b$ (burn-in), $l$ (lag)}
	initialise $\iS$;\\
	initialise $i=1$; \\
	\While{$i\leq m$}{
	    sample $z \sim \text{Bernoulli}(\beta)$\\
	    \eIf{z = 1}{
	        \tcp{Edit allocation move}
	        let $u$, $f(u,\iS)$ and $p(u|\iS)$ be as in \Cref{sec_sup:inference_edit_alloc} \;
	        sample $u$ via $p(u|\iS)$ \;
	        $(\iS^\prime,u^\prime) = f(\iS, u)$ \;
	        evaluate $\alpha = \alpha(\iS, \iS^\prime)$ of (\ref{eq_sup:sis_model_sampling_imcmc_acc_prob}) using \eqref{eq_sup:sis_model_sampling_edit_alloc_ratio}\\
	    }{
	        \tcp{Path insertion \& deletion move}
	        let $u$, $f(u,\iS)$ and $p(u|\iS)$ be as in \Cref{sec_sup:inference_trans_dim} \;
	        sample $u$ via $p(u|\iS)$ \;
	        $(\iS^\prime,u^\prime) = f(\iS, u)$ \;
	        evaluate $\alpha = \alpha(\iS, \iS^\prime)$ of (\ref{eq_sup:sis_model_sampling_imcmc_acc_prob}) using \eqref{eq_sup:sis_model_sampling_trans_dim_ratio} \\
	    }
	    \tcp{Accept/reject proposal} 
	    $\iS = \begin{cases}
	        \iS^\prime & \text{with probability } \alpha \\
	        \iS & \text{with probability } (1-\alpha)
	        \end{cases}$\\
	   \tcp{Store sample (accounting for lag and burn-in)}
	   \If{$(i>b)$ \text{ and } $(i\mod{l}=1)$}{
	        $\iS_i \leftarrow \iS$ \\
	        $i =  i + 1$  \\
	    }
	}
	\KwOut{$\{\iS_i\}_{i=1}^m$}
	\caption{SIS model iMCMC sampling}
	\label{alg_sup:SIS_model_sampling}
\end{algorithm}

\textbf{\begin{algorithm}[ht]
    \setstretch{1.1}
    \SetAlFnt{\sfdefault} 
	\SetAlgoLined
	\DontPrintSemicolon
	\KwIn{observed data $\edata$}
	initialise $(\E_0^m, \gamma_0)$ \;
	\For{$i=1,\dots,m$}{
	    \tcp{Update gamma}
	   $\gamma_i$ = \texttt{dispersion_update}$(\E^m_{i-1}$, $\gamma_{i-1})$
	   \tcp{(\Cref{alg_sup:sim_dispersion_accept_reject})}
	    \tcp{Update mode}
	    $\E^m_i$ = \texttt{mode_update($\E^m_{i-1}, \gamma_{i}$)} \tcp{(\Cref{alg_sup:sis_mode_accept_reject})}
	}
	\KwOut{sample $\{(\E^m_i, \gamma_i)\}_{i=1}^m$}
	\caption{SIM posterior component-wise MCMC}
	\label{alg_sup:sim_joint_posterior}
\end{algorithm}
}

\begin{algorithm}[ht]
    \setstretch{1.1}
	\SetAlgoLined
	\DontPrintSemicolon
	\KwIn{$(\iS^m_i,\gamma_i)$}
	\KwOut{$\gamma_{i+1}$}
	\SetKwFunction{FMain}{dispersion_update}
    \SetKwProg{Fn}{Function}{:}{}
	\Fn{\FMain{$\E^m_i$, $\gamma_i$}}{
	    \vspace{.1cm}
	    let $(\E^m,\gamma) = (\E^m_i,\gamma_i)$\;
	    sample $\gamma^\prime$ via $q(\gamma^\prime|\gamma)$ of (\ref{eq:gamma_proposal_density}) \tcp*{Sample proposal}
	    sample $\eadata$ i.i.d. from $\text{SIM}(\E^m, \gamma^\prime)$ \tcp*{Sample auxiliary data}
	    evaluate $\alpha=\alpha(\gamma, \gamma^\prime)$ of (\ref{eq_sup:sim_posterior_cond_dispersion_acc_prob}) \tcp*{Acceptance probability}
		$\gamma_{i+1} = \begin{cases}
	        \gamma^\prime & \text{with probability } \alpha \\
	        \gamma & \text{with probability } (1-\alpha)
	        \end{cases}$ \;
	   \KwRet $\gamma_{i+1}$ \tcp*{Accept/reject proposal}
	   \vspace{.1cm}
	}
	\textbf{end}
	\caption{SIM posterior dispersion conditional accept-reject}
	\label{alg_sup:sim_dispersion_accept_reject}
\end{algorithm}

\begin{algorithm}[ht]
    \setstretch{1.1}
	\SetAlgoLined
	\DontPrintSemicolon
	\KwIn{$(\E^m_i,\gamma_i)$}
	\KwOut{$\E^m_{i+1}$}
	\SetKwFunction{FMain}{mode_update}
    \SetKwProg{Fn}{function}{:}{}
	\Fn{\FMain{$\E^m_i$, $\gamma_i$}}{
	    \vspace{.1cm}
	    let $(\E^m,\gamma) = (\E^m_i,\gamma_i)$\;
	    obtain $\iS^m$ from $\E^m$ \tcp*{Place paths in arbitrary order}
	    sample $z \sim \text{Bernoulli}(\beta)$ \;
	    \eIf{z = 1}{
	        \tcp{Edit allocation move} 
	        let $u$, $f(u,\iS^m)$ and $p(u|\iS^m)$ be as in \Cref{sec_sup:inference_edit_alloc}\;
	        sample $u$ via $p(u | \iS^m)$ \tcp*{Sample auxiliary variable}
            $(\modeprop, u^\prime) = f(\iS^m, u)$ \tcp*{Invoke involution}
            obtain $\emodeprop$ from $\modeprop$ \tcp*{Disregard order}
            sample $\eadata$ i.i.d. from $\text{SIM}(\emodeprop, \gamma)$ \tcp*{Sample auxiliary data}
            $\alpha=\alpha(\emodecurr, \emodeprop)$ of \eqref{eq_sup:sim_posterior_cond_mode_acc_prob}, using ratio \eqref{eq_sup:edit_alloc_ratio} \tcp*{Acceptance probability}
	    }{  
	        \tcp{Path insertion \& deletion move}
	        let $u$, $f(u,\iS^m)$ and $p(u|\iS^m)$ be as in \Cref{sec_sup:inference_trans_dim}\;
            sample $u$ via $p(u | \iS^m)$ \tcp*{Sample auxiliary variable}
            $(\modeprop, u^\prime) = f(\iS^m, u)$ \tcp*{Invoke involution}
            obtain $\emodeprop$ from $\modeprop$ \tcp*{Disregard order}
            sample $\eadata$ i.i.d. from $\text{SIM}(\emodeprop, \gamma)$ \tcp*{Sample auxiliary data}
    	    $\alpha=\alpha(\emodecurr, \emodeprop)$ of \eqref{eq_sup:sim_posterior_cond_mode_acc_prob}, using ratio \eqref{eq_sup:trans_dim_ratio}\tcp*{Acceptance probability}
	    }
	    \vspace{.1cm}
		$\E^m_{i+1} = \begin{cases}
	        \emodeprop & \text{with probability } \alpha \\
	        \emodecurr & \text{with probability } (1-\alpha)
	        \end{cases}$ \tcp*{Accept/reject proposal}
	   \vspace{.1cm}
	   \KwRet $\E^m_{i+1}$\;
	   \vspace{.1cm}
	}
	\textbf{end}
	\caption{SIM posterior mode conditional accept-reject}
	\label{alg_sup:sim_mode_accept_reject}
\end{algorithm}

\begin{algorithm}[ht]
    \SetAlFnt{\sfdefault} 
	\SetAlgoLined
	\DontPrintSemicolon
	\KwIn{$(\E^m, \gamma)$ (model parameters)}
	\KwIn{$\nu_{\text{ed}}$, $\nu_{\text{td}}$, $p(\I|\iS)$, $\beta$ (MCMC tuning parameters)}
	\KwIn{$m$ (sample size), $b$ (burn-in), $l$ (lag)}
	initialise $\E$;\\
	initialise $i=1$; \\
	\While{$i\leq m$}{
	    obtain $\iS$ from $\E$ \tcp*{Place paths in arbitrary order}
	    sample $z \sim \text{Bernoulli}(\beta)$\\
	    \eIf{z = 1}{
	        \tcp{Edit allocation move}
	        let $u$, $f(u,\iS)$ and $p(u|\iS)$ be as in \Cref{sec_sup:inference_edit_alloc} \;
	        sample $u$ via $p(u|\iS)$ \tcp*{Sample auxiliary variable}
	        $(\iS^\prime,u^\prime) = f(\iS, u)$ \tcp*{Invoke involution}
	        obtain $\E^\prime$ from $\iS^\prime$ \tcp*{Disregard order}
	        $\alpha = \alpha(\E, \E^\prime)$ of (\ref{eq_sup:sim_model_sampling_acc_prob}) using \eqref{eq_sup:sis_model_sampling_edit_alloc_ratio}\tcp*{Acceptance probability}
	    }{
	        \tcp{Path insertion \& deletion move}
	        let $u$, $f(u,\iS)$ and $p(u|\iS)$ be as in \Cref{sec_sup:inference_trans_dim} \;
	        sample $u$ via $p(u|\iS)$ \tcp*{Sample auxiliary variable}
	        $(\iS^\prime,u^\prime) = f(\iS, u)$ \tcp*{Invoke involution}
	        obtain $\E^\prime$ from $\iS^\prime$ \tcp*{Disregard order}
	        $\alpha = \alpha(\E, \E^\prime)$ of  (\ref{eq_sup:sim_model_sampling_acc_prob}) using \eqref{eq_sup:sis_model_sampling_trans_dim_ratio} \tcp*{Acceptance probability}
	    }
	    \tcp{Accept/reject proposal} 
	    $\E = \begin{cases}
	        \E^\prime & \text{with probability } \alpha \\
	        \E & \text{with probability } (1-\alpha)
	        \end{cases}$\\
	   \tcp{Store sample (accounting for lag and burn-in)} 
	   \If{$(i>b)$ \text{ and } $(i\mod{l}=1)$}{
	        $\E_i \leftarrow \E$ \\
	        $i =  i + 1$ 
	    }
	}
	\KwOut{$\{\E_i\}_{i=1}^m$}
	\caption{SIM model iMCMC sampling}
	\label{alg_sup:SIM_model_sampling}
\end{algorithm}

\end{appendices}

\end{document}